\documentclass[onecolumn,amsmath,amssymb,nofootinbib,12pt]{article}
 
 \ifx\pdfoutput\not\undefined
 \pdfoutput=1
\fi
\usepackage{jheppub}
\usepackage{ifpdf}
\usepackage{mdframed}

\usepackage{graphicx,subcaption,comment}
\graphicspath{ {figures/} }
\usepackage{mathtools}
\usepackage{epsfig}
\usepackage{pdfpages}
\usepackage{bbm}
\usepackage{graphicx,epstopdf}
\usepackage[numbers,sort&compress]{natbib}
\usepackage[makeroom]{cancel}
\usepackage{hyperref}
\hypersetup{pdftex,colorlinks=true,allcolors=blue}
\usepackage{array}
\usepackage[export]{adjustbox}

\usepackage[normalem]{ulem}

\usepackage[utf8]{inputenc}

\usepackage{tocloft}
\usepackage{bbold}
\usepackage{tikz}
\usetikzlibrary{backgrounds}
\usetikzlibrary{patterns}
\usetikzlibrary{arrows.meta}
\usetikzlibrary{shapes,snakes}
\tikzstyle{bag} = [align=center]
\usetikzlibrary{decorations.pathmorphing}
\def\bea{\begin{eqnarray}}
\def\eea{\end{eqnarray}}

\usepackage{hyperref}
\usepackage{subcaption}

\newcommand{\eg}{{\it e.g.,}\ }
\newcommand{\ie}{{\it i.e.,}\ }
\newcommand{\mt}[1]{\textrm{\tiny #1}}
\newcommand{\reef}[1]{(\ref{#1})}

\newcommand{\ruv}[1]{r_{\mt{UV}}}
\newcommand{\thbdy}{\theta_\mt{bdy}}

 \newcommand{\badat}{\begin{alignedat}}
 \newcommand{\eadat}{\end{alignedat}}
 
 \def\be{\begin{equation}}
\def\ee{\end{equation}}

\newcommand{\rstflat}{r_b}

\usepackage{xcolor}

\newcommand{\pink}[1]{\textcolor{\pink}{#1}}

\definecolor{dblue}{rgb}{0.2,0.50,0.80}

\def\rhoo{P} 
\def\ruv{r_{\rm uv}}       

\DeclareFontFamily{OT1}{pzc}{}
\DeclareFontShape{OT1}{pzc}{m}{it}{<-> s * [1.10] pzcmi7t}{}
\DeclareMathAlphabet{\mathpzc}{OT1}{pzc}{m}{it}

\definecolor{vert}{rgb}{0.1367 0.543 0.1367}

\title{Flat Space Entanglement: A Coulomb Branch Perspective}

\author[a,b]{Eivind J\o rstad,}
\author[a]{Robert C. Myers,}
\author[a]{and Sabrina Pasterski} 

\affiliation[a]{Perimeter Institute for Theoretical Physics, Waterloo, ON N2L 2Y5, Canada}
\affiliation[b]{Dept.~of Physics $\&$ Astronomy, University of Waterloo, Waterloo, ON N2L 3G1, Canada}

\emailAdd{ejorstad@perimeterinstitute.ca}
\emailAdd{rmyers.perimeter@gmail.com}
\emailAdd{spasterski@perimeterinstitute.ca}

\date{\today}

\abstract{We study holographic entanglement entropy in Coulomb-branch solutions describing spherical shells of D$p$-branes. The corresponding throat geometries contain a flat-space bubble in the infrared region, providing a concrete top-down framework for exploring holographic entanglement of flat space. We find that the flat-space region is associated with a reduction of entanglement and of the effective infrared degrees of freedom in the dual boundary state relative to the standard vacuum. We also examine internal RT surfaces and holographic complexity, and show that they exhibit similar qualitative behaviour. Finally, we comment on the broader implications of our results for flat space holography.}

\begin{document}

\maketitle


\section{Introduction}\label{sec:intro}

The AdS/CFT correspondence has revealed surprising connections between quantum entanglement and spacetime geometry through the Ryu-Takayanagi (RT) formula \cite{Ryu:2006bv,Ryu:2006ef,Rangamani:2016dms} and its covariant and quantum generalizations \cite{Hubeny:2007xt,Faulkner:2010jy,Engelhardt:2014gca}. 
This has led to interesting new insights, \eg the island formula yielding the Page curve for black hole evaporation \cite{Almheiri:2019psf,Penington:2019npb} or the interpretation of the holography in terms of quantum error correcting codes \cite{Almheiri:2014lwa,Pastawski:2015qua}.
In its simplest form, the formula states that the von Neumann entropy of a boundary region $A$ is dual to the area of a codimension-two bulk surface $\gamma$ that is homologous to the boundary region and minimizes the area,
\begin{equation}
    S(A)= \min_{\gamma\sim A} \frac{Area(\gamma)}{4 G_N}\,.
    \label{eq:rtpre}
\end{equation}
Interestingly, one can arrive at this result through a gravitational path integral argument that makes no explicit reference to the cosmological constant or the asymptotic structure of the underlying spacetime \cite{Lewkowycz:2013nqa}. Therefore, it is natural to ask if the RT prescription can teach us something about the holographic description of asymptotically flat spacetimes.

Answering this question directly is challenging. For one, it has so far been difficult to construct top-down realizations of a flat space hologram (although, see \cite{Costello:2022jpg,Costello:2023hmi}). This has led to alternative bottom-up Carrollian~\cite{Duval:2014uva,Bagchi:2016bcd,Ciambelli:2018wre,Ciambelli:2018ojf,Bagchi:2019xfx,Bagchi:2019clu} and celestial~\cite{Strominger:2017zoo,Raclariu:2021zjz,Pasterski:2021rjz,Pasterski:2021raf,Pasterski:2023ikd} approaches to flat space holography. Within these frameworks, entanglement entropy is currently less well understood than in, \eg conventional CFTs, so the boundary interpretation of the area of the flat-space RT surfaces is not as clear; see \cite{Capone:2024oim,Apolo:2020bld,Apolo:2020qjm,Jiang:2017ecm,cmp1,Setare:2021auh} for related work. Furthermore, if one examines minimal surfaces in asymptotically flat geometries, one finds behavior that differs from AdS in puzzling ways. For example, one is not free to choose the boundary regions to which one assigns an entanglement entropy, since the surfaces generally anchor on the equator of the sphere at infinity \cite{Ghosh:2023kuy,flatstuff}. Instead, a cutoff seems to be required in order to freely specify boundary subregions. One also finds that the area diverges with a volume law rather than an area law, which, it has been noted, matches the behavior of certain nonlocal theories~\cite{Li:2010dr}. 

Here we take a different approach, using D$p$-brane holography as a controlled top-down framework for studying RT surfaces associated with flat space regions. Rather than working directly with asymptotically flat spacetimes, we use the RT prescription~\eqref{eq:rtpre} to probe a region of flat space embedded inside a D$p$-brane throat, where the holographic interpretation is inherited from the corresponding worldvolume theory. Specifically, we consider multi-center geometries sourced by D$p$-branes smeared over a sphere, such that the full back-reacted geometry contains a flat Minkowski region in the interior while coinciding with the usual D$p$-brane throat in the exterior. These solutions were examined for $p=3$ in \cite{Kraus:1998hv,Giddings:1999zu}. The bulk geometries are dual to Coulomb branch states of the boundary theory, in which the scalar fields acquire nonzero vacuum expectation values. In this way we obtain a controlled realization of flat-space RT physics equipped with a physical regulator: the brane shell provides a timelike cutoff surface for the flat space region, while the exterior throat translates data at this cutoff to observables in the dual worldvolume theory.

It will be important for us not to restrict attention to the conformal D$3$-brane case, but rather to consider general D$p$-branes. For $p\neq 3$, the corresponding holographic dualities involve nonconformal boundary theories and non-AdS bulk geometries~\cite{Itzhaki:1998dd}. Nevertheless, one can still construct spherical brane-shell geometries with a Coulomb-branch interpretation. The reason to include $p\neq 3$ is that the proper radius of the shell becomes a freely adjustable parameter, allowing us to vary the size of the flat space region. In some situations, this freedom lets us control when surfaces that enter the bubble become the dominant saddles compared to those that remain in the exterior throat. This control is unavailable in the conformal case, where the proper size of the flat space bubble is fixed, as explained below.

These shell geometries provide a top-down setting in which the rules for evaluating and interpreting holographic entanglement entropy, as well as holographic complexity, are concrete and under control. By examining extremal surfaces that enter the flat-space region, we can begin to explore holographic entanglement in flat space from this perspective. We also use holographic entanglement entropy to construct entropic c-functions~\cite{Myers:2012ed,Liu:2013una,Liu:2012eea}, which provide a useful measure of the effective number of degrees of freedom at the scale set by the size of the boundary region. As we will see, these probes reveal a depletion of infrared degrees of freedom associated with the flat space bubble. Readers interested primarily in our main results, and in what they suggest for flat space holography, may skip ahead to the discussion in section~\ref{sec:disc}.

The remainder of the paper is organized as follows:
In section~\ref{sec:start}, we make some preliminary remarks about the puzzling features of the RT prescription in asymptotically flat geometries and about D$p$-brane holography. In section~\ref{sec:EE}, we use the RT prescription to examine the entanglement entropies of strip and ball-shaped boundary-regions in the shell geometries. In section~\ref{sec:target} we examine internal RT surfaces, which are proposed to be the holographic duals of target space entanglement entropy. We then briefly comment on holographic complexity in section~\ref{sec:complex} before closing with a discussion and future outlook in section~\ref{sec:disc}. Appendix~\ref{app:bc} contains some details about the boundary conditions used to numerically integrate the minimal surface equations for the ball-shaped boundary-regions. Appendix~\ref{app:perturb} contains a perturbative calculation of the entanglement of a large radius ball region. Appendix~\ref{sec:mon} contains details about the monotonicity of the area of internal RT surfaces. Appendix~\ref{sec:particle} provides some intuition for the asymptotic behavior of internal RT surfaces by treating the minimal surface equation as a particle in a potential.

\section{Preliminary Remarks} \label{sec:start}

Here we begin with a few preliminary remarks. First, we develop some simple intuition for flat-space holography by examining RT surfaces in flat space. We then recall the basic features of D$p$-brane holography and introduce the shell geometries that describe a flat-space bubble deep in the D$p$-brane throat geometry.

\subsection{RT surfaces in flat space}\label{sec:botup}

We begin by examining holographic entanglement entropy in asymptotically flat space through a simple bottom-up picture, which illustrates some of the difficulties one encounters \cite{Robtalk}. For simplicity, we consider ($d$+1)-dimensional Minkowski space $\mathbf{R}^{1,d}$ with the metric written in polar coordinates,\footnote{Notice our slightly unusual choice of coordinates on the boundary $S^{d-1}$ where the equator lies at $\theta=0$ and the poles are at $\theta=\pm\pi/2$.} 
\begin{equation}
    ds^2=-dt^2+dr^2+r^2\left(d\theta^2+\cos^2\theta\, d\Omega_{d-2}^2\right)\,.
    \label{eq:flatmetric}
\end{equation}
A natural expectation is that the dual theory should live on the conformal boundary of the spacetime,
which includes future and past null infinity, as well as timelike and spacelike infinity.
However, following the usual lessons of AdS/CFT, it is useful to introduce a regulator, by placing a cutoff surface at some large radius, $r=r_{\rm uv}$. This defines a
$d$-dimensional holographic screen with topology $\mathbf{R}\times S^{d-1}$, which
approaches the conformal boundary as $r_{\rm uv}\to\infty$.

We can then ask what the analogue of the Ryu--Takayanagi (RT) prescription \cite{Ryu:2006bv,Ryu:2006ef} would give for a subregion on this regulated screen. We emphasize that
the derivation of the RT formula in ref.~\cite{Lewkowycz:2013nqa} was not intrinsically tied
to a negative cosmological constant or the AdS/CFT correspondence.  However, recall our intuition in the latter case: We can consider any boundary region $A$ bounded by the entangling surface $\Sigma$ and the corresponding extremal surface is pulled into the bulk by the gravitational potential created by the AdS geometry. The leading contribution to the holographic entanglement entropy comes from area associated with the near-boundary region and produces the familiar area-law divergence
\begin{equation}
    S(A)\sim c_T\,{ {\cal A}_{\Sigma} \over \delta^{d-2}}+\cdots\,.
    \label{eq:arellaw}
\end{equation}
Here ${\cal A}_{\Sigma}$ is the area of the entangling surface measured in the metric of the boundary theory and we have used the AdS radius $L$ in two important ways: first it defines the short-distance cutoff in the boundary CFT with $\delta=L^2/\ruv$ and it fixes the central charge counting the number of degrees of freedom,
$c_T\sim L^{d-1}/G_{10}$. We also note that generally the ellipsis in eq.~\eqref{eq:arellaw} contains lower order divergences as well as interesting universal contributions.

We now see that an analogous exercise in flat space gives a very different result. Let $A$ be a spherical cap on the cutoff surface, at fixed time, bounded by the ($d-2$)-sphere $\Sigma$ at $\theta=\theta_0$. It is straightforward to determine that the extremal bulk surface ending on $\Sigma$ is just the flat hyperplane. That is, there is no gravitational potential in flat space to pull the RT surfaces in towards the center of the geometry. Hence we may describe the surface as
\begin{equation}
    r\sin\theta=r_{min}\equiv r_{\rm uv}\,\sin\theta_0\,.
    \label{hyperplane}
\end{equation} 
Evaluating the corresponding area, one finds
as
\begin{equation}
    S(A)={A_V\over 4G_N}
      = \frac{\Omega_{d-2}}{4(d-1)G_N}\left(r_{\rm uv}\cos\theta_0\right)^{d-1}\,,
    \label{eq:flatRTvolume}
\end{equation}
where $\Omega_{d-2}=2\pi^{\frac{d-1}{2}}/\Gamma(\frac{d-1}{2})$ is the volume of the round unit sphere $S^{d-2}$. 

Up to order-one angular factors, this result~\eqref{eq:flatRTvolume} is  proportional to the
volume of the cap on the cutoff surface,
\begin{equation}
    S(A)\sim {r_{\rm uv}^{d-1}\,\Omega_A\over G_N}\,,
    \label{eq:vollaw}
\end{equation}
where $\Omega_A$ is the solid angle filled by the cap.\footnote{The area of the entangling surface would be ${\rm Vol}_\Sigma =\Omega_{d-2} (\ruv \cos\theta_0)^{d-2}$. Further, we note that this relation becomes precise for a small cap, with $\delta\theta=\pi/2-\theta_0\ll 1$. The previous expression~\eqref{eq:flatRTvolume} approaches $S(A)\simeq\frac{\Omega_{d-2}}{4(d-1)G_N}\left(r_{\rm uv}\delta\theta\right)^{d-1}= \frac{r_{\rm uv}^{d-1}\Omega_{A}}{4(d-1)G_N}$.}
Hence, the leading term is not an area-law divergence localized near the entangling
surface $\Sigma$, but rather we find a volume-law divergence extensive in the size of the
regulated boundary region. This is the first indication that a naive holographic dual
living on the flat-space screen would not behave like an ordinary local quantum field
theory. Rather, if such a description exists, its entanglement structure suggests that the dual theory must be intrinsically nonlocal \cite{Li:2010dr,Shiba:2013jja}. Another possible interpretation is that the flat-space vacuum is
represented by a highly excited state of the putative boundary theory, rather than by
a conventional vacuum state \cite{Li:2010dr}.

There is a second related puzzle. As noted above, in AdS/CFT, the AdS scale $L$ provides both a 
geometric length scale that plays an essential role in defining the short-distance cutoff $\delta$ and the central charge $c_T$ in the boundary theory. 
In contrast, asymptotically flat space provides no additional scale to play this role.
We might artificially introduce a macroscopic length $\ell$ to write
\begin{equation}
   \ruv \sim {\ell^2\over \delta}\,,\qquad
    c_{\rm flat}\sim {\ell^{d-1}\over G_N}\,,
\end{equation}
These definitions are guided by dimensional analysis, but remain somewhat arbitrary. The expression~\eqref{eq:vollaw} for the holographic entanglement entropy then becomes
\begin{equation}
    S(A)\sim c_{\rm flat}\,{V_A'\over \delta^{d-1}}\,,
\end{equation}
where $V_A'=\ell^{d-1}\,\Omega_A$ is the rescaled volume associated with the cap. This
looks like a volume-law divergence in a boundary theory, but the boundary volume here contains the same auxiliary scale $\ell$ used to define $\delta$ and $c_{\rm flat}$. Thus, the interpretation of the leading RT divergence from the perspective of a dual boundary theory remains somewhat perplexing.

Another challenge concerns what such observables actually probe in the bulk. If we
keep the opening angle $\theta_0$ fixed while taking $r_{\rm UV}\to\infty$, then the
corresponding flat extremal surface only reaches
\begin{equation}
    r_{\rm min}=\ruv \sin\theta_0\to\infty\, .
\end{equation}
Thus these entanglement entropies are not efficient probes of the deep interior of
flat space. Instead, information about the IR region near the origin would be encoded
only in a highly restricted set of boundary regions with
\begin{equation}
    \theta_0\sim {1\over \ruv}\,,
\end{equation}
or more loosely, in entanglement entropies with $\theta_0\sim0$ as the cutoff is
removed. This is another way in which the naive application of the RT prescription to flat space differs sharply from
the familiar AdS setting. In the latter, the entanglement of fixed-size angular regions on the boundary screen is typically dominated by RT surfaces probing the central IR region, rather than remaining in the large-radius UV region.

Further, we note that an additional puzzle emerges for asymptotically flat black holes, where one finds the holographic entanglement entropy contains state-dependent divergences, \ie divergent contributions proportional to the black hole mass \cite{flatstuff}.
Taken together, these observations show that the simple bottom-up description of holographic entanglement entropy in asymptotically flat spaces is
confusing. For example, the RT calculation gives a volume law rather than an area law and the
identification of a boundary cutoff requires a new arbitrary scale.
Rather than resolving these issues in a purely bottom-up framework, we will use the
Coulomb-branch shell geometries below as a controlled top-down arena in which we can ask
more sharply what RT surfaces (and  holographic probes, more generally) can teach us
about the flat space hologram.

\subsection{D$p$-brane holography and shell geometries} \label{sec:intro}

Multi-center supergravity solutions describing general D$p$-brane configurations are well known, \eg \cite{Stelle:1996tz,Stelle:1998xg} and, by examining the throat region of a large number of coincident D$p$-branes, one obtains a holographic description of the worldvolume  theory, at least within certain regimes of validity \cite{Itzhaki:1998dd}. This construction can be straightforwardly extended to Coulomb branch solutions describing a spherical shell of D$p$-branes, which create a flat-space bubble in the infrared region. As D$p$-brane holography is somewhat less familiar than the standard AdS/CFT correspondence, we provide a brief review of D$p$-brane holography \cite{Itzhaki:1998dd}, as well as the corresponding shell geometries. While the D3-brane case (\ie AdS$_5\times S^5$) remains our benchmark, an important feature of the nonconformal $p\ne3$ cases is that the size of the flat-space bubble becomes a freely adjustable parameter.

 Recall that configurations of parallel D$p$-branes give rise to so-called multi-center solutions of the supergravity field equations, which are parametrized by a harmonic function $H$. The metric takes the following form in Einstein frame,\footnote{Recall in the Einstein frame, the metric is rescaled by an appropriate power of $e^{\phi}$ so that the Einstein-Hilbert term in the low-energy effective action appears without any dilaton prefactor. In contrast, in the string frame the Einstein-Hilbert term carries an overall $e^{-2\phi}$ factor and the corresponding metric is the one that appears naturally in the two-dimensional string worldsheet action.}
\begin{equation}
    ds_E^2 = H^{\frac{p-7}{8}}\left(-dt^2 + dx_{\parallel}^2\right)+H^{\frac{p+1}{8}}\left(dr^2 + r^2 d\Omega^2_{8-p} \right)\,,
    \label{eq:Dp-metric}
\end{equation}
 while the dilaton is given by
\begin{equation}
    e^{\phi} = g_s H^{(3-p)/4}\,.
    \label{eq:dilaton}
\end{equation}
Of course, the solution is also supported by an RR flux  but it will not play a role in our discussion. The harmonic function $H$ is obtained by solving Laplace's equation in the $9-p$ directions orthogonal to the branes, with the D$p$-branes acting as point sources. If we consider $p\le 6$, the result for a single source of $N$ coincident branes gives
\begin{equation}
    H(r)=1+\left( \frac{r_p}{r}\right)^{7-p}\,.
    \label{eq:stack-Dp-metric}
\end{equation}
Here the ``1" is an integration constant chosen to ensure that the solution~\eqref{eq:Dp-metric} approaches flat ten-dimensional space asymptotically.
The scale $r_p$ is given in terms of the  underlying parameters by (\eg see \cite{Kanitscheider:2008kd})
\begin{equation}
r_p^{7-p}=2^{5-p}\, \pi^{\frac{5-p}2}\, \Gamma\!\left(\tfrac{7-p}{2}\right)\, g_s\, N\, \ell_s^{7-p}\,,
    \label{eq:parameterfdef}
\end{equation}
where $g_s$ and $\ell_s$ are the string coupling and string length, respectively. 

One can then isolate the near-horizon physics of this configuration
\eqref{eq:stack-Dp-metric} to obtain the holographic dualities of
\cite{Itzhaki:1998dd}. Following \cite{Itzhaki:1998dd,Kanitscheider:2008kd},
this limit is most usefully characterized by focusing on the dimensionful Yang--Mills coupling of the worldvolume theory\footnote{One might note that the Yang-Mills coupling is renormalizable for \(p<3\),
marginal for \(p=3\), and non-renormalizable for \(p>3\).}
\begin{equation}
    g_{YM}^2= (2\pi)^{p-2}\, g_s\,  \ell_s{}^{\!\!p-3}\,,
    \label{eq:gym2}
\end{equation}
along with energies measured
relative to the string scale. Along these lines, we may introduce the energy variable
\begin{equation}
    E_\mt{string} \equiv \frac{r}{\ell_s^2}\,,
    \label{eq:Estring}
\end{equation}
where the subscript indicates that this can be identified with the energy of an open string stretching from the origin to a D$p$-brane at radius $r$. The limit is arranged so that the dynamics in the throat decouples from the
asymptotically flat region, while the worldvolume gauge theory parameters are
kept fixed in the appropriate units. In practice, this amounts to dropping the
constant term in eq.~\eqref{eq:stack-Dp-metric}, leaving
\begin{equation}
    H(r)=\left( \frac{r_p}{r}\right)^{7-p}\,.
    \label{eq:throat-Dp-metric}
\end{equation}
For \(p=3\), this reproduces the AdS\(_5\times S^5\) geometry, with curvature
scale $r_{p=3}=\left(4 \pi g_s N \right)^{1/4}\ell_s=L$,
and the dual boundary theory being four-dimensional \({\cal N}=4\) super-Yang--Mills theory. For \(p\neq 3\), the corresponding holographic description
relates the throat geometry of the D\(p\)-branes to the \((p+1)\)-dimensional
worldvolume theory, namely maximally supersymmetric Yang--Mills theory, which
is nonconformal.

In order for the supergravity approximation to be valid in the throat geometry, we require that both quantum corrections and string corrections are suppressed. For the former, we need the (local) string coupling~\eqref{eq:dilaton} to be small, which requires 
\begin{equation}
\begin{rcases}
    r\gg \hat{r}_p&\ \ {\rm for}\  p<3\ \\
    r\ll\hat{r}_p&\ \ {\rm for}\ p>3\ 
\end{rcases} 
\quad {\rm where}\ \   \hat{r}_p^{3-p}\equiv  g_s^{\frac{4}{7-p}}\,r_p^{3-p}\,. 
\label{eq:weakdilaton}
\end{equation}
As an indication of when higher-derivative stringy corrections are small, we examine the Ricci scalar in the string frame and require \cite{Itzhaki:1998dd}
\begin{equation}
     {R}_\mt{string} \simeq \sqrt{\frac{r^{3-p}}{g^2_{YM}N}} \frac{1}{\ell_s^{5-p}}\ll \frac{1}{\ell_s^2}\, .
\end{equation}
This small curvature condition may be recast as
\begin{equation}
\begin{rcases}
    r\ll \tilde{r}_p&\ \ {\rm for}\  p<3\ \\
    r\gg \tilde{r}_p&\ \ {\rm for}\ p>3\ 
\end{rcases}
\quad {\rm where}\ \  \tilde{r}_p^{3-p} \equiv g^2_{YM}N\,\ell_s^{2(3-p)} \propto \left(\frac{r_p}{\ell_s}\right)^4\,r_p^{3-p}\, .
\label{eq:weakcurv}
\end{equation}
For all of these cases, $g_s\ll1$ along with $r_p/\ell_s\gg 1$ ensures that there is a large radial range in which the supergravity provides an good approximation of the throat physics. Of course, although the above analysis no longer holds for $p=3$, the same restrictions apply in this special case. In the boundary theory, these conditions may be written as a restriction on the effective 't Hooft coupling \cite{Itzhaki:1998dd},
\begin{equation}
    1 \ll g_{eff}^2\equiv g_{YM}^2 N E_\mt{string}^{p-3} \ll N^{\frac{4}{7-p}}\,.
    \label{eq:safe}
\end{equation}

Following \cite{Peet:1998wn}, one can also consider the energy of supergravity excitations at a given radius
\begin{equation}
E_{\rm sugra}
= \frac{\sqrt{\Gamma\left(\frac{7-p}2\right)}}{2^{p-3}\, \pi^{\frac{3(p-3)}4}}\,\left(\frac{r}{r_p}\right)^{(5-p)/2}\frac1{r_p}\,.
\label{Esugra}
\end{equation}
These excitations would be associated with simple operators in the low energy spectrum of the worldvolume gauge theory.\footnote{As expected, eq.~\eqref{Esugra} reduces to $E_\mt{sugra}=r/L^2$ with $p=3$. Further, note the same parametric scaling appears in the temperature of a D$p$-brane black hole: $T=\frac{7-p}{4\pi}\,
{r_H^{(5-p)/2}}/{r_p^{(7-p)/2}}$ \cite{Horowitz:1991cd,Klebanov:1996un}.} With this expression \eqref{Esugra}, we can define 
\begin{equation}
  \hat{g}_{eff}^2\equiv g_{YM}^2 N E_\mt{sugra}^{p-3} \,,
   \label{eq:sugra}
\end{equation} 
where we have distinguished this effective coupling from that in eq.~\eqref{eq:safe}. Although both expressions have the form $g_{YM}^2 N E^{p-3}$, we distinguish them because they are implicitly characterized by a different set of degrees of freedom associated with a particular radius in the bulk. Hence the corresponding physical interpretation is quite different in the two cases. For a fixed radius, it is straightforward to translate between these couplings with $\hat{g}_{eff}^2=[g^2_{eff}]^{\frac{5-p}4}$. Hence, for $p \le 4$, the constraints~\eqref{eq:safe} become
\begin{equation}
    1 \ll \hat{g}_{\mathrm{eff}}^2 \ll N^{\frac{5-p}{7-p}} \,.
    \label{eq:safe2}
\end{equation}
We note that the conversion from $g_{\mathrm{eff}}$ to $\hat{g}_{\mathrm{eff}}$ becomes subtle for $p=5$. A more careful analysis yields a logarithmic dependence \cite{Peet:1998wn}. For $p=6$, this conversion instead implies a weak coupling regime,
$\frac{1}{N} \ll \hat{g}_{\mathrm{eff}}^2 \ll 1$.
We will comment further on these subtle cases below.

Both eqs.~\eqref{eq:safe} and~\eqref{eq:safe2} indicate that there exists a restricted regime in which the boundary theory is strongly coupled, but not excessively so, where the D$p$-brane throat geometry provides a reliable holographic description.

As usual, we must also introduce a regulator surface to make sense of the holographic entanglement entropy, which we again choose at $r=\ruv$. Implicitly, we will assume $\ruv$ lies near the maximum allowed radius set by eq.~\eqref{eq:weakdilaton} or~\eqref{eq:weakcurv} for $p>3$ or $p<3$, respectively. In either case, we are setting $\ruv\gg r_p$. We can then make use of eq.~\eqref{Esugra} to define the corresponding short distance cutoff in the boundary theory
\begin{equation}
   \delta=1/E_\mt{sugra}(\ruv)=  \frac{2^{p-3} \pi^{\frac{3(p-3)}4}}{\left[\Gamma\left(\frac{7-p}2\right)\right]^{1/2}}\,\left(\frac{r_p}{\ruv}\right)^{(5-p)/2}\,r_p\,.
   \label{pcutoff}
\end{equation}
Of course, as expected, this relation reduces to the expected $\delta=L^2/\ruv$ for $p=3$ but it is more elaborate for general values of $p$. We note however that for $p\le 4$, $\delta$ is proportional to a negative power of $\ruv$. Hence, as we move the cutoff surface in the bulk to larger and larger radii, the corresponding short distance cutoff in the boundary theory becomes smaller and smaller, as expected. We comment on the problematic cases of $p=5,6$ below.

\paragraph{$\mathbf{0\le p\le 4}$:} In the following, we restrict  attention to $0 \le p \le 4$, for which  a well-defined decoupling limit exists and the holographic correspondence is under better control.

Up to this point, the discussion could in principle be extended to all $p \le 6$, for which the harmonic function is given by eq.~\eqref{eq:throat-Dp-metric}. However, as noted above, for $p \ge 4$, the theories run to strong coupling at large radius, and new microscopic degrees of freedom are expected to emerge in the ultraviolet. In the case of D4-branes, this behavior is well understood with the D4-throat geometry lifting AdS$_7 \times S^4$, with one of the boundary directions compactified on a circle, of eleven-dimensional M-theory. This reflects the UV completion of the boundary gauge theory in terms of the $(2,0)$ theory \cite{Seiberg:1996bd,Witten:1995zh,Lambert:2010iw}.

For $p=5,6$, however, no satisfactory holographic description is known. In these cases, the putative duality between bulk supergravity and a boundary gauge theory breaks down. This failure can be traced in part to the unusual energy--radius relations~\eqref{eq:sugra} for supergravity probes \cite{Peet:1998wn}, which obstruct a clean decoupling limit and blur the correspondence between radial position and energy scale. This issue manifests itself in the unexpected relation between $\delta$ and $\ruv$ in eq.~\eqref{pcutoff}. For D5-branes, the system is instead related via S-duality to NS5-branes, whose near-horizon limit leads to little string theory, a nonlocal theory without a conventional field-theoretic UV completion \cite{Aharony:1998ub,Aharony:1999ks}. For D6-branes, the situation is even more problematic: the geometry lifts to a Kaluza--Klein monopole in eleven dimensions \cite{Townsend:1995kk}, and no decoupled near-horizon region exists. As a result, a standard holographic dual in terms of a local quantum field theory is absent in both cases.

A final caveat arises for $p=0$.  The ${\cal N}=16$ D0-brane quantum mechanics has a single threshold bound state in the $SU(N)$ sector \cite{Sen:1995vr,Banks:1996vh,Polchinski:1999br}. While one might consider the classical commuting-matrix Coulomb branch, those directions are gapless asymptotic directions tied to the continuum in the quantum mechanics. Hence, the shell solution should not be regarded as a distinct normalizable vacuum.
Rather it can be considered as a semiclassical configuration on the classical Coulomb branch, or equivalently as a continuum wave packet built from separated D0-branes.

\begin{figure}[ht]
    \centering
    \includegraphics[width=0.9\linewidth]{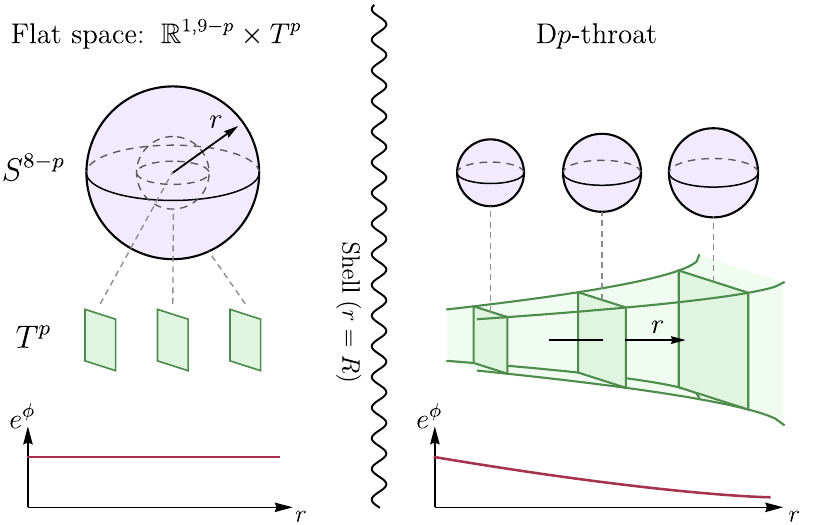}
    \caption{A depiction of the shell geometry described by eqs.~\eqref{eq:Dp-metric},~\eqref{eq:dilaton} and~\eqref{eq:Dp-metric22}. The right-hand side shows the D$p$-throat geometry: the internal sphere is either constant ($p=3$) or grows slowly ($p\neq 3$) with increasing radius. The torus factor grows radially for all $0\leq p \leq 4$. The dilaton decreases with $r$ for $p<3$, remains constant for $p=3$, and grows for $p=4$. The left-hand side shows the flat region: the sphere is incorporated in the $R^{1,9-p}$ factor with radius proportional to $r$. The torus factor has fixed size, and the dilaton is constant.  
}
    \label{fig:shelly}
\end{figure}

\paragraph{Shell geometries:} Now as described above because the D$p$-branes are BPS, they can be separated and placed at independent positions. So we now consider a Coulomb branch solution where the $N$ D$p$-branes are smeared over an $S^{8-p}$.  This leaves a bubble of empty flat space as the full back-reacted solution inside of the sphere. The corresponding throat geometry is given by eq.~\eqref{eq:Dp-metric} with
\begin{equation}
    H(r) = \begin{cases}
        \left(\frac{r_p}{r}\right)^{7-p}, \quad r\ge R\,,\\
        \left(\frac{r_p}{R}\right)^{7-p},\quad r\le R\,,
    \end{cases}
    \label{eq:Dp-metric22}
\end{equation}
where $R$ is the radial coordinate position of the shell. The resulting geometry is illustrated in figure \ref{fig:shelly}. Outside of the shell, the solutions are identical to the D$p$-brane throat for $N$ coincident branes. Inside the shell, the fully backreacted geometry has vanishing curvature and constant dilaton. (Further, the RR flux vanishes.) Hence the geometry locally corresponds to a portion of ten-dimensional flat space. In the figure, we depict this geometry as $\mathbf{R}^{1,9-p}\times T^p$, where the $p$-dimensional torus $T^p$
represents the spatial worldvolume directions. It will be convenient to compactify these directions with some large volume $V_p$. This notation also provides a useful way to distinguish the gauge-theory directions from the remaining spatial directions in the flat space region.

We note that the interface between the two regions has the geometry $\mathbf{R}\times S^{8-p}\times T^p$. In the supergravity field equations, there is a localized source at this interface corresponding to the \(N\) D\(p\)-branes smeared over the \(S^{8-p}\). For example, Einstein's equations contain a delta-function contribution, localized in the radial direction, to the stress tensor, with $S_{ab}=\rho\, g_{ab}$ where the \(a,b\) directions correspond to time and \(x^i_\parallel\) directions, \ie the directions along the D\(p\)-branes' worldvolume. Of course, the density $\rho$ is given by $N$ times the D$p$-brane tension divided by the volume of the sphere. 

Implicitly, we will regard the interface at $r=R$ as playing the role of the cutoff surface for the flat space region. From this perspective, the D$p$-brane throat geometry outside the shell serves as ``translation device,'' converting holographic data at the throat boundary into holographic flat space data at the cutoff surface. We return to this perspective in section \ref{sec:disc}.

It will be useful later on to introduce a simple example of a boundary observable that is sensitive to the presence of the shell. Recall that a stack of D$p$-branes placed at the origin $r=0$ with a single, parallel brane separated a distance $r$ in the $I$'th direction, the scalar $\Phi^I$ has a single non-zero eigenvalue $\frac{r}{2\pi l_s^2}$. Similarly, in the shell geometry, we are displacing all $N$ branes and so the scalars will have $N$ nonvanishing eigenvalues: $\Phi^I_{a}=\frac{x_a}{2\pi l_s^2}$ with $\sum x_a^2=R^2$ (in an appropriate gauge).  A simple gauge invariant expression that captures the presence of these non-zero eigenvalues is
\begin{equation}
    \langle \text{tr}\, \Phi^2 \rangle \equiv \frac{1}{N}\langle \text{Tr}\,\Phi^2\rangle = \frac{1}{N}\sum_{a=1}^N \Phi^I_a \Phi^I_a= \frac{R^2}{4\pi^2\ell_s^4} \,,
    \label{eq:Konishi}
\end{equation}
where we introduced the normalized trace `tr' to make the expectation value independent of $N$.

These shell solutions were first investigated in the context of AdS$_5\times S^5$ in \cite{Kraus:1998hv,Giddings:1999zu}. Moreover, the
Coulomb branch is well understood in the dual \(\mathcal{N}=4\) SYM
theory. The reader may therefore wonder why we wish to examine the shell geometries in the
broader setting of D$p$-brane holography. The answer comes from evaluating
the proper radius of the shell,
\begin{equation}
    R_\mt{shell}=r_p{}^{\!\!\frac{(7-p)(p+1)}{16}}\, R^{\frac{(p-3)^2}{16}}\,.
    \label{eq:flatsize}
\end{equation}
Hence for $p\ne3$, we can adjust the size of the flat space bubble by varying $R$, while the $R$-dependence vanishes for $p=3$. As noted previously, this feature will be important when we examine target space entanglement in section \ref{sec:target}.

This behaviour~\eqref{eq:flatsize} can be understood from the illustration of the geometry in figure \ref{fig:shelly}. From eq.~\eqref{eq:Dp-metric} for $p\ne3$, the $S^{8-p}$ slowly increases in size  with $r^{\frac{(p-3)^2}8}$ as we move out through the throat region. This sphere becomes the boundary sphere of the $R^{1,9-p}$ in the flat space region. Hence by connecting the two regions at a larger radius (\ie moving the shell to larger $r$), the size of the flat space region is larger for $p\ne 3$, as expected. In contrast, for $p=3$, the size of the $S^5$ is independent of $r$ and so the size of the corresponding flat space region must be frozen.

Hence the result for $p=3$ that the shell radius is fixed with $R_\mt{shell}=r_\mt{p=3}=L$ must mean that $R$ must be a spurious parameter in this case.  Indeed, by rescaling the coordinates $(t,x^i,r)\to(Lt/R,Lx^i/R,Lr/R)$ in eq.~\eqref{eq:Dp-metric} with $p=3$, we recover the same solution with $R=L$. This scaling symmetry is, of course, related to the conformal invariance of the corresponding boundary theory. In contrast, for $p\ne3$, this  scaling symmetry is absent both in the background metric and in the dual worldvolume theory.\footnote{However, for these theories, there remains a scaling `similarity', which is related to the fact that the D$p$-brane throat geometries are conformal to AdS$_{p+2}\times S^{8-p}$ \cite{Boonstra:1998mp, Skenderis:1998dq,Kanitscheider:2008kd,Biggs:2023sqw}.}

We must choose $R$ to satisfy the constraints in eqs.~\eqref{eq:weakdilaton} and~\eqref{eq:weakcurv} to ensure the shell does not reside in either the regime of strong coupling or of strong curvature. 
Note then that for $p<3$, the shell geometry fully removes the strong coupling region since within the shell the dilaton is constant with $e^\phi=g_s (R/r_p)^{(7-p)(p-3)/4}$. Similarly, for $p>3$, the high curvature region is eliminated in the shell geometry. However, for general $p\ne 3$, the shell geometry still extends to regions at large radius where the supergravity approximation breaks down (\ie because of high curvatures for $p<3$ and strong coupling for $p>3$). In principle then the extremal surfaces investigated in the following extend beyond the supergravity region for $p\ne3$. However, as discussed above, we regulate the surfaces by cutting them off at $r=\ruv$, which we choose to lie in the allowed supergravity region.

\section{Entanglement entropy} \label{sec:EE}

In this section, we will explore the entanglement entropies of strip and spherical regions in the boundary
of the shell geometries. 
In this case, the RT surfaces implicitly wrap the $S^{8-p}$, lie in some fixed $t$ slice of the geometry and are specified by some radial profile $r(x^i)$. When the entangling region on the boundary is small, the RT surfaces will remain completely in the D$p$-throat region, \ie $r>R$. However, 
if we make the entangling region large enough, the RT surface will pass through the shell of D$p$-branes into the flat space region. 
Hence, we generally divide the total area of the RT surface into two contributions coming from the throat and flat regions
\begin{equation}
    A_\mt{tot}= A_\mt{throat} + A_\mt{flat}\,.
    \label{area0}
\end{equation}
Using eqs.~\eqref{eq:Dp-metric} and~\eqref{eq:Dp-metric22}, the area functional in the exterior throat region becomes
\begin{equation}
    A_\mt{throat}= \Omega_{8-p}\, r_p^{\frac{7-p}2}\int_{B'} d^{\,p}x_{||}\  r^{\frac{9-p}2} \,\sqrt{1+\left(\tfrac{r_p}{r}\right)^{7-p} (\partial_i r)^2}\,,
    \label{area1}
\end{equation}
where $\Omega_{8-p}\equiv 2\pi^{\frac{9-p}{2}}/\Gamma(\frac{9-p}{2})$ corresponds to the volume of a round unit-sphere $S^{8-p}$. We denote the entangling region in the boundary theory as $B$ and here $B'$ is the portion of $B$ covered by this exterior portion of the RT surface. 
Similarly, the area functional in the flat space bubble  becomes
\begin{equation}
    A_\mt{flat}= \Omega_{8-p}\, \left(\tfrac{r_p}R\right)^{\frac{7-p}2}\int_{B''} d^{\,p}x_{||}\  r^{8-p} \,\sqrt{1+\left(\tfrac{r_p}{R}\right)^{7-p} (\partial_i r)^2}\,,
    \label{area3}
\end{equation}
where $B''$ is the portion of the boundary entangling region covered by this inner segment of the RT surface. As usual, the extremal RT candidates are found by making variations $\delta r(x^i)$ in eqs.~\reef{area1} and \reef{area3}. 
Furthermore, to ensure that the surfaces are extremal as they cross the shell, we must also check that  the surface terms coming from the previous variations cancel when evaluated together at the $r=R$ interface 
\begin{equation}
    \delta A_\mt{tot}=\delta A_\mt{flat} + \delta A_\mt{throat}=0\,. 
    \label{check-shell}
\end{equation}
However, as we shall see below, since the interface sits at a constant radius, this boundary condition is best determined when expressing the profile as $x^i(r)$.
With this general setup, we are ready to explore some simple examples.

\begin{figure}
    \hspace{3.cm}\includegraphics[width=0.75\linewidth]{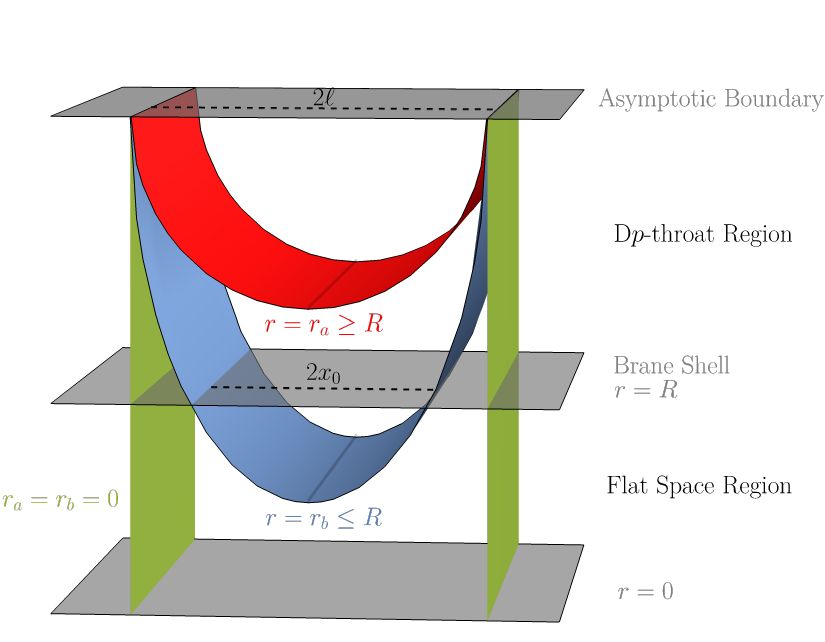}
    \caption{Diagram showing profiles of three candidate RT surfaces associated with a strip of a fixed width $2\ell$. i) $r_a\ge R$: The red surface stays in the throat region $r>R$, closing off at $r=r_a$. ii) $r_b\le R$: The blue surface probes the flat space bubble at $r<R$, closing off at $r=r_b$. iii) $r_a=r_b=0$: The green surface is comprised of two flat sheets that fall straight in from the asymptotic boundary to $r=0$. The labels and geometries are explained more fully in the text.} 
    \label{fig:flat-strip-profile}
\end{figure}

\subsection{Strip geometry} \label{sec:strip}

We begin with the strip or slab region on the boundary,
\begin{equation}
    -\ell \leq x_1 \leq \ell \,,
\end{equation}
so that the total width of the strip is $2\ell$. Since we need at least one spatial gauge theory direction, we do not consider $p=0$ here and only examine the shell geometries with $1\le p\le 4$.\footnote{The strip entanglement entropy was examined in the D$p$-brane throat geometry (without a shell) for $1\le p\leq4$  in \cite{vanNiekerk:2011yi}. It was also considered for the shell geometry with $p=3$ in \cite{Bhattacharya:2012mi}.} We denote by $V_{p-1}$ the regulated
volume of the remaining spatial directions $x_i$ with $i=2,\cdots,p$. Hence the total
area of the entangling surface on the boundary, consisting of the two planes at
$x_1=\pm \ell$, is $2V_{p-1}$. By translation symmetry along these $p-1$ directions,
the bulk profile of the RT surface is described by a single function $r(x_1)$, or
equivalently $x_1(r)$. In the following, we drop the subscript on $x_1$ and write the
profile as $x(r)$. 
The general structure of the candidate surfaces is illustrated in
figure~\ref{fig:flat-strip-profile}.

We first consider the portion of the surface lying outside the shell, \ie in the
D$p$-throat region with $r>R$. Given the Einstein-frame metric
\eqref{eq:Dp-metric} with eq.~\eqref{eq:Dp-metric22}, the area functional in this
region is
\begin{equation}
    A_\mt{throat}
    =
    \Omega_{8-p}\, V_{p-1}\, r_p^{7-p}
    \int dr\, r\,
    \sqrt{
    1+\left(\frac{r}{r_p}\right)^{7-p}
    \left(\frac{dx}{dr}\right)^2
    }\, .
    \label{eq:stripAbrane}
\end{equation}
Extremizing this functional gives the first-order equation
\begin{equation}
    \frac{dx}{dr}
    =
    \left(\frac{r_p}{r}\right)^{\frac{7-p}{2}}
    \frac{r_a^{\frac{9-p}{2}}}{\sqrt{r^{9-p}-r_a^{9-p}}}\,,
    \label{eq:stripBraneEOM}
\end{equation}
where $r_a$ is an integration constant. When $r_a\geq R$, the surface closes off in the
throat region at $r=r_a$, where $dr/dx=0$. In this case, the half-width of the strip is
related to $r_a$ by
\begin{equation}
    \ell
    =
    \int_{r_a}^{\infty} dr\, \frac{dx}{dr}
    =
    \frac{2\sqrt{\pi}}{5-p}\,
    \frac{\Gamma\!\left(\frac{7-p}{9-p}\right)}
         {\Gamma\!\left(\frac{5-p}{2(9-p)}\right)}
    \left(\frac{r_p}{r_a}\right)^{\frac{5-p}{2}} r_p\, .
    \label{eq:stripBraneWidth}
\end{equation}
For $1\leq p\leq4$ which we are examining here, $\ell$ decreases monotonically as $r_a$ increases, as expected, \ie narrower strips are described by surfaces that remain closer to the asymptotic
boundary.

When $r_a<R$, the surface passes through the shell and enters the flat space bubble.
In the flat region, the area functional becomes
\begin{equation}
    A_\mt{flat}
    =
    \Omega_{8-p} V_{p-1}
    \left(\frac{r_p}{R}\right)^{7-p}
    \int dr\, r^{8-p}
    \sqrt{
    1+
    \left(\frac{R}{r_p}\right)^{7-p}
    \left(\frac{dx}{dr}\right)^2
    }\, .
    \label{eq:stripAflat}
\end{equation}
The corresponding first-order equation is
\begin{equation}
    \frac{dx}{dr}
    =
    \left(\frac{r_p}{R}\right)^{\frac{7-p}{2}}
    \frac{r_b^{8-p}}
         {\sqrt{r^{2(8-p)}-r_b^{2(8-p)}}}\,,
    \label{eq:stripFlatEOM}
\end{equation}
where $r_b$ is the minimum radius reached by the surface in the flat region. The case
$r_b=0$ corresponds to a surface which falls straight to the origin at fixed $x$.

\begin{figure}
    \centering
    \includegraphics[width=0.6\linewidth]{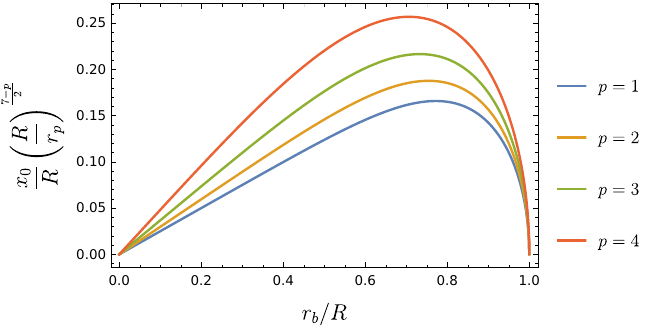}
    \caption{Half-width $x_0$ of the strip at the shell, $r=R$, as a function of the
    minimum radius $r_b$ of the surface in the flat region.}
    \label{fig:WidthsDp}
\end{figure}

Let $x_0$ denote the half-width of the surface at the shell, \ie the surface crosses
$r=R$ at $x=\pm x_0$. Integrating eq.~\eqref{eq:stripFlatEOM} from $r=r_b$ to $r=R$
gives
\begin{equation}
\begin{split}
    x_0 &= \int _{r_b}^R \frac{dx}{dr} =r_b\left(\frac{r_p}{R}\right)^{\frac{7-p}{2}}\Bigg[\frac{\sqrt{\pi}\, \Gamma\left(\tfrac{7-p}{2(8-p)}\right)}{2(8-p)\,\Gamma\left(\tfrac{15-2p}{2(8-p)}\right)} \\
    &\qquad\qquad\qquad\quad-\frac{1}{7-p}\,\left(\frac{\rstflat}{R}\right)^{7-p}\,  \,_2 F_1\left(\tfrac{1}{2},\tfrac{7-p}{2(8-p)},\tfrac{23-3p}{2(8-p)},\left(\frac{\rstflat}{R}\right)^{2(8-p)}\right)\Bigg]\,. \label{eq:stripx0Dp}
\end{split}
\end{equation}
This expression vanishes at both $r_b=0$ and $r_b=R$, and reaches a maximum at an
intermediate value of $r_b$, as shown in figure~\ref{fig:WidthsDp}. Hence for a range
of values of $x_0$, there are two nontrivial extremal surfaces in the flat region, a
deeper one and a shallower one. For sufficiently large $x_0$, however, there is no
nontrivial surface closing off at finite $r_b$, and the only candidate in the flat region is
the flat-sheeted surface with $r_b=0$.\footnote{The position of the maximum is approximately  $(r_b/R)_{\max}\approx 0.78474-0.01049\,p-0.0022824\,p^2$, which illustrates the trend shown in figure \ref{fig:WidthsDp}, namely that the position of the maximum decreases with increasing $p$.}

In order to glue together the extremal surfaces in the two regions, we need matching conditions that relate the  integration constants, $r_a$ and $r_b$, controlling their profiles. First, for the surface to be continuous, we ensure that the two segments reach the interface at $x=x_0$. Then, the differential equations in eqs.~\eqref{eq:stripBraneEOM} and~\eqref{eq:stripFlatEOM} ensure that the surfaces are extremal everywhere away from the $r=R$ shell. However, we impose an extra matching condition at the shell to ensure that the total area functional is extremal, as described around eq.~\eqref{check-shell}, \ie, to ensure that the boundary terms, which arise from varying $A_\mt{throat}$ and $A_\mt{flat}$ separately, cancel. From eqs.~\eqref{eq:stripAbrane} and~\eqref{eq:stripAflat}, we find 
\begin{equation}
 \Bigg(\frac{r^{8-p}\,\dot x}{\sqrt{1+\left(\tfrac{r}{r_p}\right)^{7-p}\,\dot x ^2}}\,\delta x \Bigg)\Bigg|^\mt{throat}_{r=R}=   \Bigg(\frac{r^{8-p}\,\dot x}{\sqrt{1+\left(\tfrac{R}{r_p}\right)^{7-p}\,\dot x ^2}}\,\delta x \Bigg)\Bigg|^\mt{flat}_{r=R}\,.
   \label{eq:bound1}
\end{equation}
Here $\delta x$ is the same on both sides as we are varying both surfaces to move the intersection point $x=x_0$ along the shell, and hence eq.~\eqref{eq:bound1} is satisfied if 
\begin{equation}
    \dot x \big|^\mt{throat}_{r=R}=\dot x \big|^\mt{flat}_{r=R}\,. 
    \label{eq:smooth}
\end{equation}
One finds that eq.~\eqref{eq:smooth} is satisfied when 
\begin{equation}
    \left(\frac{r_a}{R}\right)^{9-p}
    =
    \left(\frac{r_b}{R}\right)^{2(8-p)}
    \label{eq:stripMatchDp}
\end{equation}
giving us the desired relation between the two integration constants. With this matching condition, the candidate RT surfaces are continuous and smooth as they cross the brane shell.

Combining eqs.~\eqref{eq:stripx0Dp} and~\eqref{eq:stripMatchDp}, we can express the
boundary half-width $\ell$ in terms of $r_b$:
\begin{equation}
\begin{split}
    \ell = &
\frac{\left(\frac{r_p}{R}\right)^{\tfrac{7-p}{2}} \, r_b}{(7-p)}
\Bigg[
\frac{\sqrt{\pi}\, \,
\Gamma\!\left(\tfrac{23-3p}{2(8-p)}\right)}
{\Gamma\!\left(\tfrac{15-2p}{2(8-p)}\right)}
+
\left(\frac{r_b}{R}\right)^{\,7-p}
\bigg(
{}_2F_1\!\left(
\tfrac{1}{2},
\tfrac{7-p}{9-p},
\tfrac{2(8-p)}{9-p},
\left(\frac{r_b}{R}\right)^{2(8-p)}
\right)\\
&\qquad\qquad\qquad\qquad\qquad-
{}_2F_1\!\left(
\tfrac{7-p}{2(8-p)},
\tfrac{1}{2},
\tfrac{23-3p}{2(8-p)},
\left(\frac{r_b}{R}\right)^{2(8-p)}
\right)
\bigg)
\Bigg]\,.
\end{split}
\label{eq:stripWidthDp}
\end{equation}
For the range of our inquiry (\ie $1\leq p\leq4$), this expression starts at $\ell=0$ when $r_b=0$ and increases
monotonically as $r_b$ approaches the shell. In the limit $r_b\to R$, we find
\begin{equation}
    \ell \to \ell_0
    \equiv
    \frac{
    2\sqrt{\pi}\,
    \Gamma\!\left(\frac{7-p}{9-p}\right)}
    {(5-p)\,
    \Gamma\!\left(\frac{5-p}{2(9-p)}\right)}
    \left(\frac{r_p}{R}\right)^{\frac{7-p}{2}} R\, .
    \label{eq:stripEllZeroDp}
\end{equation}
This agrees with the $r_a\to R$ limit of eq.~\eqref{eq:stripBraneWidth}, as it must since
both limits describe a surface that just closes off at the brane shell.

We have therefore identified three families of candidate surfaces: First, there are
surfaces that remain entirely in the brane region and close off at $r=r_a\geq R$.
Second, there are surfaces that pass through the shell and close off in the flat bubble
at $0<r_b\leq R$, with $r_a$ related to $r_b$ by eq.~\eqref{eq:stripMatchDp}. These
first two families exist only for $0\leq \ell\leq \ell_0$. Third, there are the flat-sheeted
surfaces with $r_a=r_b=0$, which exist for any $\ell$.

Hence we must compare the areas of all of these candidates to determine the true RT surface. However, all areas have a universal UV divergence. It is simplest to evaluate this divergence by considering the surfaces which fall straight in from the boundary. Substituting $dx/dr=0$ in this case, eq.~\eqref{eq:stripAbrane} yields\footnote{The extra factor of 2 comes because the full surface is comprised of two flat sheets, one for each of the strip boundaries. As a related remark, we recall for the following discussion that the total area of the entangling surface is $2 V_{p-1}$.}
\begin{equation}
    A_\mt{throat}= 2\Omega_{8-p}\, V_{p-1}\, r_p^{7-p}\int_R^{\ruv} dr\,r\simeq \Omega_{8-p} V_{p-1}\, r_p^{7-p}\, \ruv^{\,2}\equiv A_\mt{UV}\,,
    \label{eq:StripDiv}
\end{equation}
where we introduced the UV cutoff $\ruv$ for the radial integral. It may seem somewhat surprising this divergence is proportional to the fixed power $\ruv^2$ for all values of $p$. However, applying this result to evaluate the leading contribution to the entanglement entropy in the boundary theory yields\footnote{Here, we use $G_{10}= 8\pi^6 g_s^2 \ell_s^8$ for the ten-dimensional Newton's constant.}
\begin{equation}
    S_\mt{EE}
    =
    \frac{A}{4G_{10}}
    =
    \frac{N^2}{2^{\frac{3-p}{5-p}}\pi(7-p)}
    \left[\hat g_\mt{eff}(1/\delta)\right]^{\frac{2(p-3)}{5-p}}
    \frac{2V_{p-1}}{\delta^{p-1}}
    +\cdots\, ,
    \label{eq:stripAreaLawDp}
\end{equation}
where the short-distance cutoff $\delta$ is related to $r_\mt{UV}$ as in
eq.~\eqref{pcutoff}. Thus we recover the expected area law contribution proportional to $2V_{p-1}/\delta^{p-1}$. 

For $p=3$, the prefactor above reduces to $N^2/4\pi$, which we recognize as providing a measure of the number of degrees of freedom in the boundary theory, \ie it is proportional to $N^2$. Of course, the normalization of this coefficient is not universal here, depending on the precise choice of the cutoff. For $p\neq3$, the prefactor contains additional factors (and in particular cutoff dependence). However, as shown in eq.~\eqref{eq:stripAreaLawDp}, this can be assembled as $N^2
    \left[\hat g_\mt{eff}(1/\delta)\right]^{\frac{2(p-3)}{5-p}}$,
which we again interpret as a measure of the number of ultraviolet degrees of freedom in the nonconformal D$p$-brane worldvolume theory \cite{vanNiekerk:2011yi}. We will discuss this combination further below, but here we emphasize for $p\ne3$ this measure of the degrees of freedom runs with the energy scale. Hence in eq.~\eqref{eq:stripAreaLawDp}, where the relevant energy scale is $E_\mt{string}=1/\delta$, it measures the number of degrees of freedom at the cutoff scale.

In principle, we might have expected further subleading divergences in the entanglement entropy. In a CFT (\eg for $p=3$), these would be  proportional to the curvature of the entangling surface \cite{Solodukhin:2008dh,Hung:2011xb} and so they vanish here since the strip boundaries are flat. It is not obvious that the same structure should arise for the nonconformal dual theories for $p\ne3$, but our holographic calculations indicate that no additional divergences arise in this case.

We now subtract the common divergence~\eqref{eq:StripDiv} and compare the regulated
areas. For surfaces that close off in the brane region, evaluating
eq.~\eqref{eq:stripAbrane} gives
\begin{equation}
    A_\mt{reg}
    =
    -\Omega_{8-p} V_{p-1}\,
    \frac{
    \sqrt{\pi}\,
    \Gamma\!\left(\frac{7-p}{9-p}\right)}
    {\Gamma\!\left(\frac{5-p}{2(9-p)}\right)}
    r_p^{7-p} r_a^2\, .
    \label{eq:stripAregBrane}
\end{equation}
This regulated area is negative for $1\leq p\leq4$, and becomes more negative as
$r_a$ decreases. Equivalently, by eq.~\eqref{eq:stripBraneWidth}, the entropy grows as
the width of the strip increases.
For the surfaces that pass through the shell and close off in the flat region at
$r=r_b$, the regulated area can also be evaluated analytically
\begin{eqnarray}
    A_{reg} &= \frac{2\Omega_{8-p}V_{p-1} }{(9-p)}\,r_p^{7-p}\,R^2 \Bigg[\frac{\sqrt{\pi}\, \Gamma\!\left(\frac{7-p}{2(8-p)}\right)}{2(8-p)\, \Gamma\!\left(\frac{15-2p}{2(8-p)}\right)}\left(\frac{r_b}{R}\right)^{9-p}
    + {}_2F_1\!\left(-\tfrac{9-p}{2(8-p)},\tfrac{1}{2},\tfrac{7-p}{2(8-p)},\left(\frac{\rstflat}{R}\right)^{2(8-p)}\right)
    \nonumber\\
    &\qquad\qquad\qquad\qquad-\frac{(9-p)}{2}\,{}_2F_1\! \left(\tfrac{1}{2},-\tfrac{2}{9-p},\tfrac{7-p}{9-p},\left(\frac{\rstflat}{R}\right)^{2(8-p)}\right) \Bigg]\,.
    \label{eq:stripAregBubble}
\end{eqnarray}
At $r_b=R$, this expression joins smoothly onto
eq.~\eqref{eq:stripAregBrane} with $r_a=R$, as required by the matching condition.
Finally, for the flat-sheeted surfaces with $r_a=r_b=0$, the regulated area is
\begin{equation}
    A_\mt{reg}
    =
    -\Omega_{8-p} V_{p-1}\,
    \frac{7-p}{9-p}\,
    r_p^{7-p}R^2\, .
    \label{eq:stripAregFlat}
\end{equation}

\begin{figure}
    \centering
    \includegraphics[width=0.6\linewidth]{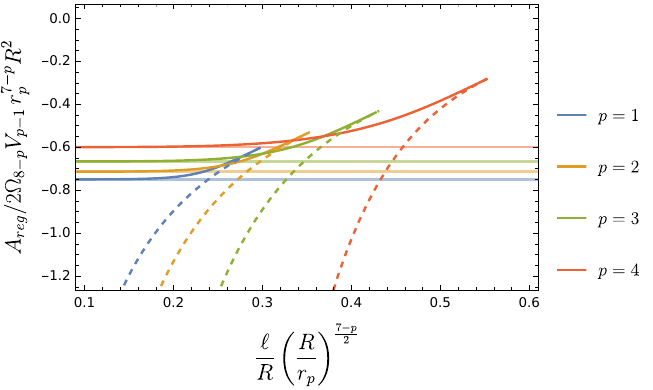}
    \caption{The regulated areas of the three classes of candidate RT surfaces in the shell geometry as functions of the boundary strip width. The nontrivial surfaces that enter the flat
    bubble are shown as solid curves (with the flat-sheeted surfaces being the horizontal lines) and the surfaces that remain outside the shell are shown as dashed curves.}
    \label{fig:p4plot}
\end{figure}

Figure~\ref{fig:p4plot} shows the regulated areas for $1\leq p\leq4$. The result has
the familiar `swallowtail' structure found, \eg in first-order phase transitions
\cite{gilmore1993catastrophe,Chamblin:1999tk}. The surfaces that enter the flat space bubble and close off at $r_b<R$ connect the two other branches: they merge with the throat region surfaces at $\ell=\ell_0$ and with the flat-sheeted surfaces as $\ell\to0$. However, these are never the minimal-area surfaces and only play the role of the unstable saddlepoint in the swallowtail. The regulated areas of the other two candidate surfaces cross at
\begin{equation}
    \ell_c
    =
    \frac{
    2\,
    \pi^{\frac{9-p}{8}}\,
    (9-p)^{\frac{5-p}{4}}\,
    \Gamma\!\left(\frac{7-p}{9-p}\right)^{\frac{9-p}{4}}
    }
    {
    (5-p)\,
    (7-p)^{\frac{5-p}{4}}\,
    \Gamma\!\left(\frac{5-p}{2(9-p)}\right)^{\frac{9-p}{4}}
    }
    \left(\frac{r_p}{R}\right)^{\frac{7-p}{2}} R\, .
    \label{eq:stripTransitionWidthDp}
\end{equation}
Of course, as seen in the figure, $\ell_c<\ell_0$ for $1\leq p\leq4$. Hence we find for $\ell\leq\ell_c$, the dominant RT surfaces are those lying entirely in the throat region, while the flat-sheeted surfaces dominate for
$\ell\geq\ell_c$. 

The physical conclusion is therefore simple. For $\ell<\ell_c$, the RT surfaces probe only the D$p$-throat region, and the entanglement entropy grows with the width of the strip. At $\ell=\ell_c$, there is a first-order transition to the flat-sheeted surfaces.
For $\ell>\ell_c$, the regulated area is independent of $\ell$, and the entanglement entropy remains constant. The only dominant RT surfaces which reach the flat space
bubble are therefore these flat sheets that fall directly to $r=0$. This behaviour is qualitatively similar to what happens in confining holographic backgrounds
\cite{Nishioka:2006gr,Klebanov:2007ws}, despite the fact that the states considered here are Coulomb-branch states (which exhibit screening) rather than confining vacua.

\paragraph{c-function:} In the context of the AdS/CFT correspondence, it has been understood that entanglement entropy is an interesting probe of RG flows \cite{Myers:2010xs,Myers:2010tj} and in fact, this observation extends beyond holography to discussions of general quantum field theories \cite{Casini:2006es,Casini:2012ei,Casini:2015woa}. Building on the result for two dimensions \cite{Casini:2006es}, a general expression was formulated in higher dimensions in the context of the strip geometry  \cite{Myers:2012ed,Liu:2013una}. Hence we begin here by focusing on $p=3$ (where the boundary dual is conformal) for which the latter expression becomes
\begin{equation}
\tilde{c}_3 = \beta\, \frac{\ell^{3}}{V_2}\,\frac{\partial S_\mt{EE}}{\partial \ell}
    \label{c1}
\end{equation}
 where $\beta$ is a numerical factor
\begin{equation}
    \beta=\frac{1}{\sqrt{\pi}}\,\left(\frac{\Gamma\left(1/6\right)}{\Gamma\left({2}/{3}\right)}\right)^{3}\,.
    \label{eq:beta}
\end{equation}
This constant $\tilde{c}_3$ plays the role of a central charge giving a measure of the number of degrees of freedom at energy scales of the order $1/(2\ell)$ in the underlying RG flow.
In the context of AdS$_5\times S^5$ (or $p=3$), one finds
\begin{equation}
    \tilde{c}_3=\frac{\pi^{4}}{2}\,\frac{L^8}{G_{10}}={N^2}=4\,C_T\,,
    \label{eq:ctilde}
\end{equation}
where $C_T$ is the central charge appearing in the two-point correlator of the stress-energy tensor, \eg see \cite{Buchel:2009sk}.\footnote{Further, in the present case with $\mathcal{N}=4$ SYM as the boundary dual, $C_T=c=a$ where $c$ and $a$ are the central charges appearing in the boundary trace anomaly.}

In the case of the $p=3$ shell geometry, for small separations (\ie $\ell\le\ell_c$), the RT surfaces simply correspond to the same surfaces that would be found in the AdS vacuum. Hence, the c-function is constant with $\tilde{c}_3=4C_T$. However, with large separations (\ie $\ell\ge\ell_c$), one has $\partial S/\partial \ell=0$ and hence the corresponding central charge~\eqref{c1} vanishes in this regime. So using this probe, it seems that there are no  degrees of freedom  in the infrared \ie associated with the flat space region -- or rather less than $O(N^2)$ in the large $N$ expansion. Again, this IR behaviour is reminiscent of that found in confining geometries, \eg \cite{Nishioka:2006gr,Klebanov:2007ws}.

Let us now turn to the case of general $p$. Above, we found that the entanglement entropy of the strip only contained an area law  divergence~\eqref{eq:stripAreaLawDp}, just as in the conformal case. Hence this divergence is removed by a simple divergence $\partial S_\mt{EE}/\partial\ell$ and we can try to generalize eq.~\eqref{c1} to construct a strip c-function for general $p$ with
\begin{equation}
    \tilde c_p
    =
    \beta_p\,
    \frac{\ell^p}{V_{p-1}}\,
    \frac{dS_\mt{EE}}{d\ell}\, ,
    \label{eq:stripcpDef}
\end{equation}
where $\beta_p$ is a numerical normalization factor. As in eq.~\eqref{c1}, we introduced the $1/V_{p-1}$ factor to remove the dependence on the area of the entangling surface and then the $\ell^p$ factor ensures that $\tilde c_p$ is dimensionless.
Evaluating this expression for the
surfaces that remain entirely in the D$p$-brane throat gives
\begin{equation}
    \tilde c_p
    =
    N^2
    \left[\hat g_\mt{eff}(1/2\ell)\right]^{\frac{2(p-3)}{5-p}}
    =
    N^2
    \left(g_{YM}^2N\right)^{\frac{p-3}{5-p}}
    (2\ell)^{-\frac{(p-3)^2}{5-p}}\, ,
    \label{eq:stripcpDp}
\end{equation}
where we have chosen
\begin{equation}
    \beta_p
    =
    \frac{
    (7-p)(5-p)^{\frac{9-p}{5-p}}
    }
    {
    2^{\frac{p^2-12p+37}{5-p}}\,
    \pi^{\frac{17-5p}{2(5-p)}}
    }\,
    \frac{
    \Gamma\!\left(\frac{5-p}{2(9-p)}\right)^{\frac{9-p}{5-p}}
    }
    {
    \Gamma\!\left(\frac{7-p}{2}\right)^{\frac{2}{5-p}}\,
    \Gamma\!\left(\frac{7-p}{9-p}\right)^{\frac{9-p}{5-p}}
    }\, .
    \label{eq:stripBetaDp}
\end{equation}
With this normalization, $\tilde c_p$ reproduces the same combination of $N$ and
$\hat g_\mt{eff}$ that appears in the coefficient of the UV area law in
eq.~\eqref{eq:stripAreaLawDp}. The difference is that the effective coupling is now
evaluated at the scale $1/(2\ell)$, set by the width of the strip, rather than at the UV
cutoff scale $1/\delta$. Thus $\tilde c_p$ measures the number of degrees of freedom
associated with excitations of wavelength of order $2\ell$.
While $\tilde c_{p=3}$ remains a constant for the conformal case, it runs with the energy scale when $p\neq3$. In fact, we see from  eq.~\eqref{eq:stripcpDp} that $\tilde c_{p}$
decreases with increasing $\ell$ for all $p\ne3$, irrespective of whether the Yang-Mills coupling is relevant, as for $p=1,2$, or irrelevant, as for $p=4$. Of course, this is precisely the desired behaviour since a proper c-function should decrease (or remain constant), as we move toward the IR irrespective of these details.

Above, we compared the form of $\tilde{c}_p$ to the prefactor appearing in the area law term~\eqref{eq:stripAreaLawDp} in the holographic entanglement entropy. However, we now recall that for $p=3$,  $\tilde c_3$ was also related to $C_T$ is the central charge appearing in the two-point correlator of the boundary stress-energy tensor \cite{Buchel:2009sk}. The analogous two-point functions were calculated holographically for the D$p$-brane worldvolume theories with $p=0,1,2$ in \cite{Kanitscheider:2008kd}. In particular, the leading transverse traceless component was shown to take the form 
\begin{equation}
    \langle T_{ij}(x) T_{kl}(y) \rangle_{TT}=  A_p\,\Pi^{TT}_{ijkl}\, \frac{ N^2\, [\hat{g}^2_{eff}(1/|x-y|)]^{\frac{p-3}{5-p}}}{|x-y|^{2d}},
    \label{eq:stress}
\end{equation}
where $\Pi^{TT}_{ijkl}$ projects on to the transverse traceless tensor structure and $A_p$ is a positive numerical constant. Here again, we see that the effective coupling $\hat{g}^2_{eff}$ appears with precisely the same power as in eq.~\eqref{eq:stripcpDp}. Hence the effective central charge appearing in the stress tensor correlator counts the degrees of freedom in the same way as seen in $\tilde{c}_p$ above. In this case, the relevant energy scale is set by the separation of the two stress tensor operators, \ie $E=1/ |x-y|$.

Returning to the shell geometries for general $p$, the transition at $\ell=\ell_c$ implies that this
effective central charge also undergoes a sharp transition. For $\ell<\ell_c$, the RT
surface remains in the brane region and $\tilde c_p$ is given by
eq.~\eqref{eq:stripcpDp}. For $\ell>\ell_c$, the dominant surface is the flat-sheeted
surface, whose regulated area~\eqref{eq:stripAregFlat} is independent of $\ell$. Hence
$dS_\mt{EE}/d\ell=0$, and hence the effective central charge vanishes in this regime.

Hence, if we are following central charge as a function of the width of the strip for general $p$, it will begin for narrow strips (\ie $\ell<\ell_c$) with $\tilde{c}_p$ decreasing as $(2\ell)^{-\frac{(p-3)^2}{5-p}}$ as the width increases. But then, $\tilde{c}_p$ will suddenly jump to zero and stay there as the width increases beyond $2\ell_c$. Of course, this is the regime where the holographic entanglement entropy is probing the flat space region. So just as we found in section \ref{sec:strip} for the conformal case with $p=3$, this probe suggests there are `no'  degrees of freedom in the boundary theory associated with the flat-space region. To be more precise, we expect the number of degrees of freedom is only order one in the large $N$ expansion, compared to $O(N^2)$ in the brane region.

\subsection{Spherical geometry} \label{sec:sphere}

Spherical entangling surfaces have proved useful in examining holographic RG flows \cite{Myers:2010xs,Myers:2010tj}. However, analyzing holographic entanglement entropy for spherical entangling surfaces across different $p$ is challenging because the structure of the UV divergences in the entanglement entropy changes as we move between different dimensions. These divergent contributions were evaluated for the D$p$-throat geometries in \cite{vanNiekerk:2011yi}\footnote{Note that the case of an $S^0$ entangling surface for $p=1$ corresponds precisely to the strip geometry examined in the previous section.}
\begin{equation}
\begin{split}
 S^1\ {\rm for}\  p=2:\ &\quad S_\mt{EE}=\frac{N^2}{5 \pi\, 2^{1/3}} \, \left[\hat g_\mt{eff}(1/\delta)\right]^{-2/3}\,\frac{2\pi P}{\delta } + \cdots\,,  \\
  S^2\ {\rm for}\  p=3:\  &\quad S_\mt{EE} =\frac{N^2}{4\pi}\,\frac{4\pi \rhoo^2} {\delta^2}-N^2\,\log\!\left(\rhoo/\delta\right) +\cdots\,,\\
S^3\ {\rm for}\  p=4:\ & \quad S_\mt{EE}=\frac{2\,N^2}{3 \pi} \, \left[\hat g_\mt{eff}(1/\delta)\right]^{2}\frac{2\pi^2 P^3}{\delta^3 } -  N^2 \left[\hat g_\mt{eff}(1/\delta)\right]^{2}\,\frac{9\,P}{16\,\delta}\\
&\qquad\qquad\qquad+
\frac{45 N^2}{2^{10}\pi} \left[\hat g_\mt{eff}(1/P)\right]^{2}\,\log\!\left(\rhoo/\delta\right)
+\cdots\,,
\end{split}
    \label{eq:SphereDiv}
\end{equation}
for a sphere of radius $P$ in the boundary.
The leading contribution takes the form of the expected area law divergence for all values of $p$, but the subleading contributions take a different form for each dimension. To deal with these singularities, one must study each dimension separately, \eg to isolate the universal coefficients which might appear in the corresponding c-function \cite{Liu:2012eea}. Note that the ellipses indicate finite contributions and these determine the universal contribution for $p=2$. In the following, we consider how the flat space region of the shell geometry is probed by spherical RT surfaces first for an $S^2$ entangling surface for $p=3$, which corresponds to the  familiar AdS/CFT setting. Then, we examine the nonconformal example with $S^1$ for  $p=2$.
We expect that similar results can be derived for the $S^3$ entangling surface for $p=4$, but we leave this for future work.

\begin{figure}
    \hspace{3cm}\includegraphics[width=0.9\linewidth]{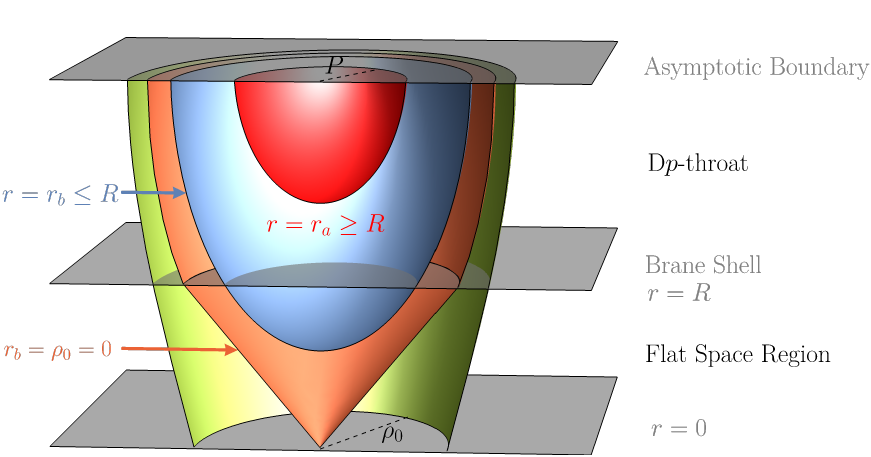}
    \caption{Diagram showing profiles of three RT surfaces with different boundary radii $P$. i) $P<P_{\mt{shell}}$: The red surface stays in the D$p$-throat region $r>R$, closing off at $r=r_a$. ii) $P_{\mt{shell}}\leq P\leq P_c$: The blue surface closes in the flat space bubble at $r<R$, closing off at $r=r_b$. iii) $P=P_c$: the pink surface closes off at $r_b=\rho_0=0$. The surface is exactly a cone in the flat space region. iv) $P > P_c$: The green surface extends all the way to $r=0$ where it has radius $\rho_0$. } 
    \label{fig:spherediagram}
\end{figure}

Following the above discussion for a general D$p$-throat, we take the entangling surface in the boundary to be a $(p-1)$-sphere. In this case, it is natural to introduce spherical polar coordinates in the gauge theory directions, cf.~eq.~\eqref{eq:Dp-metric}, 
\begin{equation}
    dx_{||}^2= d\rho^2 + \rho^2 d\Omega_{p-1} ^2\,.
    \label{eq:polar}
\end{equation}
Following our notation above, we denote the radius of the entangling surface as  $\rho=\rhoo$. Due to spherical symmetry, the profile of the bulk surfaces is given by $\rho(r)$ or equivalently $r(\rho)$.  For sufficiently large $\rhoo$, the extremal surfaces will intersect the brane shell and extend into the flat space bubble. As we will see, the behavior of RT surfaces here is qualitatively different from that seen in the previous section with the strip geometry. Here, there is a unique extremal bulk surface for each value of $P$. Hence, they exhibit a smooth transition between surfaces that close off in the D$p$-throat region and in the flat space bubble, and between surfaces that close off in the flat region and those that reach $r=0$. The radius at $r=0$ is $\rho_0$. Figure \ref{fig:spherediagram} illustrates the setup and the various RT surfaces that arise in our discussion below.

In the D$p$-throat, the area functional~\eqref{area1} can be written as 
\begin{equation}
    A_\mt{throat}= \Omega_{8-p}\, \Omega_{p-1}\, r_p^{7-p} \int d\sigma\, r\, \rho^{p-1} \sqrt{\left(\frac{dr}{d\sigma}\right)^2 + \left(\frac{r}{r_p}\right)^{7-p} \left(\frac{d\rho}{d\sigma}\right)^2}   \, ,
    \label{eq:zzzzx}
\end{equation}
where $\sigma$ is a radial worldvolume coordinate.
In the flat region, the area functional~\eqref{area3} takes the form
 \begin{equation}
    A_\mt{flat}= \Omega_{8-p}\, \Omega_{p-1}\, \left(\frac{r_p}{R}\right)^{7-p} \int d\sigma\, r^{8-p}\, \rho^{p-1} \sqrt{\left(\frac{dr}{d\sigma}\right)^2 + \left(\frac{R}{r_p}\right)^{7-p} \left(\frac{d\rho}{d\sigma}\right)^2}    \, .
    \label{eq:yyyyx}
\end{equation}
In general, we are not able to solve the resulting equations of motions analytically due to their nonlinearity. For this reason, we will need to resort to numerics below. The corresponding equations of motion are shown in appendix \ref{app:bc}.

The surfaces that probe the flat region must satisfy matching conditions as they cross the shell at $r=R$. Continuity of the surface requires that the solutions in the two regions meet the shell at the same value of $\rho=\rho_R$. In addition, we require that the boundary contributions from the variation of the flat-region and D$p$-throat area functionals cancel. Fixing the worldvolume gauge with $\sigma=r$ in eqs.~\eqref{eq:zzzzx} and~\eqref{eq:yyyyx}, the profile is given by $\rho=\rho(r)$ and considering a variation $\delta \rho$  yields 
\begin{equation}
  \Bigg(\frac{r^{8-p}\,\rho\   \frac{d\rho}{d r}}{\sqrt{1+\big(\tfrac{r}{r_p}\big)^{7-p}\,\big(\frac{d\rho}{d r}\big)^2}}\delta \rho \Bigg)\Bigg|^\mt{throat}_{r=R}=   \Bigg( \frac{ r^{8-p}\,\rho\   \frac{d\rho}{d r}}{\sqrt{1+\big(\frac{R}{r_p}\big)^{7-p}\big(\frac{d\rho}{d r}\big)^2}}\,\delta \rho \Bigg)\Bigg|^\mt{flat}_{r=R}\,.
\end{equation}
This implies the continuity of the derivative across the shell,
\begin{equation}
    \frac{d\rho}{d r} \Big|^\mt{throat}_{r=R}=\frac{d\rho}{d r} \Big|^\mt{flat}_{r=R}\,,
    \label{eq:p2-bdy8}
\end{equation}
as we also saw in the previous example of the strip geometry.

 In order to solve the equations of motion numerically, it is useful to briefly study the solutions near the closing-off points. There are four cases to consider, as illustrated in figure \ref{fig:spherediagram}: For surfaces that close in the shell exterior ($r_a> R$), the closing-off point is at $(r=r_a, \rho=0$). For surfaces that probe the shell with $0 < r_b \leq R$ it is $(r=r_b,\rho=0)$, and for surfaces that extend to the origin with $\rho_0>0$, it is $(r=0,\rho=\rho_0)$. Finally, there is the special case $r_b=\rho_0=0$. Some details about the solution of the equations of motion near these closing-off points can be found in appendix \ref{app:bc}. The summary is that the boundary conditions at these points are
\begin{eqnarray}
a)& R<r_a\ &:\quad r =  r_a + \frac{(9-p)}{4 p} \left(\frac{r_a}{r_p}\right)^{6-p} \frac{\rho^2}{r_p} + \cdots\,,
\nonumber\\
b)&\ 0<r_b\le R\ &:\quad r = r_b + \frac{(8-p) }{ 2\,  p }\left(\frac{R}{r_p}\right)^{7-p} \frac{\rho^2}{r_b} + \cdots\,,
\nonumber\\
c)&\ r_b=\rho_0=0\ &:\quad r = \sqrt{\frac{8-p}{p-1}} \left(\frac{R}{r_p}\right)^{\frac{7-p}{2}} \rho \,,
\label{eq:p2-bdycond}\\
d)&\rho_0>0\ &:\quad \rho = \rho_0 +\frac{p-1}{2(9-p)} \left(\frac{r_p}{R}\right)^{7-p} \frac{r^2}{\rho_0}+ \cdots\, .
\nonumber
\end{eqnarray}
With these boundary conditions, we start our numerical integration at either small $\rho$ or small $r$,\footnote{Practically, $\rho=0$ and $r=0$ are avoided in the numerics due to the singularities in the equations of motion.} and evolve the corresponding equation of motion, \ie eqs.~\eqref{eq:r-Dp-eom},~\eqref{eq:r-flat-eom} and~\eqref{eq:rho-flat-eom} for cases $a$, $b$ and $d$ above, while either eq.~\eqref{eq:r-flat-eom} or~\eqref{eq:rho-flat-eom} applies for case $c$. However, note that for case $c$, eq.~\eqref{eq:p2-bdycond} gives the full analytic solution for the segment of the surface in the flat space region. Generally, for the surfaces that probe the flat space bubble, we evolve out to the shell at $r=R$, where the numerical solution determines the values of $\rho=\rho_{\mt{R}}$ and $d\rho/dr$ at the shell, and together with eq.~\eqref{eq:p2-bdy8}, provides the initial data required to integrate eq.~\eqref{eq:r-Dp-eom} out to the asymptotic boundary.

\subsubsection*{$S^2$ entangling surface for $p=3$}

Recall that in this case, $R$ is a spurious parameter which can be changed by a simple scaling of the AdS$_5$ coordinates. Hence, we simplify the following discussion by setting $R=L=r_3$. Now it is well known that for an $S^2$ entangling surface in the boundary, the bulk RT surface is a hemisphere in Poincare coordinates~\cite{Ryu:2006bv,Ryu:2006ef}. Hence, these solutions are the relevant surfaces here that close off smoothly in the AdS region. We take a closer look at this result by introducing the usual $z=L^2/r$ and then the RT surfaces are given by
\begin{equation}
    \rho=\sqrt{\rhoo^2-z^2}= \sqrt{\rhoo^2-\frac{L^4}{r^2}}\,.
    \label{eq:prof5}
\end{equation}
Hence, they smoothly close off at $r= r_a=L^2/\rhoo$, which requires $\rhoo\le L$ in order for these surfaces not to extend into the flat space bubble. 

Having these analytic solutions is also useful to identify the divergent contribution to the area coming from large radii. Given the profile~\eqref{eq:prof5}, we have $\dot{\rho}=-L^4/\rho\, r^3$
and $r^2\rho^2 =\rhoo^2\,r^2-L^4$, and then the area~\eqref{eq:zzzzx} yields
\begin{equation}
\begin{split}
    A_\mt{AdS}&= 4\pi^4 L^4\int_{r_a}^{r_\mt{UV}} \!\!\!dr\,  r\,\rho^2\sqrt{1+\frac{L^4}{r^2\rho^2}} =
    4\pi^4 L^4\rhoo\int_{L^2/\rhoo}^{r_\mt{UV}} \!\!dr\,\sqrt{P^2r^2-L^4}\\
    &= 2\pi^4 L^4\left[\,\rhoo\ruv\sqrt{\rhoo^2\ruv^2-L^{4}} -L^{4}\log\!\left(\frac{\rhoo\ruv+\sqrt{\rhoo^2\ruv^2-L^{4}}}{L^2}\right)\,\right] \\
    &\simeq 2\pi^4 L^4\rhoo^2\ruv^2 -2\pi^4 L^{8}\log\!\left(\frac{\rhoo\ruv}{L^2}\right) \equiv A_\mt{UV}
    \,,
\end{split}
    \label{eq:zz22}
\end{equation}
where, as before, we introduced $r_\mt{UV}$ as the UV cutoff  for the radial integral. In this case, we have both an area law divergence and a logarithmic divergence \cite{Solodukhin:2008dh,Hung:2011xb} in the corresponding entanglement entropy shown in eq.~\eqref{eq:SphereDiv}.
As usual, this result is derived by setting $\delta=L^2/r_\mt{UV}$ for the short distance cutoff in the boundary theory. The divergence~\eqref{eq:zz22} was derived here from the explicit solutions~\eqref{eq:prof5} which are valid for $\rhoo\le L$. However, we know that this form of the divergence will be valid for any value of $\rhoo$. Hence we will subtract $A_\mt{UV}$ for all of the RT surfaces in the following to produce the regulated area $A_\mt{reg}$. One can verify this with a general analysis of the asymptotic behaviour of the spherical RT surfaces without referring to the AdS solution~\eqref{eq:prof5}. We note that from eq.~\eqref{eq:zz22} for $\rhoo<L$, 
\begin{equation}
    A_\mt{reg}=A_\mt{AdS}-A_\mt{UV}=-2\pi^4L^8\,\log2+O(L^{12}/\rhoo^2\ruv^2)\,.\label{eq:aregads}
\end{equation}

\begin{figure}
    \hspace{2cm}\includegraphics[width=0.9\linewidth]{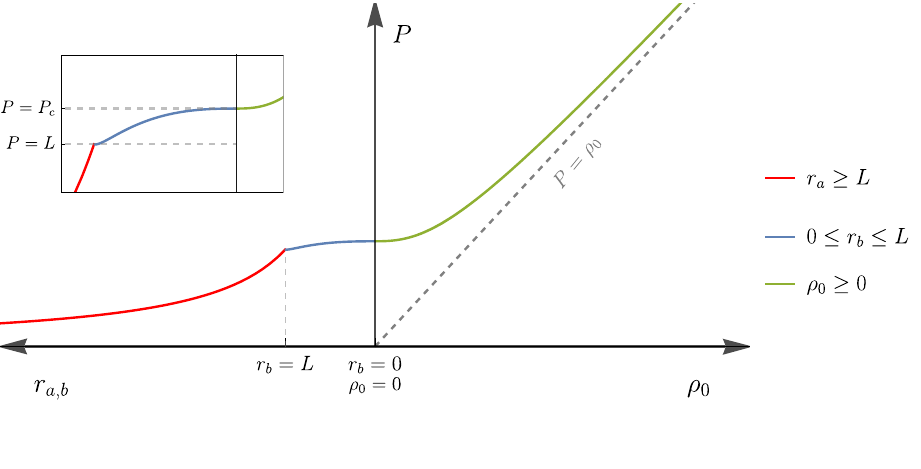}
    \caption{The radius $P$ of the boundary circle as a function of $r_a$, $r_b$ and $\rho_0$. i)~{$r_a>L$}: The RT surface is the usual hemisphere probing only the AdS region and the boundary radius is given $P=L^2/r_a$, as shown in red. ii) $0\leq r_b \leq L$: The RT surface closes off in the flat region and the blue segment shows the numerically determined function $\rhoo(r_b)$.  Note  there is a kink where the red and blue curves meet.  iii) $\rho_0>0$: The RT surfaces reaches $r=0$ at  $\rho=\rho_0$. The corresponding boundary radius is shown in green. }
    \label{fig:Pplot}
\end{figure}

Before turning to the area, we first comment on the behaviour of the radius $\rhoo$ in the boundary, as illustrated in figure \ref{fig:Pplot}. As we noted above, there is a unique extremal bulk surface for each value of the boundary radius $P$. Hence in contrast to the case of the strip geometry, there is no phase transition in the behaviour of the corresponding area or entanglement entropy. There are two special values of $P$: The first is $P=P_\mt{shell}$, where the RT surfaces just touch the shell, \ie $r_a=L=r_b$. Given the analytic solution \eqref{eq:prof5}, we find $P_\mt{shell}=L$. The second special radius is $P=P_c$, where the extremal surface just reaches the center of the flat space region, \ie $r_b=0=\rho_0$. In the bubble, this corresponds to the critical solution in case $c$ of eq.~\eqref{eq:p2-bdycond} and we must solve for the evolution in the throat region numerically, which yields $\rhoo_{c} \simeq 1.09\,L$.

Hence as we scan through the various solutions, we find:
From eq.~\eqref{eq:prof5}, we have $r_a = L^{2}/P$, and therefore, as the boundary radius ranges from $\rhoo = 0$ to $P_\mt{shell}=L$, we scan through the AdS solutions with $r_{a}$ decreasing monotonically from $\infty$ to $L$,  as shown with the red curve in figure \ref{fig:Pplot}. For the bubble solutions, our numerical analysis shows that the boundary radius continues to increase monotonically from $P_\mt{shell}= L$ to $\rhoo_{c} \simeq 1.09\,L$ as $r_{b}$ decreases from $L$ to $0$. This corresponds to the blue segment in the figure.
Finally, for the family of bubble solutions with $\rho_{0}>0$, the boundary radius grows monotonically beyond $\rhoo_{c}$ as $\rho_{0}$ increases, asymptotically approaching $\rhoo\simeq \rho_0$ for very large $\rho_0$, as shown with the green curve. Hence, as noted above, there is a unique extremal surface for any given value of $\rhoo$ and so there is no phase transition of the form that we found in the previous section. 

\begin{figure}
    \hspace{1cm}\includegraphics[width=1\linewidth]{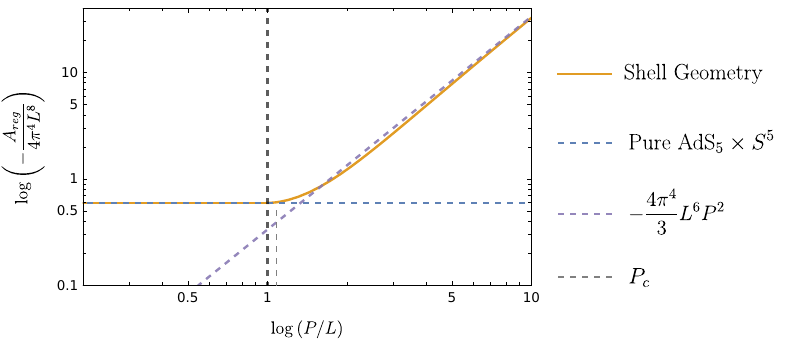}
    \caption{A log-log plot of the regularized area obtained by subtracting of the UV divergent area $A_\mt{UV}$ given in~\eqref{eq:zz22}. In this figure, we used $\ruv/L=50$. The dashed purple curve shows the quadratic power-law behavior of $A_{reg}$~\eqref{eq:AregApprox} for large $P/L$. Note that the area and hence the entanglement entropy is less for the shell geometry than for pure AdS$_5\times S^5$ for $P>L$, where the RT surfaces are probing the flat space bubble. 
    }   
    \label{fig:totareaplotreg}
\end{figure}

Recall that we regulate the total area of the RT surfaces by subtracting the divergent contribution $A_{\mt{UV}}$ given in eq.~\eqref{eq:zz22}. The regulated area is shown as a function of the boundary radius $\rhoo$ in figure~\ref{fig:totareaplotreg}. As noted above, there are no sharp phase transitions in which $\partial_{\rhoo} A_{\mt{reg}}$ exhibits a discontinuous jump. Instead, $A_{\mt{reg}}$ varies smoothly at $\rhoo = L$ and $\rhoo = \rhoo_{c}$ where the RT surface transitions from one family to the next. 
Furthermore, for $P\gg L$, a perturbative calculation in $P/L$ (see appendix \ref{app:perturb} for details) shows that the regulated area decreases as 
\begin{equation}
    A_{reg}= 4\pi^4L^8 \left(-\frac{P^2}{3L^2}\, +\frac{1}{2} \log{\frac{P}{L}}- \frac{11}{48}-\frac{61}{3240} \frac{L^2}{P^2}+\mathcal{O}(P^{-4})\right).
    \label{eq:AregApprox}
\end{equation}
The leading result corresponds to the area of a cylindrical surface $S^2\times R$ that falls straight in from the boundary to $r=0$ (minus the UV contribution~\eqref{eq:zz22}). 
We can write this expression as $A_{reg}\simeq - \frac{\pi^3}{3}\,L^6\,A_{ent}$, where $A_{ent}$ is the area of the entangling surface in the boundary theory. In this form, we see that this leading result matches that in eq.~\eqref{eq:stripAregFlat} for the strip geometry when the width of the strip is large.
The logarithmic term arises from the subtraction of the UV contribution~\eqref{eq:zz22}, which cancels a cutoff dependent logarithm $\log \ruv/{L}$ in the unregulated area.
From figure \ref{fig:totareaplotreg}, we observe that the RT surfaces have smaller area when they enter the flat space bubble compared to the corresponding surfaces in pure AdS$_5\times S^5$, and hence the corresponding entanglement entropies are smaller.

\paragraph{c-function:} It was argued in  \cite{Myers:2010xs,Myers:2010tj} that  spherical entangling surfaces are good probes of RG flows. Refs.~\cite{Liu:2012eea,Liu:2013una} proposed that the associated c-function can be extracted with the following geometric expression
\begin{equation}\label{eq:dof}
    {\cal S}_4(\rhoo)=\frac18\,\rhoo\,\frac{d\ }{d\rhoo}\left(\rhoo\,\frac{d\ }{d\rhoo}-2\right)S_\mt{EE}(\rhoo)=\frac18\,\left(\rhoo^2\,\frac{\partial^2S\mt{EE}}{\partial \rhoo^2}-\rhoo\,\frac{\partial S_\mt{EE}}{\partial \rhoo}\right) \,.
\end{equation}
for a four-dimensional boundary theory. We note that on the right hand side, one is differentiating the full holographic entanglement entropy including the UV divergences appearing in eq.~\eqref{eq:SphereDiv}. The first differential operator (\ie $\rhoo\,\partial_\rhoo-2$) removes the area law divergence while the second differentiation (\ie $\rhoo\,\partial_\rhoo$) isolates the universal coefficient of the logarithmic term. In pure AdS$_5\times S^5$, this refined entanglement entropy \eqref{eq:dof} can be computed exactly, giving 
\begin{equation}
    \mathcal{S}_4 ^{AdS} = \frac{\pi^4}{8} \frac{L^8}{G_{10}}=\frac{N^2}4=a\,.
    \label{eq:adsa}
\end{equation}
As indicated, this corresponds to the $a$ central charge appearing in the trace anomaly of the dual $N=4$ super-Yang-Mills theory \cite{Myers:2010xs,Myers:2010tj}.\footnote{Of course, $a=c$ for ${\cal N}=4$ super-Yang-Mills.}

\begin{figure}[htbp]
  \centering
  \begin{subfigure}{\linewidth}
     \hspace{3cm}\includegraphics[width=0.83\linewidth]{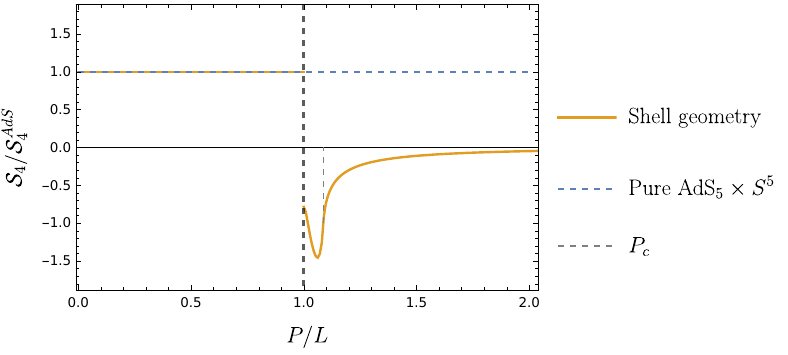}
    \caption{}
  \end{subfigure}

  \vspace{0.5em}

  \begin{subfigure}{\linewidth}
      \hspace{3cm}\includegraphics[width=0.8\linewidth]{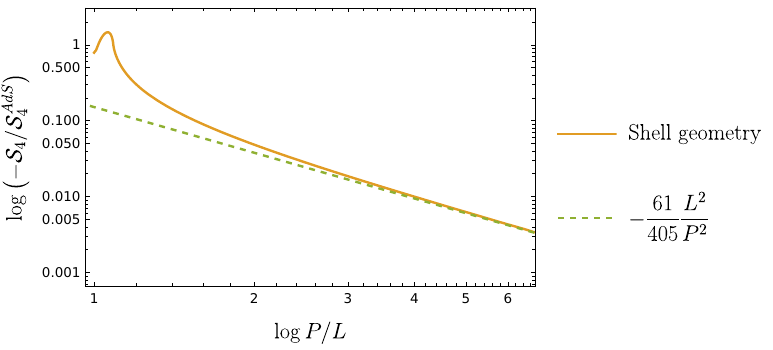}
    \caption{}
  \end{subfigure}  
  \caption{(a) Plot of the c-function~\eqref{eq:dof} as a function of the radius $\rhoo$ of the spherical entangling surface on the boundary. (b) A log-log plot of the c-function for $P\geq L$. The green dashed curve shows the leading contribution from the perturbative result~\eqref{eq:ctail} for large $P/L$. 
    }
    \label{fig:spheredof}
\end{figure}
We evaluated $\mathcal{S}_4$ numerically in the shell geometry, giving the result shown in figure \ref{fig:spheredof}. For small radii $\rhoo$ where the corresponding RT surface only probes the AdS region, $\mathcal{S}_4$ is constant as a function of $\rhoo$ and we reproduce the expected result in eq.~\eqref{eq:adsa}. When we reach $\rhoo=L$, $\mathcal{S}_4$ jumps down to the negative value $\mathcal{S}_4\simeq -0.81\,\mathcal{S}_4^{AdS} $. However, after reaching the minimum value  $\mathcal{S}_4\simeq -1.45 \,\mathcal{S}_4^\mt{AdS}$ at $\rhoo\simeq 1.06\,L$, the c-function rises and approaches zero asymptotically as $\rhoo\to\infty$. The tail in the c-function is related to the order $1/P^2$ corrections appearing for the regulated area in eq.~\eqref{eq:AregApprox}. The differential operator acting on the entanglement entropy in eq.~\eqref{eq:dof} removes the UV divergent terms in eq.~\eqref{eq:SphereDiv}, and also the $P^2$, $\log P$ and constant terms in the regulated area~\eqref{eq:AregApprox}. Hence for large $P$, we are left with
\begin{equation}
   \frac{\mathcal{S}_4}{\mathcal{S}_4 ^{AdS}} = - \frac{61}{405}\,\frac{L^2}{P^2}+\mathcal{O}(P^{-4})\,.
   \label{eq:ctail}
\end{equation}

As shown in the figure, the minimum of the c-function lies between $\rhoo=L$ and $\rhoo_c$, and therefore the corresponding RT surface is in the middle of the family of surfaces closing off at $\rho=0$ with $0<r_b\le L$. In particular, this does not occur for the critical surface on the third line of~\eqref{eq:p2-bdycond} with $\rhoo=\rhoo_c$. We also note that a discontinuity appearing at $P=L$ but not at $P=P_c$ seems to be related to the fact that there is a kink at $P=L$ in figure \ref{fig:Pplot} and no kink at $P=P_c$.

It is noteworthy that the behaviour in figure \ref{fig:spheredof}, where the c-function is not monotonic or even positive everywhere, is similar to that found in certain bulk geometries describing gapped phases or sharp RG flows in \cite{Liu:2012eea}. Let us also observe that the asymptotic limit where $\mathcal{S}_4$ vanishes seems to agree with the results for the c-function in the strip geometry where $\tilde c$ vanishes when $\ell$ was large enough and the RT surfaces probe the flat space bubble.  In any event, both c-functions in eqs.~\eqref{c1} and~\eqref{eq:dof}  indicate that there are far fewer degrees of freedom in the infrared of our Coulomb branch geometry, \ie associated with the flat space  bubble. 

\subsubsection*{$S^1$ entangling surface for $p=2$}

For general $p\neq 3$, even focusing on the throat region, there are no known analytic solutions for the RT surfaces corresponding to spherical entangling surfaces in the boundary theory. Hence we must rely on numerics in this case. Nevertheless, we can understand some properties of these solutions by noting a scaling similarity in eq.~\eqref{eq:zzzzx}. Making a constant scaling of the extremal solutions: $\rho(\sigma) \to \tilde \rho =\lambda\rho$ and  $r(\sigma)\to \tilde r  = \lambda^{-\frac{2}{5-p}}r $, we find that\footnote{We can also scale the worldvolume coordinate $\sigma$, but it has no effect on the final scaling in eq.~\eqref{eq:elacs}.}
\begin{equation}
      \mathcal{L}(\tilde \rho,\, \partial_\sigma\tilde \rho,\, \tilde r,\, \partial_\sigma\tilde r)= \lambda^{-\frac{(p-3)^2}{5-p\ }} \mathcal{L}(\rho,\,\partial_\sigma\rho,\,r,\,\partial_\sigma r)\,,
      \label{eq:elacs}
\end{equation}
where the Lagrangian $\mathcal{L}$ is the integrand in eq.~\eqref{eq:zzzzx}. As a consequence, this transformation maps solutions to solutions.  More precisely, we see that for the extremal surfaces remaining in the throat region, solutions at different $P$ can be related to one another by the above scaling which takes $P \to \lambda P$. This immediately tells us that the relationship between the boundary radius $P$ and the minimal bulk radius $r_a$ takes the form 
\begin{equation}
    r_a = a_p\, r_p\left(\frac{r_p}{P}\right)^{\frac{2}{5-p}}\, ,
\end{equation}
where the constant $a_p$ is to be fixed by numerics. For $p=3$ with $r_3=L$ and setting $a_3=1$, we recover $r_a=L^2/P$ as found from the analytic solution in eq.~\eqref{eq:prof5}. If now we turn to $p=2$, which will be the focus of the present discussion, we solved the equation of motion \eqref{eq:r-Dp-eom} for multiple values of $r_a$. Then extracting the corresponding $P$ and fitting, we find $a_2 \simeq 0.72$.

In \cite{vanNiekerk:2011yi}, the profile $\rho(r)$ was solved by series expansion  around $r=\infty$ to obtain the UV divergent contributions to the  disk entanglement for general $p$. Again focusing on $p=2$, the divergent structure was found to take the form
\begin{equation}
    A=\frac{16\pi^4}{15}\,r_2^5\, \ruv^2\,P + \text{finite terms}\,,
    \label{CircDiv}
\end{equation}
showing the universal leading area-law divergence with no subleading divergences. Dividing by $4G_N$ and using eqs.~\eqref{eq:sugra} and \eqref{pcutoff}, we can rewrite the above expression in the form given by the first line of eq.~\eqref{eq:SphereDiv}. Comparing the above result with eq.~\eqref{eq:StripDiv} (for $p=2$), we see that as expected the leading divergence takes the form of an area law with $A\simeq\frac{8\pi^3}{15}\,r_2^5\, \ruv^2\,A_{ent}$, where $A_{ent}$ is the area of the entangling surface. Above, $A_{ent}=2\pi P$ while in eq.~\eqref{eq:StripDiv}, we have $A_{ent}=2V_1$. Of course, we expect that this form applies for entangling surfaces with general geometries. 

Now denoting the divergent term in eq.~\eqref{CircDiv} as $A_{\mt{UV}}$, we define $A_{reg} \equiv A- A_{\mt{UV}}$. This regulated area will contain terms proportional to inverse powers of the cutoff radius $\ruv$, but we are only interested in the finite contributions which independent of of this cutoff, \ie which remain in the limit $\ruv\to\infty$. This finite contribution is a function of $P$ and $r_p$ alone but the form is completely fixed by the scaling similarity introduced above. Eq.~\eqref{eq:elacs} indicates 
$A_{reg}\to \lambda^{-1/3} A_{reg}$ and hence we find
\begin{equation}
    A_{reg}
    = -b_2 \ \frac{32\,\pi^4}{15}\  r_2^8 \, \left(\frac{r_2}{P}\right)^{\frac{1}{3}}\,, \label{eq:p2-Areg-1}
\end{equation}
where $b_2$ is a numerical constant which must be fixed by numerically solving for the extremal surfaces. Integrating the aforementioned numerical solutions using eq.~\eqref{eq:r-Dp-eom}, we find $b_2 \simeq 0.30$.

As in the previous section, there is again a unique extremal bulk surface for each value of the boundary radius $P$. As before, there are two special values of $P$ but these must now both be determined numerically. First, we have $P=P_\mt{shell}$, where the RT surface just touches the shell of D2-branes, \ie $r_a=L=r_b$, for which we find
\begin{equation}
    P_\mt{shell}=a_2^{\frac32}\,\left(\frac{ r_2}{R}\right)^{\frac{5}{2}}R\simeq 0.61\,\left(\frac{ r_2}{R}\right)^{\frac{5}{2}}R\,.
    \label{eq:Pshell}
\end{equation}
Again, the second special radius is $P=P_c$, where the extremal surface just reaches the center of the the flat space region, \ie $r_b=0=\rho_0$. In the bubble, this corresponds to the critical solution in case $c$ of eq.~\eqref{eq:p2-bdycond} and we must still solve numerically for the surface as it extends out in the throat region. In this case, we find
\begin{equation}
    P_c \simeq 0.62 \,\left(\frac{ r_2}{R}\right)^{\frac{5}{2}}R\,.
\end{equation}
Hence, these two special radii are very close with $P_{\mt{shell}}/P_c \approx 0.98$. Further, we note that both of these expressions show the parametric same scaling with $R$ and $r_2$ as the critical width of the strip in eq.~\eqref{eq:stripTransitionWidthDp}.

\begin{figure}
    \centering
    \includegraphics[width=0.7\linewidth]{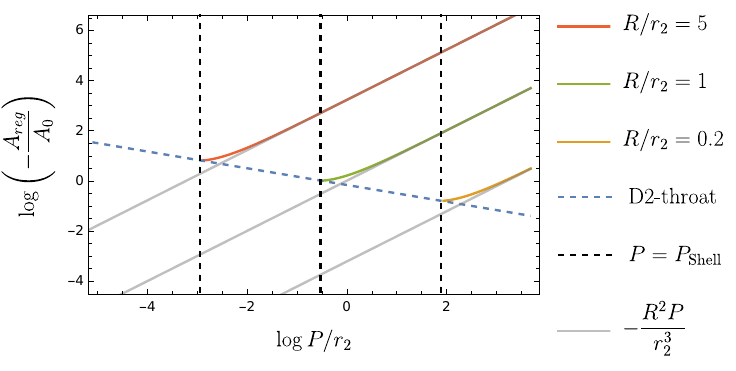}
    \caption{A log-log plot of the regulated area against $P/r_2$. In this figure, we used $\ruv/r_2 = 1000$. The gray opaque lines show the leading large $P/r_2$ behavior given in eq.~\eqref{eq:p2-A-pert}. For notational conciseness, we have introduced the constant $A_0=\frac{16 \pi^4 r_2^8}{21}$. 
    Note that the areas for surfaces probing the flat space region (\ie with $P>P_\mt{shell}$) are smaller than in the D$2$-throat. }
    \label{fig:p2Areg}
\end{figure}

The areas for three choices of $R/r_2$ are plotted in figure~\ref{fig:p2Areg}. 
 The general result is as follows: For $P \leq P_\mt{shell}$, the regulated area is described by eq.~\eqref{eq:p2-Areg-1}, where $b_2 \simeq 0.30$. We note that in this regime, $A_{reg}$ is negative and increasing towards zero with increasing $P$. There is then a continuous transition to a different class of surfaces for the range $P=P_\mt{shell}$ to $P_c$. We note however that while $A_{reg}$ is continuous, $\partial_P A$ is discontinuous at $P=P_\mt{shell}$ -- see discussion of the c-function below. There is another, smooth transition at $P=P_c$ where $A$ and $\partial_P A$ are both continuous. We see from the figure~\ref{fig:p2Areg} that in this regime $P>P_\mt{shell}$, $A_{reg}$ is negative and decreases (\ie becomes more negative) with increasing $P$. A perturbative calculation (see appendix \ref{app:perturb}) for $P\gg R$ yields
\begin{equation}
    A_{reg} = -\frac{16 \pi^4 r_2^8}{21}\left( \frac{R^2 P}{r_2^3} + \frac{2}{9} \frac{r_2^2}{R P}  + \mathcal{O}\left(P^{-3}\right)\right)\,. \label{eq:p2-A-pert}
\end{equation}
where the first term corresponds to a cylinder that falls straight into the origin and the second term is the leading correction due to the surfaces curving in towards $\rho=0$. In this regime, the regulated area decreases linearly with $P$. Thus in comparison to eq.~\eqref{eq:p2-Areg-1}, we see that at large $P/r_2$ the shell geometry has less entanglement than the pure D$2$-throat. Figure \ref{fig:p2Areg}, shows that this conclusion extends to all values of $P>P_{\mt{shell}}$. We conclude that the entanglement entropies are smaller whenever the RT surfaces are probing the flat space region. 

\paragraph{c-function:} 
Following \cite{Liu:2012eea}, we construct a c-function  using the refined entropy for three boundary dimensions
\begin{equation}
    \mathcal{S}_3(P)= \gamma_3 \left(P \frac{d}{dP} -1 \right)S_{EE}(P)\,, \label{eq:S3def}
\end{equation}
where we have introduced a normalization constant
\begin{equation}
    \gamma_3= \frac{5 }{8 \pi\, b_2 }\left(\frac{3}{ 4 \pi}\right)^{\frac{1}{3}}\,.
\end{equation}
Recall $b_2 \simeq 0.30$, from above.
By design, the differential operator removes the universal UV divergent piece in eq.~\eqref{CircDiv}, leaving a finite result, which however may still depend on $P$. 
In the pure D$2$-brane geometry (or for $P\leq P_\mt{shell}$), this finite contribution may be determined using the regulated area~\eqref{eq:p2-Areg-1}. Then using eq.~\eqref{eq:sugra} for the effective coupling, we have
\begin{equation}
    \mathcal{S}_3(P) = \gamma_3\,b_2\, \frac{32\pi^4 r_2^8}{45 G_{10}} \left(\frac{r_2}{P}\right)^{\frac{1}{3}}= \,\frac{N^2}{ \hat g _{eff} \left(1/P\right)^{{2}/{3}}}=\frac{N^2}{(g_{YM}^2 N)^{1/3}}\,\frac1{P^{1/3}}\,.
    \label{eq:S3throat}
\end{equation}
We observe that the refined entropy decreases as $P$ increases and the entanglement entropy probes longer wavelengths. Since the corresponding boundary theory is a nonconformal QFT, this behaviour is expected, \ie the c-function decreases as we  flow to IR.

\begin{figure}
    \centering
   \includegraphics[width=0.75\linewidth]{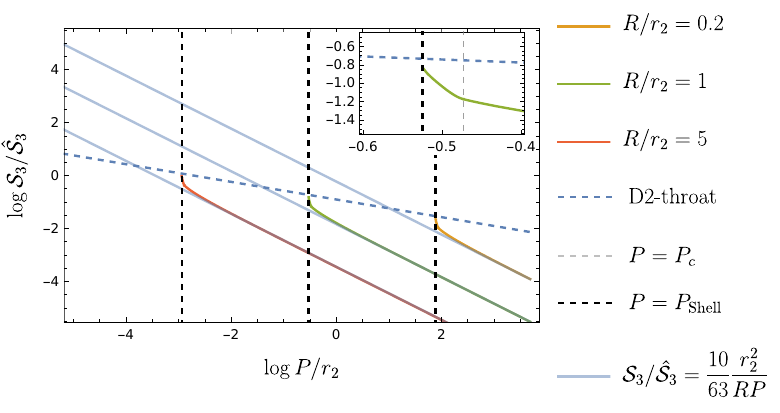}
    \caption{A log-log plot of the dof $\mathcal{S}_3$ against $P/r_2$. The gray opaque lines show the large $P/r_2$ behavior~\eqref{eq:S3-pert}. For notational conciseness we have introduced the constant $\hat{\mathcal{S}}_3= \frac{8 \pi^4}{15}\frac{r_2^8}{G_{10}}$. Note that $\mathcal{S}_3$ goes to zero faster in the shell-geometry than in the D$2$-throat.}
    \label{fig:S3}
\end{figure}

In the shell geometry, the result for large $P/r_2$ can be obtained from eq.~\eqref{eq:p2-A-pert}. Again the differential operator kills the leading linear term in $P$, leaving 
\begin{equation}
    \mathcal{S}_3(P)  = \gamma_3\,\frac{16 \pi^4 \,r_2^8}{189\, G_{10}}\,\frac{r_2^2}{R P}  + \mathcal{O}\left(P^{-3}\right) = \frac{5}{14\, (36 \pi)^\frac{1}{3}\,b_2}\,  \frac{N^2}{ \sqrt{\langle \text{tr}\, \Phi^2 \rangle} \, P} + \mathcal{O}(P^{-3}). 
    \label{eq:S3-pert}
\end{equation}
Hence we see that at large $P/r_2$, $\mathcal{S}_3$ decays to zero faster in the shell geometry than in the D$2$-throat geometry. Figure \ref{fig:S3} shows that this holds also for general $P>P_\mt{shell}$. Thus, we find that the shell-geometry leads to a reduction in the number of degrees of freedom in the infrared, where the RT surfaces are probing the flat space region.

Examining the full running of the c-function in figure \ref{fig:S3}, we see that the c-function is everywhere positive and monotonically decreasing, but it is not continuous at $P=P_{\mt {Shell}}$.
The jump there is small (see the inset) with $\Delta \mathcal S_3/\hat{\mathcal{S}}_3\simeq 0.054,\, 0.120 , \,0.270 $ for $R/r_2= 1/5 ,\, 1 ,\, 5$, respectively. This seems to follow the scaling $\Delta \mathcal S_3/\hat{\mathcal{S}}_3\simeq 0.12   \left(R/r_2 \right)^{1/2}$. Note that the exponent is the expected scaling with $R$ obtained by substituting eq.~\eqref{eq:Pshell} into eq.~\eqref{eq:S3throat}. We note that the monotonicity of $\mathcal{S}_3$ is guaranteed from the field theory perspective as it is a consequence of strong-sub-additivity \cite{Casini:2012ei}. Meanwhile, strong sub-additivity does not guarantee the monotonicity of $\mathcal{S}_4$ in the previous section, which is related to the fact that $\mathcal{S}_4$ involves a second derivative.

\section{Target space entanglement} 
\label{sec:target}

In this section, we examine a different class of extremal codimension-two surfaces
which end on non-trivial profiles on the internal sphere $S^{8-p}$. These surfaces
have been proposed as holographic duals of target space entanglement entropy
\cite{Das:2022njy,Mollabashi:2014qfa,Karch:2014pma}.
Target space entanglement entropy measures entanglement across a partition of the
target space, where the underlying fields take their values, rather than across a
partition of the base space on which the fields live. While this notion is somewhat
unfamiliar from the conventional perspective of local quantum field theory, it arises
naturally in string theory, where the base space is the worldsheet and the target
space is the ambient spacetime \cite{Balasubramanian:2018axm,Hartnoll:2015fca,Prudenziati:2018jcf} --
see also \cite{Mazenc:2019ety} for a general discussion.\footnote{See also
\cite{Ahmadain:2022eso,Ahmadain:2022tew,Ahmadain:2025pox} for progress in
understanding entanglement entropy in off-shell string field theory.}

In the context of AdS$_5\times S^5$, target space entanglement has been argued to be captured by minimal codimension-two surfaces which have a non-trivial profile on
the internal space $S^5$ \cite{Das:2022njy,Mollabashi:2014qfa,Karch:2014pma}. Following \cite{Das:2022njy}, we will refer to these as
internal RT surfaces. Such surfaces were originally studied in multi-center
geometries \cite{Karch:2014pma,Mollabashi:2014qfa}, where the underlying D3-branes are separated into two stacks of $n$ and $m$ branes. From the boundary perspective, this corresponds to a Coulomb branch solution where the $U(n+m)$ gauge theory decomposes into separate $U(n)$ and $U(m)$ sectors.

There is another motivation for considering these internal RT surfaces in the present
setting. Inside the shell, the geometry contains a flat-space region with an
$\mathbb{R}^{1,9-p}$ factor. The angular directions of the internal $S^{8-p}$ in the
throat become the angular directions in this component of this flat region,
and hence the $S^{8-p}$ plays the role of the celestial sphere for the flat space bubble. Thus internal RT surfaces anchored on non-trivial profiles on the $S^{8-p}$ provide a natural way to make contact with the bottom-up picture of holographic entanglement entropy in flat space developed in section \ref{sec:botup}. In particular, these surfaces are anchored on
regions of the sphere at the edge of the throat and can, when they enter the shell, be interpreted as probing angular subregions of the celestial sphere in the flat space region. Our goal will be to understand how such surfaces actually probe the flat space bubble and, importantly, when they are the dominant saddles.
As alluded to in section \ref{sec:intro}, a key feature in the latter will be the size of the flat space region~\eqref{eq:flatsize} and our ability to vary this size for $p\ne3$.

Before proceeding, we wish to recall a result by Graham and
Karch~\cite{Graham:2014iya}. The latter
examined the asymptotic behaviour of minimal area surfaces in product spacetimes of the form  asymptotically anti-de Sitter space times a compact internal manifold, such as $AdS_5\times S^5$. Their interesting result was that the internal part of such surfaces  must be anchored on an extremal submanifold in the internal space. 
Intuitively, this behaviour occurs because, near the boundary, the internal geometry is constant
decoupling from the radial direction, and the extremality condition forces the surface to approach an extremal submanifold of the internal space. 
Hence their result constrains any extremal internal RT surface in $AdS_5\times S^5$ to approach an equator of the $S^5$ as the asymptotic boundary is approached. Consequently,
in order to anchor such internal RT surfaces to a more general codimension-one surfaces on the $S^5$, one must introduce a cutoff surface at a finite radius and specify the anchoring surface there \cite{Mollabashi:2014qfa}. While the Graham-Karch theorem only applies to our shell geometries for $p=3$, we will find analogous behaviour arises even with $p\ne 3$.

Now, we begin with the D$p$-brane throat geometries for $0\le p\le4$. We introduce
coordinates on the internal sphere appearing in eq.~\eqref{eq:Dp-metric},
\begin{equation}
    d\Omega_{8-p}^2=d\theta^2+\cos^2\theta\,d\Omega_{7-p}^2\,,
    \label{eq:internalSphereCoords}
\end{equation}
so that the equator lies at $\theta=0$ and the poles at
$\theta=\pm\pi/2$. We consider surfaces lying in a fixed time slice which wrap all $p$ spatial worldvolume directions of the D$p$-brane throat and have a
non-trivial profile $\theta=\theta(r)$ on the internal sphere. As we did previously, the regulated volume of the $x^i_\parallel$ directions is denoted $V_p$. The surfaces of 
interest cap off smoothly at some radius $r=r_c$, and without loss of generality, we choose $\theta(r_c)=\pi/2$.  

Let us first consider surfaces which lie entirely in the throat region,
$r\ge R$. The corresponding area functional is
\begin{equation}
    {\cal A}_\mt{throat}
    =
    V_p\,\Omega_{7-p}\, r_p^{\,{7-p\over2}}
    \int dr\,r^{\,{7-p\over2}}\,
    (\cos\theta)^{7-p}\,
    \sqrt{1+r^2\,\dot\theta^{\,2}}\,,
    \label{eq:DpbraneArea}
\end{equation}
where again, $\Omega_{7-p}=2\pi^{\frac{8-p}{2}}/\Gamma(\frac{8-p}{2})$ is the volume of the round unit sphere $S^{7-p}$, and dots denote derivatives with respect
to $r$. Extremizing eq.~\eqref{eq:DpbraneArea} gives
\begin{equation}
    \ddot\theta
    =
    -{(7-p)\tan\theta\over r^2}
    -{(11-p)\dot\theta\over 2r}
    -(7-p)\tan\theta\,\dot\theta^{\,2}
    -{(9-p)\over2}\,r\,\dot\theta^{\,3}\,.
    \label{eq:DpThetaEOM}
\end{equation}
Equivalently, writing the surface as $r=r(\theta)$ gives
\begin{equation}
    r''
    =
    {9-p\over2}\,r
    +(7-p)\tan\theta\, r'
    +{(11-p)r'^{\,2}\over2r}
    +{(7-p)\tan\theta\, r'^{\,3}\over r^2}\,,
    \label{eq:DprThetaEOM}
\end{equation}
where primes denote derivatives with respect to $\theta$. Now we observe that in either form, the equation of motion is left invariant if we scale $r\to\lambda r$ (up to an overall factor of $\lambda^{-2}$ and $\lambda$ in eqs.~\eqref{eq:DpThetaEOM} and~\eqref{eq:DprThetaEOM}, respectively). 
This implies that this scaling maps a given solution of the equation of motion to a new solution. This can also be inferred from the fact that, up to subtleties having to do with the UV cutoff that we explain below, the action~\eqref{eq:DpbraneArea} scales with a factor of $\lambda^3$ under the radial rescaling. 

Examining eq.~\eqref{eq:DprThetaEOM}, we see that for the acceleration $r''$ to remain finite as the surface closes off (\ie $\theta\to\pi/2$), we must have $ r' =0$ at $r=r_c$. The power series expansion near the closing-off point then takes the form
\begin{equation}
    r(\theta)
    =
    r_c\left[
    1+{9-p\over4(8-p)}
    \left({\pi\over2}-\theta\right)^2+\cdots
    \right]\,.
    \label{eq:DpClosingExpansion}
\end{equation}
Higher order terms above are completely fixed in terms of $r_c$, \ie these solutions that close off with $r_c>R$ are completely specified by the single parameter $r_c$. Since as noted above, we can change $r_c$ by rescaling the radial coordinate, we see that all such solutions are related by such a scaling. In other words, all solutions with $r_c > R$ can be written in terms of a universal profile $\theta(r)=\theta_0 (r/r_c)$, where $\theta_0$ is obtained by solving the equations of motion~\eqref{eq:DpThetaEOM} in terms of the dimensionless coordinate $r/r_c$ with the closing-off condition $\theta_0(r/r_c=1) = \pi/2$. 

\begin{figure}[htbp]
  \centering
  \begin{subfigure}{\linewidth}
    \hspace{2.8cm}\includegraphics[width=.8\linewidth]{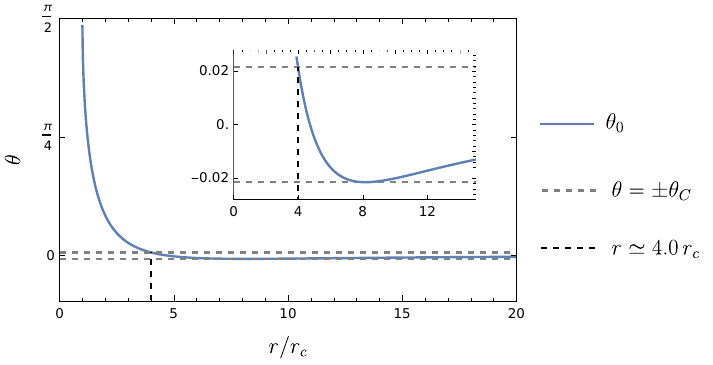}
    \caption{}
  \end{subfigure}

  \vspace{0.5em}

  \begin{subfigure}{\linewidth}
    \hspace{3.5cm}\includegraphics[width=.8\linewidth]{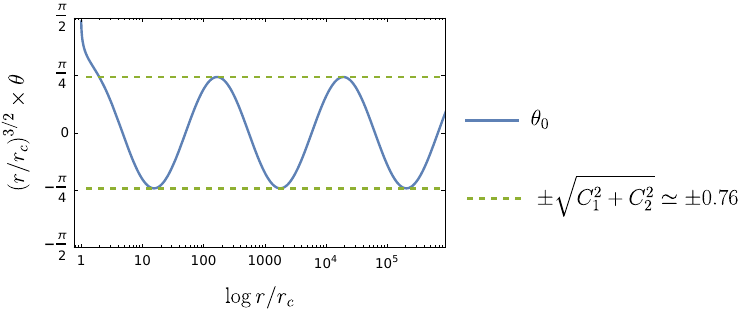}
    \caption{}
  \end{subfigure}
   
  \caption{(a) A plot of the universal solution $\theta_0(r/r_c)$ for $p=3$, valid for $r_c \geq R$ in the shell geometry. The behaviour exhibited here for $p=3$ is typical of what we find for general $p$. (b) A plot of $\theta_0$ for $p=3$ with the $\theta$ axis rescaled to clearly show the oscillatory behavior. The $r/r_c$ axis is scaled logarithmically. The green dotted lines show the amplitude of the oscillation as calculated in the text.}
  \label{fig:theta0}
\end{figure}
We illustrate this universal profile for $p=3$ in figure \ref{fig:theta0}, but this behaviour is typical for all values of $p$. Starting from \(\theta_0(1)=\pi/2\), the profile rapidly decreases to a global minimum below the equator, and subsequently undergoes damped oscillations (as described below). Asymptotically, the universal profile approaches $\theta_0\sim0$ for large $r/r_c$. The latter behaviour is dictated by the Graham-Karch theorem \cite{Graham:2014iya} for $p=3$, but this structure is typical of what we find for general $p$. 

Suppose that we now anchor the surface at $\theta(\ruv)=\thbdy$ on the cutoff surface. Let us first consider the case $\thbdy\sim O(1)$. From figure~\ref{fig:theta0}, it is clear that there is a unique surface realizing this boundary condition, with $r_c\simeq \ruv$. More generally, there is a single candidate surface for all $|\thbdy|>\theta_C$, where $\theta_C$ corresponds to the global minimum described above. Consequently, for boundary data $\thbdy\in[-\theta_C,\theta_C]$, there are always at least two candidate surfaces: one which closes off at $\theta=\pi/2$, and another which closes off at $\theta=-\pi/2$, with a slightly shifted value of $r_c$ relative to the first.
Table~\ref{tab:DpthetaC} gives our numerical estimates of $\theta_C$ for different values of $p$. 

The table also shows the first zero of the universal profile, which we present in terms of the inverse ratio ${\cal R}=r_c/r_0$. Across the full range of $p$, we find that ${\cal R}$ is approximately $1/5$. The significance of this ratio is that, as $\thbdy$ is swept from $\pi/2$ to $0$, one can always find a candidate internal RT surface with $r_c\ge{\cal R}\,\ruv$. In other words, there are always candidate surfaces which remain very close to the UV cutoff surface for any value of $\thbdy$.
\begin{table}[h]
\centering
\begin{tabular}{c|c|c|c|c|c}
    $p$ & $0$ & $1$ & $2$ & $3$ & $4$ \\
    \hline
    $\theta_C$ & $0.003$ & $0.007$ & $0.012$ & $0.022$  & $0.036$ \\
    ${\cal R}=r_c/r_0$ & $0.144$ & $0.170$ & $0.192$ & $0.211$  & $0.225$ \\
\end{tabular}
    \caption{The critical value $\theta_C$ such that multiple solutions arise when {$|\thbdy|\leq\theta_C$}.}
    \label{tab:DpthetaC}
\end{table} 

As noted above, when $\thbdy<\theta_C$, there are multiple candidates for the internal RT surface. To understand this feature better, we study the asymptotic behaviour of $\theta_0(r/r_c)$ in the regime $r/r_c\gg 1$. In this regime, $\theta_0$ may be analyzed by linearizing the equations of motion around $\theta_0(r/r_c)\simeq0$.
As noted in~\cite{Karch:2014pma,Das:2022njy}, the internal RT surfaces approach
the equator asymptotically with a characteristic oscillatory behaviour,
\begin{equation}
    \theta_0(r/r_c)
    \simeq
    \left({r_c\over r}\right)^{(9-p)/4}
    \left[
    C^{(p)}_1
    \cos\!\left(
    {\sqrt{31+2p-p^2}\over4}\log {r\over r_c}
    \right)
    +
    C^{(p)}_2
    \sin\!\left(
    {\sqrt{31+2p-p^2}\over4}\log {r\over r_c}
    \right)
    \right]\, .
    \label{eq:Dptheta0}
\end{equation}
The constants $C^{(p)}_1$ and $C^{(p)}_2$ are fixed by the full nonlinear solution near the cap. Note that two coefficients $C^{(p)}_i$ are independent of $r_c$. Numerically solving for $\theta_0$ in the dimensonless coordinate $r/r_c$ and fitting the asymptotic tail to eq.~\eqref{eq:Dptheta0}, gives the numerical values in Table \ref{tab:cp}, with the right most column showing the  overall amplitude. This damped oscillatory behaviour shows us that the smaller we choose $\thbdy$ (\ie closer to the equator in $S^5$), the more candidate surfaces we will have. In particular, in the pure throat geometry $\thbdy=0$ would give infinitely many candidate surfaces. In the shell geometry, we will see that there is always a finite number of surfaces for given boundary profile. 
\begin{table}[h]
\centering
\begin{tabular}{c|c|c|c}

$p$ & $C_1^{(p)}$  & $C_2^{(p)} $ & $\sqrt{(C_1^{(p)})^2+(C^{(p)}_2)^2}$ \\
\hline

0 & 0.48 & 1.02 & 1.12\\

1 & 0.56 & 0.76 & 0.94\\

2 & 0.62 & 0.55 & 0.83\\

3 & 0.67 & 0.35 & 0.76\\

4 & 0.72 & 0.16 & 0.74\\



\end{tabular}
\caption{Numerical results for the coefficients appearing in eq.~\eqref{eq:Dptheta0}.}
\label{tab:cp}
\end{table}

Let us now consider surfaces that close off inside the shell, \ie with $r_c < R$. We know that the surface will intersect the brane shell at $r=R$ on some line of latitude. As in the discussion of the bottom-up hologram in section \ref{sec:botup}, the extremal surface in the flat region is then simply a flat plane anchored on this line of latitude. In other words, for $r\leq R$ we have 
\begin{equation}
    \theta(r)= \arcsin\left(\frac{r_c}{r}\right),
    \label{eq:flatplane}
\end{equation}
where as above, we have chosen the surface to close off at $\theta=\pi/2$ pole. Geometrically $r_c$ is simply the shortest distance between the origin $r=0$ and the extremal surface. In order to extend the solution into the throat region, we briefly confirm that the usual matching condition holds. The area functional in the flat space region is 
\begin{equation}
    A_\mt{flat} =  V_p\,\Omega_{7-p}\left(\frac{r_p}{R}\right)^{\frac{7-p}{2}} \int dr \, r^{7-p}\, \cos^{7-p}\theta \sqrt{1+r^2\,\dot \theta ^2}\,.
    \label{Aflat}
\end{equation}
As usual, we demand that the boundary terms coming from the variation of eqs.~\eqref{eq:DpbraneArea} and~\eqref{Aflat}  cancel at the shell, which yields 
\begin{equation}
      \frac{r_p^{\frac{7-p}{2}}\,r^{\frac{11-p}{2}}\,\cos^{\frac{7-p}{2}}\! \theta \,  \dot \theta }{\sqrt{1+r^2 \dot \theta^2}}\delta \theta \bigg|^\mt{throat}_{r=R}=
     \left(\frac{r_p}{R}\right)^{\frac{7-p}{2}} \frac{r^{9-p} \cos^{\frac{7-p}{2}}\!\theta\, \dot \theta}{\sqrt{1+r^2 \dot \theta ^2}} \delta \theta\bigg|^\mt{flat}_{r=R}\,.
      \label{eq:matchy}
\end{equation}
Hence, we find that $\dot \theta $ is continuous across the shell. Given the profile~\eqref{eq:flatplane}, we then have the initial conditions to solve for the profile in the exterior region, \ie
\begin{equation}
    r=R\ :\quad \theta=\arcsin \left(\frac{r_c}{R}\right)\,,\quad
    \dot\theta= -\frac{1}{R \sqrt{\frac{R^2}{r_c^2}-1}}\,.
    \label{eq:intmatch}
\end{equation}
Extending the solution to the asymptotic region $r\gg R$, the surface will have the same asymptotic form~\eqref{eq:Dptheta0} as the surfaces that close off outside the shell. However, the coefficients that appear in front of the trigonometric functions will differ from the $r_c > R$ case. Let us call the new factors $\tilde C_i^{(p)}(r_c)$ in this  case. As indicated, the constants will depend on $r_c$, since the shell breaks the scaling symmetry (or rather similarity) that was present in the throat region. 

In the regime $r_c \ll R$, we can estimate the $\tilde C_i^{(p)} (r_c)$ to leading order in $r_c$, because the boundary conditions in eq.~\eqref{eq:intmatch} reduce to $\theta(R)\simeq R\dot\theta(R)\simeq\tfrac{r_c}{R}\ll1$. Therefore $\theta(r)$ is small everywhere outside of the shell and the profile takes the same form as in eq.~\eqref{eq:Dptheta0}. Solving for the profile with the preceding boundary conditions, one finds
\begin{equation}
\begin{split}
    \tilde C^{(p)}_1 \simeq  &\left(\tfrac{R}{r_c}\right)^{\frac{5-p}{4}}\Bigg(\tfrac{(5-p)}{\sqrt{31+2p-p^{2}}}\cos\!\left(\tfrac{\sqrt{31+2p-p^{2}}}{4}\,\log\!\tfrac{R}{r_c}\right) +\sin\!\left(\tfrac{\sqrt{31+2p-p^{2}}}{4}\,\log\!\tfrac{R}{r_c}\right)\Bigg)\,,\\
    \tilde C^{(p)}_2  \simeq & 
   \left(\tfrac{R}{r_c}\right)^{\frac{5-p}{4}}\Bigg(\cos\!\left(\tfrac{\sqrt{31+2p-p^{2}}}{4}\,\log\!\tfrac{R}{r_c}\right) -\tfrac{(5-p)}{\sqrt{31+2p-p^{2}}}\sin\!\left(\tfrac{\sqrt{31+2p-p^{2}}}{4}\,\log\!\tfrac{R}{r_c}\right)\Bigg)\,.
\end{split}
\label{eq:tilCapprox}
\end{equation}
However, for general $r_c\leq R$, one must resort to numerics to determine $\tilde C^p _i(r_c)$. 
We also observe that the presence of the flat space bubble means that there are only a finite number of candidate surfaces for a given $\thbdy$ (including $\thbdy=0$), and usually fewer candidate surfaces than in the pure D$p$-throat geometry. This is because eq.~\eqref{eq:flatplane} is strictly monotonic, so oscillations can only occur for $r > R$.

Let us now turn to the areas of these surfaces. First, for surfaces that remain
entirely in the brane throat, the scaling symmetry implies that the total area scales
as $r_{\rm uv}^{(9-p)/2}$ if the anchoring angle $\theta(r_{\rm uv})=\thbdy$ is kept fixed
while the cutoff is moved. Thus, as emphasized in \cite{Das:2022njy}, one
cannot isolate a universal UV-divergent contribution to the target-space entropy in
the same way as for ordinary spatial entanglement entropy. The entire area scales
with the cutoff. This feature persists in the shell geometry: if $r_{\rm uv}$ is taken
large enough at fixed $\thbdy$, the corresponding surface is eventually pushed outside the shell, and the scaling argument for the pure D$p$-throat applies.

 The case of surfaces closing off in the throat region, \ie, $r_c\geq R$, was examined in \cite{Das:2022njy}, where they concluded that the area decreases monotonically with increasing $r_c$ for fixed $\ruv$. In other words, for a given $\thbdy$ with multiple candidate surfaces, the one with largest $r_c$ is the dominant saddle point. Since $\ruv$ is taken large, this means that these internal RT surfaces do not probe deeply into the bulk. In appendix \ref{sec:mon}, we note that, despite still appearing monotonic, the area appears to have a large number of saddlepoints in $r_c$ that were not previously noted. This occurs due to a surprisingly simple relationship between the coefficients $C^{(p)}_{1,2}$ and another constant that appears in the area depending on the full profile of $\theta_0$ -- see eq.~\eqref{eq:DpA}. According to this result, if $C^{(p)}_1$ or $C^{(p)}_2$ were only slightly larger in magnitude, the area would have been (slightly) non-monotonic. As long as the relationship is (approximately) true, as we confirm numerically, the conclusion that surfaces with large $r_c$ are favored is still valid.

Let us then consider the areas of surfaces with $r_c\leq R$. The area of the minimal surface in the flat region is just that of the flat plane anchored at $\theta(R) = \arcsin(r_c/R)$. The resulting area is 
 \begin{equation}
     A_{\mt{flat}} = V_p\Omega_{7-p} \left(\frac{r_p}{R}\right)^\frac{7-p}{2} \frac{\left(R^2 - r_c^2\right)^{\frac{8-p}{2}}}{8-p}\,.
     \label{eq:Aflat2}
 \end{equation}
In general, we need numerics to estimate the total area including the contribution form the throat region. However, before discussing the numerical result, we note that some qualitative features can be gleaned from the leading order result for $r_c \ll R$, namely, using eq.~\eqref{eq:tilCapprox}, 
    \begin{eqnarray}\label{eq:A-Dp-approx}
        A\simeq&  V_p \Omega_{7-p}& r_p^{\frac{7-p}{2}}\,R^{\frac{9-p}{2}}  \Bigg[\frac{2\, }{9-p}\,\left(\frac{\ruv}{R}\right)^{\frac{9-p}{2}}  - \frac{7-p}{(8-p)(9-p)} \\
        && + \frac{(r_c/R)^2}{ (31+2p-p^2)}\Bigg( (16-9p+p^2) \cos\!\left( \tfrac{\sqrt{31+2p-p^2}}{2}\log \tfrac{R}{\ruv}\right) \nonumber\\
        &&+ (6-p)\sqrt{31+2p-p^2}\sin\!\left( \tfrac{\sqrt{31+2p-p^2}}{2}\log \tfrac{R}{\ruv}\right)\nonumber\\
        &&+\frac{(9-p)(7-p)}{2} \Bigg)\Bigg]. \nonumber
    \end{eqnarray}
The first two terms give the result for $r_c= 0$, \ie the surface lying at~$\theta=0$. The fact that the second term is negative means that the area of this surface is reduced compared to the pure D$p$-throat geometry (\ie with $R=0$). Furthermore, the area decreases further as we increase the radius of the shell. The second order term in $r_c$ includes two oscillating terms and one non-oscillating term. The combined magnitude of the oscillating terms is $2(7-p)^\frac{3}{2}$, which is larger than the absolute value of the non-oscillating terms on last line for all $0\leq p \leq 4$. This shows, for instance, that the solution with $r_c =0$ can be a local minimum or maximum as a function of $r_c$ with fixed $\ruv$, depending on the choice of $\ruv$ and $R$. The numerics confirm that the reduction in area holds for
larger values of $r_c$.

\begin{figure}
    \centering
    \includegraphics[width=1\linewidth]{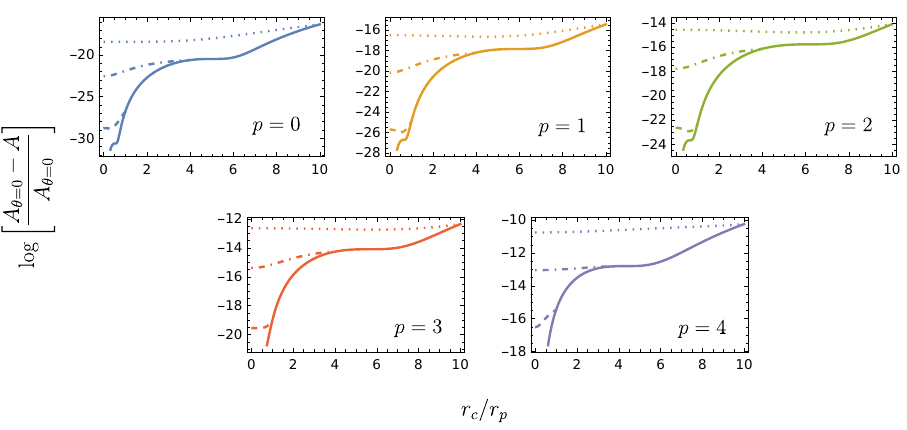}
    \caption{Logarithmic plots of the areas as function of $r_c/r_p$ with $R/\ruv=1/500$ (dashed line), $R/\ruv= 1 /125$ (dot-dashed line) and $R/\ruv= 1/50$ (dotted line). We have chosen $\ruv/r_p=100$. The solid curve shows the result in the pure brane geometry while the dashed, dot-dashed and dotted curves shows the result in the shell-geometry when $r_c\leq R$. Note that the $y$-axis is independent of $r_p$.}
    \label{fig:Dprc}
\end{figure}

\begin{figure}
    \centering
    \includegraphics[width=1\linewidth]{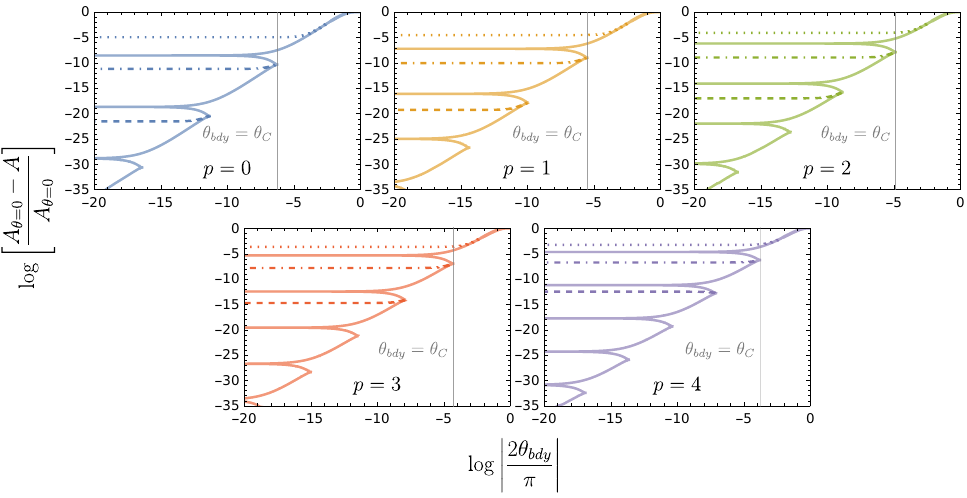}
        \caption{Logarithmic plot of the areas as function of $\thbdy$ 
        with $R/\ruv=1/100$ (dashed curve), $1/10 $ (dot-dashed curve) and $2/5 $ (dotted curve). We have chosen $\ruv/r_p=100$. The solid curve shows the result in the pure throat geometry while the `dashed' curves show the result in the shell-geometry when $r_c\leq R$. Note that the $y$-axis is independent of $r_p$. The gray lines show the critical angle $\theta_C$ listed in table~\ref{tab:DpthetaC}.
        }
    \label{fig:Dptheta}
\end{figure}

Figure~\ref{fig:Dprc} shows the logarithm of the relative area compared to that of the $\theta=0$ surface in the pure brane geometry as a function of $r_c$. The area of the $\theta=0$ surface in the pure brane geometry is given by$A_{\theta=0}=\frac{2V_p\Omega_{7-p}}{9-p}\, r_p^{\frac{7-p}2}\,\ruv^{\frac{9-p}2}$, \ie the first term on the first line of eq.~\eqref{eq:A-Dp-approx}. The solid curves show the result in the pure throat geometry. As discussed in appendix \ref{sec:mon}, there appears to be several approximate saddlepoints where the area is approximately stationary with respect to $r_c$. The dashed curves indicate the areas of surfaces probing the shell for three choices of the ratio $R/\ruv=1/500,1/125,1/50$. We see that increasing the ratio decreases the area, \ie
the corresponding curve appears higher in the plot.

Figure~\ref{fig:Dptheta} shows the same relative area as a function of $\thbdy$, now presented in a log-log plot. The solid curves correspond to the pure brane geometry. The stationary points in the area with respect to $r_c$ discussed above appear as the right-pointing cusps, \ie the first cusp corresponds to $\thbdy=\theta_C$, the critical angle given in table \ref{tab:DpthetaC}.  Meanwhile, the oscillatory dependence of $\thbdy$ on $r_c$ gives rise to the horizontal branches extending to $\thbdy=0$. Hence, for $0<|\thbdy|<\theta_C$, there is a finite discrete set of candidate surfaces realizing the same fixed boundary angle. As $|\thbdy|$ becomes smaller, the number of such candidate surfaces increases, becoming infinite at $\thbdy=0$.

Note that $r_c$ increases toward $\ruv$ in the upper right of figure~\ref{fig:Dptheta}, where $\thbdy=\pi/2$. The topmost horizontal branch extends $\thbdy\to0$ to the left, where $r_c\to r_0$ where $r_0$ was the first zero of the universal profile. Using the ratio in table \ref{tab:DpthetaC}, we can express this as $r_c\to {\cal R}\,\ruv$ with ${\cal R}\sim 1/5$ across all values of $p$. These candidate surfaces near the top of the plots are the dominant internal RT surfaces, and hence all have $r_c\sim \ruv$, \ie target space entanglement does not probe the IR of the throat region, as observed by \cite{Das:2022njy}. The plot also shows the areas of surfaces that probe the shell for three representative choices of the ratio $R/\ruv=1/100,1/10,2/5$. Generally, introducing the shell reduces the area and caps the number of candidate surfaces available at fixed $\thbdy$. Further, increasing the shell radius $R$ decreases the overall area of the surfaces probing the flat region. As illustrated by the curves with $R=\frac25\,\ruv$, for sufficiently large $R>r_0$,\footnote{Recall $R$ is a spurious metric parameter for $p=3$, \ie the size of the shell is independent of $R$. However, as we increase the ratio $R/\ruv$, we are moving the cutoff surface closer to the flat space region, which has a fixed size.} the surfaces probing the flat region become the dominant saddlepoints for sufficiently small $\thbdy$. Note that while the boundary values of $\thbdy$ remain small, the corresponding surfaces still sweep through the entire range $0\le\theta(R)\le \pi/2$. However, if $R$ lies close to $r_0$, there can still be oscillations in $\thbdy$ such that the dominant surfaces that probe the shell do not sweep through all of $0\leq r_c \leq R$. This situation is avoided for $R/\ruv \gtrsim1/4$ for all $p$, though this is not a tight lower bound.

In summary, internal RT surfaces provide a natural bridge between target space
entanglement in D$p$-brane holography and the naive bottom-up picture of
holographic entanglement entropy in flat space. Their anchoring data lives on the
internal $S^{8-p}$, which becomes the celestial sphere of the flat-space bubble.
For all $0\le p\le4$, introducing the shell lowers the areas of surfaces that enter
the flat region relative to their counterparts in the pure throat geometry. Generally, the internal surfaces probing the flat space region are not the dominant saddles. However, by making this region sufficiently large, \ie $R>{\cal R}\,\ruv$, the
dominant internal RT surfaces are forced to probe the flat space bubble. For $p\ne3$, this can be accomplished with a fixed cutoff surface and increasing $R$ to increase the physical size of the bubble. In contrast, for $p=3$, this physical size is fixed and so one must lower the cutoff surface deep into the IR of the throat so that it is close to the flat space region.

\section{Holographic complexity}\label{sec:complex}

We close here with a few observations about holographic complexity in the shell geometries described by eqs.~\eqref{eq:Dp-metric} and~\eqref{eq:Dp-metric22}.
Holographic complexity aims to extend holography beyond entanglement entropy to more refined probes of quantum entanglement \cite{Susskind:2018pmk,Baiguera:2025dkc}.
The proposal is that the quantum circuit complexity of a quantum state in a boundary CFT, \ie roughly, the minimum number of simple quantum gates needed to prepare the state from a reference state, is dual to a geometric quantity in the bulk gravitational spacetime. In the literature, the discussion of holographic complexity has focused on three proposals: Complexity=Volume \cite{Susskind:2014rva,Stanford:2014jda}, Complexity=Action \cite{Brown:2015bva,Brown:2015lvg} and Complexity=Spacetime Volume \cite{Couch:2016exn}. However, a broad range of gravitational observables have recently been shown to exhibit the key properties expected of holographic complexity in the boundary theory \cite{Belin:2021bga,Belin:2022xmt,Jorstad:2023kmq}. 

For simplicity here, we focus on probing the shell geometry using the Complexity=Volume approach,\footnote{Ref.~\cite{Couch:2017yil} examines Complexity=Action in  D$p$-brane throats with nontrivial NS-NS $B$ fields to study holographic complexity for noncommutative field theories.} where the holographic complexity is determined by the volume of an extremal codimension-one surface anchored on a Cauchy surface in the boundary \cite{Susskind:2014rva,Stanford:2014jda}, 
\begin{equation}
    {\cal C}_{\mt V} = \frac1{G_\mt{10}\, r_p}\,\int d^9x\, \sqrt{g}\,.
    \label{CV1}
\end{equation}
Here, we introduced a factor of $1/r_p$  in the prefactor to produce a dimensionless complexity. For the case of AdS$_5\times S^5$ (\ie $p=3$), this corresponds the conventional factor of $1/L$, found in most discussions. While most discussions of holographic complexity focus on an asymptotically AdS bulk geometry, we are assuming here that the extremal surface is codimension-one in the full ten-dimensional spacetime and so it fully wraps or fills the internal $S^{8-p}$ in the D$p$-brane geometries.

If we anchor the surfaces to a $t=\,$constant slice in the boundary, the extremal surface will be the $t=\,$constant surface throughout the entire bulk. 
Integrating over the entire time slice produces a divergent result because the complexity contains UV divergences \cite{Carmi:2016wjl,Jefferson:2017sdb}. However, we can compare the complexity of the Coulomb branch background with that of the D$p$-brane throat vacuum. Both configurations have the same asymptotic geometry and hence the same UV divergences. Therefore, the difference of the two complexities will be finite. In fact, the geometries are identical for $r\ge R$ and so we need only evaluate eq.~\eqref{CV1} on the range $0\le r\le R$. The flat space region produces
\begin{eqnarray}
    {\cal C}_{\mt V}({\rm flat}) &=& \frac{1}{G_\mt{10} r_p} \int d\Omega_{8-p} \int d^p x_\parallel \int^{R}_0 dr\, r^{8-p} \, \left(\frac{r_p}{R}\right)^{\frac{(7-p)(9+p)}{16}}\nonumber \\
    &=& \frac{1}{9-p}\, \frac{\Omega_{8-p}\,V_p}{G_\mt{10}}\left(\frac{r_p}{R}\right)^{3-\frac{(p+1)^2}{16}}\,R^{8-p}\,,
    \label{CVpflat}
\end{eqnarray}
where we have used that $H(r)=(r_p/R)^{7-p}$ is a constant in this region, as given in eq.~\eqref{eq:Dp-metric22}. Further, we adopt the notation introduced in section \ref{sec:start}: $V_p$ is the volume of the spatial gauge-theory directions and $\Omega_{8-p}=2\pi^{\frac{9-p}{2}}/\Gamma(\frac{9-p}{2})$ is the volume of the round unit-sphere $S^{8-p}$. 
Turning to the empty D$p$-brane throat, we have $H(r)=(r_p/r)^{7-p}$ and hence integrating over the same radial range gives
\begin{eqnarray}
    {\cal C}_{\mt V}({\rm throat}) &=& \frac{1}{G_\mt{10} r_p} \int d\Omega_{8-p} \int d^p x_\parallel \int^{R}_0 dr\, r_p^{\frac{(7-p)(9+p)}{16}}r^{\frac{65-14p+p^2}{16}} \nonumber \\
    &=& \frac{16}{32+(7-p)^2}\, \frac{\Omega_{8-p}\,V_p}{G_\mt{10}}\,\left(\frac{r_p}{R}\right)^{3-\frac{(p+1)^2}{16}}\,R^{8-p}\,.
    \label{CVpthroat}
\end{eqnarray}
Taking the difference then yields
\begin{align}
     \Delta{\cal C}_\mt{V}&={\cal C}_{\mt V}({\rm flat})-{\cal C}_{\mt V}({\rm throat})
     \label{CVpform}\\
     &=-\frac{(7-p)(9+p)}{(9-p)\left(32+(7-p)^2\right)}\frac{\Omega_{8-p}\,V_p}{G_\mt{10}}\,\left(\frac{r_p}{R}\right)^{3-\frac{(p+1)^2}{16}}\,R^{8-p}\,.
     \nonumber
\end{align}
This difference $\Delta{\cal C}_\mt{V}$ is also known as the complexity of formation \cite{Chapman:2016hwi}. 

We observe that the individual expressions in eqs.~\eqref{CVpflat} and~\eqref{CVpthroat} are both positive, as expected, but the difference~\eqref{CVpform} is negative for $0\leq p \leq 4$. The fact that the complexity of formation is negative indicates that the Coulomb branch solutions are easier to construct or require less entanglement than the corresponding vacuum solutions. This is in keeping with our results for holographic entanglement entropy in sections \ref{sec:EE} and \ref{sec:target}. Further, we observe that eq.~\eqref{CVpform} contains an overall factor of $R^{2+\frac{ (7-p)^2}{16}}$.
Since this exponent is positive for all values of $p$, 
$\Delta{\cal C}_\mt{V}$ becomes increasingly negative as the radius of the shell is increased, as one would expect.
We also note that the exponent of $r_p$ is positive for $0\le p\le4$ and hence the complexity of formation becomes more negative as $r_p$ increases (\eg by increasing $N$ in eq.~\eqref{eq:parameterfdef}).

It would be interesting to investigate the robustness of our results in the section for other holographic complexity observables. For instance, does Complexity=Action yield a negative result for the complexity of formation?

\section{Discussion}
\label{sec:disc}

In the preceding sections, we have used holographic entanglement entropy and
holographic complexity to probe Coulomb-branch shell geometries which contain a
flat space bubble in the interior of a D$p$-brane throat. Our motivation was to
develop a controlled top-down setting in which questions about flat space holography
can be sharpened using the standard tools of AdS/CFT holography and its D$p$-brane
generalizations. The shell geometries provide precisely such a setting. The exterior
region is an ordinary D$p$-brane throat with its familiar holographic interpretation,
while the interior is locally ten-dimensional Minkowski space. The D$p$-brane shell
therefore supplies both a physical cutoff for the flat-space region and a UV completion
which translates the observables at this cutoff into the language of the boundary
worldvolume theory.

A recurring theme in our results is that the flat-space bubble is associated with a
substantial reduction in the effective number of degrees of freedom probed by entanglement observables. In the strip geometry, the entanglement entropy
undergoes a first-order transition at a critical width, after which the dominant RT
surface consists of two disconnected sheets falling into the flat-space region. Beyond
this transition, the regulated entropy is independent of the strip width and the
corresponding entropic c-function vanishes. For spherical entangling regions,\footnote{We studied this case in detail for  $p=2$ and 3, but we expect similar results to arise for $p=4$.} the
transition is smooth rather than first order, but the refined entropy again
approaches zero at large radius. Internal RT surfaces, which are supposed to probe target-space
entanglement, also exhibit a reduction in area when they enter the
flat-space bubble. Finally, the complexity of formation in the Complexity=Volume
proposal is negative throughout the family of shell geometries considered here. Taken together, these results suggest that the Coulomb-branch state contains far fewer infrared degrees of freedom than the corresponding D$p$-brane throat vacuum. In the following, we examine this interpretation and discuss how it connects to the more naive bottom-up picture of holographic entanglement in flat space.

\subsection*{The shell as a cutoff flat-space hologram}

Let us first return to the bottom-up picture described in section~\ref{sec:botup}.
There, one introduces a large-radius cutoff surface in Minkowski space and asks what
the RT prescription would assign to angular subregions on this holographic screen. The
result is strikingly different from the corresponding calculations in AdS/CFT. Rather than
finding an area-law divergence localized near the entangling surface in the boundary, the leading
contribution scales (roughly) as the volume of the boundary region on the cutoff surface. In addition, there is no intrinsic analogue of the AdS radius with which to define either a UV cutoff or a central charge in the boundary theory. Instead, such an identification requires introducing an auxiliary macroscopic length scale by hand. Finally, fixed angular regions on the
cutoff screen do not probe the deep interior of flat space as the cutoff is removed.
Instead, information about the region near the origin is encoded only in the entropies of a restricted set of regions with opening angle tuned toward the equator, \ie
$\theta_0\sim 1/\ruv$.

The Coulomb-branch construction ameliorates this picture in several important ways. Rather
than placing an artificial cutoff surface in an asymptotically flat spacetime, the shell of
D$p$-branes provides a physical cutoff surface for the flat space region. The interior
geometry is locally flat, and the cutoff surface has topology
\[
        {\mathbf R}\times S^{8-p}\times T^p\, ,
\]
where the $T^p$ denotes the compactified spatial directions on the
D$p$-brane worldvolume. The exterior D$p$-brane throat then acts as a translation device mapping
data on this cutoff surface into observables in the worldvolume theory
living at the asymptotic boundary of the throat. This is the sense in which the shell
geometries realize a top-down version of the flat space hologram.

Of course, the D$p$-brane background introduces an intrinsic scale $r_p$ defined in eq.~\eqref{eq:parameterfdef}. This scale allows us to define a short distance cutoff in the boundary theory in terms of the position of the cutoff surface, as in eq.~\eqref{pcutoff}. However, we emphasize again that for $p\ne3$, this relation is more elaborate than the simple $\delta\simeq \ell^2/\ruv$ relation suggested in the bottom-up discussion. The scale $r_p$ also plays an essential role in defining the entropic c-functions (\eg as in eq.~\eqref{eq:stripcpDef}) which count the effective number of degrees of freedom at different energy scales in the boundary theory -- see further discussion below. We refer to these quantities as c-functions here, rather than central charges as in ${\cal N}=4$ SYM, because this counting is scale dependent for the nonconformal theories with $p\ne3$, even before the shell is introduced. As noted above, the Coulomb-branch solution~\eqref{eq:Dp-metric22} introduces a second scale $R$, which provides the physical cutoff for the flat-space region. As reviewed around eq.~\eqref{eq:Konishi}, this scale sets the vacuum expectation value of the scalars in the boundary theory. 

Our top-down approach also clarifies the role of the scale associated with the cutoff $r=R$ in the corresponding flat space hologram. In the
D$p$-brane throat, radial positions are related to a corresponding energy scale of the boundary theory.
Using the energy-radius relation in eq.~\eqref{eq:sugra}, the shell radius determines
an effective nonlocality scale
\begin{equation}
    \ell_\mt{nonlocal}
    \simeq {1\over E_{\rm sugra}(R)}
    =
    \frac{2^{p-3} \pi^{\frac{3(p-3)}4}}{\left[\Gamma\left(\tfrac{7-p}2\right)\right]^{1/2}}\,
    \left({r_p\over R}\right)^{5-p\over 2} r_p\, .
    \label{eq:disc_nonlocal}
\end{equation}
One might think of this as the scale of nonlocality that arises in the effective theory at the shell obtained
by integrating out the UV degrees of freedom in the D$p$-brane worldvolume theory 
down to the scale associated with $r=R$ \cite{Heemskerk:2010hk,Faulkner:2010jy}.
For $p=3$, this might alternatively be related to the nonlocality of the finite-cutoff theory resulting from a 
$T\bar T$ deformation, \eg \cite{McGough:2016lol,Taylor:2018xcy,Hartman:2018tkw}.
As might be expected, this scale $\ell_\mt{nonlocal}$ decreases as the radius $R$ of the flat space region is increased.

We emphasize that the dual theory lives on the $T^p$ factor of the geometry (as well as the time direction), rather than on the $S^{8-p}$, contrary to what is envisioned in the bottom-up hologram -- see further comments below. Hence the nonlocality in eq.~\eqref{eq:disc_nonlocal} applies to the field theory directions along the torus. Therefore, for this description to be effective, these directions should be large compared to the nonlocality scale, \ie $V_p \gg \ell_\mt{nonlocal}^{\,p}$.

The nonlocality of the effective theory on the cutoff surface is also reflected in our results for holographic entanglement entropy. For the strip geometry, the RT surfaces penetrate the flat-space region only when the strip exceeds a critical width $2\ell_c$, with $\ell_c\sim\left({r_p\over R}\right)^{5-p\over 2} r_p$ as given in eq.~\eqref{eq:stripTransitionWidthDp}. Up to a $p$-dependent numerical coefficient, this is precisely the scaling of the
nonlocality scale in eq.~\eqref{eq:disc_nonlocal}. Hence the RT surfaces begin to
probe the flat-space bubble only when the boundary region is larger than the intrinsic nonlocality scale of the effective theory on the shell.
The analysis of spherical entangling regions in section~\ref{sec:sphere} leads to a similar conclusion. For example,  with $p=2$,\footnote{The analysis for $p=3$ is less illuminating since we set $r_3=L=R$. However, we found the critical radius was indeed $P_\mt{shell}=L$, in line with our discussion here.} we found $P_\mt{shell} \sim\left({r_2\over R}\right)^{3\over 2} r_2$ in eq.~\eqref{eq:Pshell} as the critical radius at which the corresponding RT surfaces first reach the flat-space bubble. Hence this critical radius exhibits the same parametric scaling as the nonlocality scale in eq.~\eqref{eq:disc_nonlocal}, again up to a numerical coefficient.
Thus, both the strip and spherical entangling regions indicate that the flat-space bubble becomes visible to holographic entanglement entropy only at boundary length scales comparable to the intrinsic nonlocality scale of the effective theory on the shell.

The same scale appears naturally through the geometry. Consider a massless excitation moving radially outward from the origin. In the bulk description, it can pass through the cutoff surface (\ie the shell) and continue travelling out through the throat region. However, it will eventually reflect from the asymptotic boundary and return to the shell. From the viewpoint of the effective theory on the shell, this process induces a response spread over a time interval set by the round-trip travel time through the throat region, 
\begin{equation}
    \Delta t_\mt{throat}
    =
    2\int_R^\infty dr\,\left({r_p\over r}\right)^{7-p\over 2}
    =
    {4\over 5-p}\left({r_p\over R}\right)^{5-p\over 2}r_p\sim
    \ell_\mt{nonlocal}\, ,
    \label{eq:disc_throat_time}
\end{equation}
for $0\le p\le4$. Thus the nonlocality scale of the effective theory appears here in the time direction, whereas in the discussion of the entanglement entropy above, it appeared in the spatial directions. 

While the above nonlocality in time, \ie the return time for null signals reflected from the asymptotic boundary, is a feature of our top-down model, it is difficult to connect directly with our bottom-up hologram or with more traditional approaches to flat space holography. In the latter, massless excitations ultimately leave or enter the spacetime through null infinity. It may therefore be necessary to modify the boundary conditions in the D$p$-brane throat so that excitations reaching the asymptotic boundary are absorbed rather than reflected.\footnote{We also emphasize that the D$p$-brane shell itself supports partially absorbing boundary conditions, since excitations crossing the shell may be partially absorbed by the D$p$-branes \cite{Giddings:1999zu}.}

\subsection*{Target space entanglement and the bottom-up hologram}

As just discussed, the Coulomb branch geometries provide a flat space hologram with an intrinsic nonlocality, a conclusion similar to that suggested by the bottom-up flat-space RT calculations. There is, however, an important difference between the present construction and our naive bottom-up flat space hologram. In the latter picture, the regulated screen in ten-dimensional flat space would have topology ${\mathbf R}\times S^8$ and the putative boundary theory is expected to live on this geometry. As noted above, in the D$p$-brane shell geometries, the cutoff surface (\ie the shell) instead has topology
${\mathbf R}\times S^{8-p}\times T^p$.\footnote{These boundaries agree for $p=0$ but our subsequent comments still apply for this special case.} Thus the different spatial directions are not treated democratically. The ${\mathbf R}\times T^p$ directions are the directions in which the worldvolume theory lives, while the $S^{8-p}$ is a geometric manifestation of the internal structure and symmetries of this dual theory, \ie the global $SO(9-p)$ symmetry rotating the scalar fields among one another.

The internal RT surfaces studied in section~\ref{sec:target} provide the most
direct bridge between the bottom-up hologram of section \ref{sec:botup} and our top-down D$p$-shell construction. In the D$p$-throat, these surfaces wrap the spatial worldvolume directions and have a nontrivial profile on the internal sphere $S^{8-p}$. As discussed in section \ref{sec:target}, from the boundary theory perspective, the internal RT surfaces are naturally interpreted as probes of target space entanglement \cite{Mollabashi:2014qfa,Karch:2014pma,Das:2022njy}. 
Inside the shell, however, the $S^{8-p}$ becomes the angular or ``celestial" sphere of the $R^{1,9-p}$ factor of the flat space region. Thus the internal RT surface which we considered as anchored asymptotically at $\theta(r_{\rm uv})=\thbdy$ becomes, from the viewpoint of the flat space bubble, anchored on an angular subregion of the celestial sphere on the cutoff surface provided by the shell. This is precisely the sort of subregion considered in the bottom-up model of section~\ref{sec:botup}. In fact, once the surface enters the flat space bubble, the interior segment is simply a flat hyperplane. Comparing with eqs.~\eqref{hyperplane} and~\eqref{eq:flatplane}, we see that we need only identify $r_c=r_\mt{min}$ to match the surfaces in the two approaches. That is, the bottom-up RT surface appears as the portion of the full top-down extremal surface lying inside the shell.

It is interesting to compare the corresponding areas and entropies in eq.~\eqref{eq:flatRTvolume} for the bottom-up model with eq.~\eqref{eq:Aflat2} for the shell geometry. The latter area looks somewhat more involved than the corresponding expression for the bottom-up model. However, introducing $\theta(R) = \arcsin(r_c/R)$ and the physical radius~\eqref{eq:flatsize} of the shell, the corresponding contribution to the holographic entanglement entropy becomes
\begin{equation}
\begin{split}
    S_\mt{flat}={A_\mt{flat}\over 4G_{10}}
      &= \frac{V_p\,\Omega_{7-p}}{4(8-p)G_{10}}\,\left(\frac{r_p}{R}\right)^{-\frac{p}{16}(p-7)^2}\,\left(R_\mt{shell}\,\cos\theta(R)\right)^{8-p}\\
      &= \frac{\Omega_{7-p}}{4(8-p)G_{10-p}}\,\left(R_\mt{shell}\,\cos\theta(R)\right)^{8-p}\,,
    \label{eq:flatEE99}
\end{split}  
\end{equation}
where we have defined the effective Newton's constant $G_{10-p}$, upon making a Kaluza-Klein reduction on the $T^p$ in the flat space region, 
\begin{equation}
    \frac1{G_{10-p}}=\frac{V_\mt{KK}}{G_{10}}=\left(\frac{r_p}{R}\right)^{-\frac{p}{16}(p-7)^2} \frac{V_p}{G_{10}}\,.
\end{equation}
Hence this contribution precisely matches that in eq.~\eqref{eq:Aflat2} when we identify the parameters $(d,\theta_0,\ruv,G_{10})$ with the bottom-up hologram with
$(9-p,\theta(R),R_\mt{shell},G_{10-p})$ in the top-down hologram provided by the shell geometries. 

However, our construction still relies on the exterior D$p$-throat region, which determines how this flat surface is extended to the boundary, how it is anchored at the UV cutoff, and whether it is the dominant saddle. The latter is an important point, as we found that for most choices
of parameters, the dominant internal RT surface stays near the UV cutoff, with
$r_c\sim r_{\rm uv}$, and therefore does not probe the flat space region. Of course, this agrees with the observation of refs.~\cite{Karch:2014pma,Das:2022njy} that target space entanglement in the standard D$p$-brane holography is typically dominated by UV data. In our
shell geometries, the surfaces which enter the flat space bubble have smaller area than their counterparts in the pure D$p$-throat geometry, but they are typically not the minimal area saddles. Only when the shell is sufficiently large to be close to the cutoff with
roughly $R\gtrsim {\cal R}\,r_{\rm uv}$ with ${\cal R}\sim 1/5$ (see table \ref{tab:DpthetaC}), do the dominant
internal RT surfaces enter  the flat space region.

Recall that in our discussion of the bottom-up hologram in section~\ref{sec:botup}, we observed that information about the IR region near the origin is only encoded in boundary regions with $\theta_0\sim 1/\ruv$. Given the close parallel noted above, the same is true of the flat space region in the shell geometry with 
\begin{equation}
    \theta(R)\sim \frac{1}{R_\mt{shell}}=r_p{}^{\!\!-\frac{(7-p)(p+1)}{16}}\, R^{-\frac{(p-3)^2}{16}}\,.
\end{equation}
However, we note that the D$p$-throat compresses these angles slightly more. For $r_c\ll1$, we combine the universal profile~\eqref{eq:Dptheta0} with the coefficients~\eqref{eq:tilCapprox}, and considering the overall envelope of the profile, we find 
\begin{equation}
    \thbdy\sim \left(\frac{R }{\ruv}\right)^{(9-p)/4} \theta(R)\,.
\end{equation}
As noted above, in the regime of interest, the prefactor here is a fraction of order 1/5 now raised to some power greater than one. Hence the internal RT surfaces which probe deep in the interior of the flat space bubble
are anchored at very small values of
$\thbdy$. Hence the special role of $\theta_0\sim0$ surfaces in the bottom-up model has an imprint in our top-down hologram constructed with the shell geometries.

Let us reiterate the key conceptual divergence between the bottom-up flat space hologram and that emerging from the Coulomb-branch geometries. In the former, the putative dual boundary theory lives on the celestial sphere together with the time direction. In this setting, RT surfaces anchored on the celestial sphere compute the holographic entanglement entropy for spatial partitions of this boundary theory. In contrast, in our top-down construction, the boundary theory lives in the time and $T^p$ directions. In this case, the $S^{8-p}$ is an internal space associated with the worldvolume scalar fields and their rotational global symmetry. Hence the internal RT surfaces anchored on the sphere are naturally interpreted as computing target space entanglement for partitions of the internal degrees of freedom. From this perspective, the two approaches may seem to have little in common. However, one should not necessarily conclude that one is correct and the other incorrect. The underlying microscopic degrees of freedom are clearly very different in the two models, but they may simply provide two distinct holographic representations of the same bulk gravitational physics.

\subsection*{Effective number of degrees of freedom} 

Let us review in more detail what our results tell us about the number of degrees of freedom that describe the flat region. Consider first the strip geometry discussed in section \ref{sec:strip}. The strip entanglement entropy defines the c-function in eq.~\eqref{eq:stripcpDef}, and in the absence of the shell, the result is given in eq.~\eqref{eq:stripcpDp},
\begin{equation}
    \tilde c_p
    =
    N^2\left[\hat g_{\rm eff}^2(1/2\ell)\right]^{p-3\over 5-p}
    =
    N^2\left(g_{\rm YM}^2 N\right)^{p-3\over 5-p}
    (2\ell)^{-\frac{(p-3)^2}{5-p}}\, .
    \label{eq:disc_cp}
\end{equation}
In the conformal case $p=3$, this is simply $N^2$, proportional
to central charges of ${\cal N}=4$ super-Yang--Mills theory, as discussed above. In the nonconformal case $p\ne3$, the
effective number of degrees of freedom runs with scale, as expected of the
nonconformal worldvolume theories. In particular, $\tilde c_p$ decreases as $\ell$ is
increased for all $p\ne3$ in the range considered here, irrespective of whether the
Yang-Mills coupling is relevant or irrelevant. The scaling with $N,\ g_{YM}$ and $\ell$ matches the numerator of the stress tensor two-point function in eq.~\eqref{eq:stress}.

The shell geometry behaves identically to the pure D$p$-throat in the UV, but has modified behavior in the infrared below the scale set by the radius of the shell. That is, for strips
narrower than the critical width $\ell_c$ in eq.~\eqref{eq:stripTransitionWidthDp},
the dominant RT surface lies entirely in the D$p$-brane throat and the c-function is
unchanged from the pure-throat result \eqref{eq:disc_cp}. For wider strips, the dominant saddle becomes
the disconnected flat-sheeted surfaces. The (regulated) area is then independent of
$\ell$, so that $\partial S/\partial\ell=0$ and the strip c-function vanishes. Hence the
flat-space bubble appears, from the perspective of this entropic probe, to carry less than
order $N^2$ infrared degrees of freedom.

Let us then move on to the case of a spherical entangling surface. Recall first the pure D$p$-throat geometry. In the conformal case $p=3$, the holographic c-function~\eqref{eq:dof} is constant and is given by eq.~\eqref{eq:adsa}
\begin{equation}
    \mathcal{S}_4=\frac{N^2}{4}\,,
\end{equation}
which again equals the central charges of $\mathcal{N}=4$ SYM.
Meanwhile, for the nonconformal case $p=2$, we have the c-function defined in eq.~\eqref{eq:S3def}, which yields 
\begin{equation}
    \mathcal{S}_3 = \frac{N^2}{\hat{g}_{eff}\left(1/P\right)^{\frac{2}/{3}}} = \frac{N^2}{(g^2_{YM}N)^{{1}/{3}}}\,\frac1{P^{{1}/{3}}}\,,
\end{equation}
as shown in eq.~\eqref{eq:S3throat}. The c-function decays to zero in the infrared, as expected. The parametric scaling, in particular with $P$, matches the scaling of the stress tensor two-point function~\eqref{eq:stress}, as for the strip.

In the shell geometry, we found a qualitatively different behavior for this entanglement probe in the IR, compared with the strip. In the conformal case, once the RT surfaces enter the flat-space bubble, the
c-function jumps discontinuously and becomes non-monotonic and negative
before approaching zero at large radius (see figure~\ref{fig:spheredof}). The non-monotonicity is not surprising:
similar behavior has been seen in sharp holographic RG flows and in gapped phases \cite{Liu:2012eea}. 
The important point for the present discussion is to compare the UV and IR limit. As for the strip, the entropic probe indicates that the deep infrared of the Coulomb-branch state contains only order one degrees of freedom,\footnote{Or perhaps as much as order $N$ -- see below.} compared to the order-$N^2$ degrees of freedom present in the UV throat region. Meanwhile, for the nonconformal case $p=2$, the circle c-function \eqref{eq:S3def} is better behaved. It decays monotonically in the exterior of the shell, jumps down slightly at the shell, and then continues decreasing monotonically into the IR. The rate of decrease in the IR is $\mathcal{S}_3 \propto P^{-1}$ for the shell geometry compared to $\mathcal{S}_3\propto P^{-1/3}$ for the D$2$-throat. Hence we still find a reduction in the number of degrees of freedom, consistent with the above discussion. 

As an aside, note that for both of the entropic c-functions examined above, when the coupling is renormalizable ($p=1,2$) the c-functions become large in the UV, \ie for small values of $\ell$ and $P$. This behaviour can only be trusted in so far as the parts of the RT surface that contribute to the c-function lie in the regime where the supergravity approximation is valid (cf.~eq.~\eqref{eq:weakcurv}). For very small boundary length scales, we move out of this regime and into the strong curvature regime in the geometry where supergravity breaks down. In this limit, effective coupling $\hat{g}_{eff}$ vanishes, cf.~eq.~\eqref{eq:safe}, and so the boundary theories become free. Hence, we expect the number of effective degrees of freedom to be a constant proportional to the number of fields, $\tilde c_p \propto N^2$. It would be interesting to evaluate the proportionality constant and confirm that $\tilde c_p$ is larger in the UV limit than the largest value found in the geometric or strongly-coupled regime described by supergravity. 

The conclusion that there are less than order-$N^2$ IR degrees of freedom in the state dual to the shell geometry has a simple interpretation in the D$p$-brane worldvolume theory.
Moving onto the Coulomb branch separates the branes such that strings stretched between
different branes become massive. The off-diagonal matrix degrees of freedom
which are responsible for the order-$N^2$ entropy of the coincident-brane state are
lifted at low energies. At scales below the separation set by the shell, only the
approximately diagonal degrees of freedom remain light, which is only order-$N$. Thus the vanishing of the
entropic c-function should not be interpreted as the complete absence of degrees of
freedom, but rather as the absence of a large number of entangled degrees of freedom
visible at leading order in the large-$N$ expansion. The strip c-function shows this reduction by the $\mathcal{O}(N^2)$ term discontinuously jumping to zero near the shell-scale, while the sphere c-function showed the reduction through an $\mathcal{O}(N^2)$ term rapidly decaying to zero at large radius.

The conclusion that number of degrees of freedom is reduced in the shell geometry is reinforced by the complexity calculation in section~\ref{sec:complex}.
The complexity of formation in eq.~\eqref{CVpform} is negative for all
$0\le p\le4$. Since the shell and pure-throat geometries share the same UV
asymptotics, this finite difference measures the change in the bulk volume associated
with replacing the D$p$-brane throat by the flat-space bubble in the IR. The negative sign
again indicates that the Coulomb-branch state is simpler, in the sense of the
Complexity=Volume proposal
, than the corresponding vacuum throat geometry. While
complexity is a different observable from entanglement entropy, it points in the same
qualitative direction: the shell removes a large number of infrared degrees of freedom.

It is worth commenting on how this result relates to another approach to getting flat space physics from AdS/CFT, namely approaching flat space with the limit $L_\mt{AdS}\to\infty$. Since this flat limit involves taking the AdS scale to infinity, which implies taking $N$ to infinity, one might naively expect that there is a large number of degrees of freedom associated with flat space. However, in the usual flat space limit
of AdS/CFT, one isolates a local scattering region by zooming into a parametrically
small region near the center of AdS and simultaneously focusing on special high-energy
boundary kinematics \cite{Susskind:1998vk,Polchinski:1999ry,Hijano:2019qmi}. Most of the CFT degrees of freedom are not involved in the description of the corresponding local flat space processes. Hence, it is possible that the effective number of degrees of freedom describing the flat space physics is still smaller than that associated with an AdS geometry, which would be consistent with our result. It would be interesting to understand better the number of effective degrees of freedom arising from the flat space limit.

There is a related, more speculative, comparison with celestial and Carrollian
approaches to flat-space holography. In those settings, one often expects the infrared structure of massless fields and soft modes to play an important role
\cite{Strominger:2017zoo,Raclariu:2021zjz,Pasterski:2021rjz,Pasterski:2021raf}.
Some recent analyses suggest that effective central charges associated with these descriptions may even be divergent \cite{Ciambelli:2024swv,cmp1,cmp2}. This again appears to be in tension with our results, where the entropic central charge associated with the flat space bubble vanishes. Here the resolution is less clear, but it may simply reflect the fact that our flat space region has only a finite extent, so the would-be infrared degrees of freedom are lifted in the shell geometry.

\subsection*{Shell geometry as a finite cavity}

The shell geometry shares some interesting features with confining backgrounds, though the shell geometry is not itself confining.\footnote{For example, Wilson line observables exhibit screening at separations larger than the shell radius~\cite{Giddings:1999zu}.}  As noted for the strip geometry in section~\ref{sec:strip}, in both settings, the entropy saturates for sufficiently wide strips (\ie $\ell>\ell_c$), with the corresponding RT surfaces splitting into two disconnected components~\cite{Klebanov:2007ws,Nishioka:2006gr}. Consequently, the associated entropic $c$-function~\eqref{c1} vanishes at large separations in this setting as well. This similarity suggests an alternative, intuitive interpretation of the reduction in the effective number of degrees of freedom.

In confining geometries,  the absence of infrared degrees of freedom is realized geometrically through the spacetime smoothly closing off at a finite radial position. Further, this feature is reflected in the causal structure: a null ray sent radially inward from the boundary returns after a finite coordinate time set by the confinement scale. For example, in the AdS soliton in $d+1$ bulk dimensions~\cite{Witten:1998zw,Horowitz:1998ha}, this return time is given by
\begin{equation}
    \Delta t = \int_{r_0}^\infty\!dr\,\frac{L^2}{r^2}\frac1{\sqrt{1-(r_0/r)^d}}=\frac{2\sqrt{\pi}}{d}\,\frac{\Gamma\!\left(\frac{1}{d}\right)}{\Gamma\!\left(\frac{d+2}{2d}\right)}\,\frac{L^2}{r_0}\,,
\end{equation}
where $r_0$ is the radius at which the geometry caps off and $L$ is the AdS curvature scale. This behavior is to be contrasted to, say, pure AdS where the light-ray would simply exit the Poincare patch, only returning to the boundary in a different patch.  Comparing with the confinement scale,
\begin{equation}
    \Lambda_{\mathrm{conf}} = \frac{d\,r_0}{4\pi L^2}\,,
\end{equation}
we find $\Delta t \sim 1/\Lambda_{\mathrm{conf}}$. Similarly, the critical width at which the entanglement entropy for the strip geometry saturates also scales inversely to the confinement scale and hence up to numerical factors, we have $\Delta t \sim 2\ell_c$.

A similar relation arises in the shell geometry. The return time for a null ray sent in from the asymptotic boundary is, using eqs.~\eqref{eq:Dp-metric} and~\eqref{eq:Dp-metric22}, 
\begin{equation}
    \Delta t =  2\left(\frac{r_p}{R}\right)^{\frac{7-p}{2}}\left(\int_0^R dr \, + \, \int_R^\infty \left(\frac{R}{r}\right)^{\frac{7-p}{2}} dr\right)
    = \frac{2(7-p)}{5-p} \left(\frac{r_p}{R}\right)^{\frac{7-p}{2}} R
    \label{fence}
\end{equation} 
for $0\le p\leq 4$. The finite result for $\Delta t$ here indicates that the flat space bubble behaves as a finite cavity or effectively closes off the geometry at a finite distance, in close analogy with confining backgrounds. 
Comparing the above result to eq.~\eqref{eq:stripTransitionWidthDp}, we see that the critical width scales in precisely the same way as the return time, \ie $2\ell_c\sim \left(\frac{r_p}{R}\right)^{\frac{7-p}{2}} R\sim \Delta t$. This result strengthens the analogy between confining backgrounds and our shell geometries. Further, from this perspective, the fact that the shell geometry behaves like a finite cavity provides a natural explanation for the depletion of infrared degrees of freedom in the shell geometries.
Combined with our previous observations about the parametric scaling of the nonlocality scale~\eqref{eq:disc_nonlocal}, which matches that of the critical strip width and of the return time here, suggests that they are all different manifestations of a single underlying scale governing the effective dynamics at the cutoff surface.

\subsection*{Future Outlook}

There are many directions in which the present analysis could be extended. We
highlight several of these below.

\paragraph{Correlation functions and spectra:}
It would be useful to study more conventional correlation functions in the shell
geometries. For instance, in the pure D$p$-brane throat, the stress-tensor two-point function defines an
effective ``central charge" with the same dependence on $N$ and $\hat g_{\rm eff}$ as the
entropic c-functions. One would like to understand how this correlator is modified in the
Coulomb-branch shell state. Also, it would be interesting to see how the depletion of infrared degrees of
freedom seen by the entanglement probes is reflected in spectral densities and response functions.
This question is closely related to the analysis of absorption and scattering by
D3-brane shells in ref.~\cite{Giddings:1999zu}. 


\paragraph{Refined entanglement probes:}
The entanglement probes considered here could also be refined. For instance, the RT entropy
captures only the leading order in the large-$N$ expansion. It would be interesting to
include bulk quantum corrections and to study the generalized entropy.  Also,
because the shell itself carries localized D$p$-brane degrees of freedom, one should account not only for the bulk excitations but also for possible
contributions from the brane degrees of freedom localized at the interface. This may be necessary to capture the $\mathcal{O}(N)$ diagonal degrees expected in the Coulomb branch state and not just the $\mathcal{O}(1)$ degrees of freedom expected from subleading corrections to the RT formula. Other quantities such as mutual information may also provide
more sensitive probes of the flat-space region. Such quantities may also distinguish the finite-cavity physics
of the shell from the corresponding behavior in confining geometries.

\paragraph{Target space entanglement:} There are many open questions that would be interesting to answer in order to put the relationship between target space entanglement and internal RT surfaces on firmer footing. For instance, one natural objective would be to simply match the target space entanglement's dependence on the UV cutoff in a holographic theory to the divergent power law obtained from the holographic calculation above and in ref.~\cite{Das:2022njy}. Once the target space entanglement is better understood holographically, it would be interesting to understand the boundary interpretation of other features of the internal RT surfaces, such as the role of the discrete number of subdominant saddles.

\paragraph{Flat space dynamics in a finite cavity:}
One could hope to use the shell geometry as a laboratory to study more interesting flat space dynamics. 
For example, one could study scattering processes inside the flat space bubble and their encoding in boundary correlators.
The possible absorption by
the D$p$-branes should play an important role~\cite{Giddings:1999zu}. Similarly, one could introduce black
holes or other localized excitations inside the bubble and test to what extent their physics is
visible in the boundary description. Note that finite energy excitations, such as a black hole, will cause the shell to collapse in general. However, one should be able to choose parameters such that the shell remains large and approximately stationary for a long time, \eg by considering a black hole with a small mass.

The key point here is that such excitations will generically break the supersymmetry of the Coulomb-branch solutions. This then provides a natural place to note a related subtlety: computing holographic Renyi entropies as an intermediate step in the Lewkowycz-Maldacena replica construction of holographic entanglement entropy \cite{Lewkowycz:2013nqa,Rangamani:2016dms} may introduce related technical complications. In particular, the replica geometries need not preserve the supersymmetry of the original Coulomb-branch background, which could in turn modify the corresponding Renyi entropies. However, this breaking should become parametrically small in the limit $n\to 1$, relevant for the entanglement entropy. We therefore expect that the entropies and related quantities presented here will remain unaffected at leading order.

\paragraph{Kaluza--Klein reductions:}
A related direction is to consider holography for Kaluza--Klein reductions of flat space more broadly, \eg in celestial holography. In the present construction, the flat-space bubble has the form
${\mathbb R}^{1,9-p}\times T^p$, and the cutoff surface has topology
${\mathbb R}\times S^{8-p}\times T^p$. This differs from the more symmetric
large-radius cutoff in ten-dimensional Minkowski space, whose spatial sections are simply spheres $S^8$. It would also be interesting to understand how the holographic interpretation changes with different compactifications, and how the effective boundary conditions for the flat-space region depend on the choice of internal manifold.

\paragraph{Connections to celestial and Carrollian holography:}
The relation of our Coulomb-branch shell constructions to celestial and Carrollian holography remains an open question. It is not immediately straightforward to connect to these approaches, as the shell construction does not provide a hologram for an asymptotically
flat spacetime with past and future null infinities, but rather a flat region with a timelike cutoff. However, the shell construction may provide a useful complementary perspective.
In particular, as discussed above, the fact that partitions of the spherical cutoff surface of the flat-space bubble are target-space partitions in the D$p$-brane theory suggests an interesting way in which the
degrees of freedom of flat-space holography may be reorganized in the top-down
embedding. That is, what appears to be the celestial sphere from the bottom up perspective appears to be the internal directions in the top-down hologram.

\acknowledgments

We thank Pinaki Banerjee, Anna Biggs, Luca Ciambelli, Sumit Das, 
Roberto Emparan, Johanna Erdmenger, Ben Freivogel, Sean Hartnoll, Leonardo Pipolo de Gioia, Shiraz Minwalla, Simon Ross, Shigeki Sugimoto, Sumati Surya, Tomonori Ugajin, Tadashi Takayanagi and Pedro Vieira  for useful discussions and feedback. 
RCM would like thank the organizers and participants at the 2025 workshop, {\it Gravity: New quantum and string perspectives}, in Benasque for creating a stimulating atmosphere where some of this work was carried out. Research at Perimeter Institute is supported in part by the Government of Canada through the Department of Innovation, Science and Economic Development Canada and by the Province of Ontario through the Ministry of Colleges and Universities. 
EJ and RCM are also supported in part by a Discovery Grant from the Natural Sciences and Engineering Research Council of Canada, and by funding from the BMO Financial Group. SP is also supported by the Celestial Holography Initiative at the Perimeter Institute for Theoretical Physics and by the Simons Collaboration on Celestial Holography.


\appendix
\section{Boundary Conditions for Spherical Entangling Surface}
\label{app:bc}

Here we briefly consider the equations of motion derived from~\eqref{eq:zzzzx} and~\eqref{eq:yyyyx} and the form of their solutions near $r=0$ or $\rho=0$. The approximate solution is used to set the initial values for the numerical integration close to these boundaries. Setting initial conditions directly on the boundaries is avoided, since equations are singular there, as we shall see below. We will write the equations of motions for $r(\rho)$ or $\rho(r)$. These are obtained by gauge-fixing $\sigma$ in eqs.~\eqref{eq:zzzzx} and~\eqref{eq:yyyyx} to $\sigma=\rho$ and $\sigma = r$, respectively. As explained in section~\ref{sec:sphere}, there are four cases to consider, see figure~\ref{fig:spherediagram}. The first case is when the surface closes off outside the shell at $r_a >R$. The second case is a surface that closes off inside the shell at $0 < r_b \leq R$. The third case is a special solution with $r_b=\rho_0=0$. The fourth case is a surface that hits the origin at $\rho_0>0$. 

For surfaces closing off at $r_a\geq R$ the area functional is~\eqref{eq:zzzzx}. For our purpose it is most straightforward to consider the equation of motion for $r(\rho)$, which takes the form
\begin{equation}
\frac{d^2r}{d\rho^2} = \frac{(9-p)}{2\, r_p}\, \left(\frac{r}{r_p}\right)^{6-p} - \frac{(p-1)}{\rho} \,\frac{d r}{d\rho}+ \frac{(8-p)}{r}\,\left(\frac{d r}{d\rho}\right)^2 - \frac{(p-1)}{\,\rho}\left(\frac{r_p}{r}\right)^{7-p}\,\left(\frac{d r}{d\rho}\right)^3\, .
\label{eq:r-Dp-eom}
\end{equation}
The form of the solution near $\rho=0$ is fixed by the requirement that $d^2r/d\rho^2$ is finite. This occurs when $ dr/d\rho\sim c \rho $, which cancels the potential divergences from the terms proportional to $\rho^{-1}$. In other words, the series expansion around $\rho=0$ is to second order $r\sim r_a + c \rho^2/2$. Taking this as an ansatz, the equation of motion fixes the constant $c$ and the small $\rho$ expansion takes the form 
 \begin{equation}
     r \simeq r_a + \frac{(9-p)}{4 p} \left(\frac{r_a}{r_p}\right)^{6-p} \frac{\rho^2}{r_p}+\dots.\label{eq:p2-r-Dp-ser}
 \end{equation}
All higher order terms in the series expansion are fixed in terms of $r_a$. That is, fixing $r_a$ fully specifies the solution. 

Though not necessary for the numerics, it will be useful to also write the D$p$-brane equation of motion for $\rho(r)$, which takes the form 
\begin{equation}
    \frac{d^2\rho}{dr^2} = -\frac{(8-p)}{r}\,\frac{d\rho}{dr}- \frac{(9-p)}{2\,r_p}\, \left(\frac{r}{r_p}\right)^{6-p}\, \left(\frac{d\rho}{dr}\right)^3 + \frac{(p-1)\left(\left(\frac{r}{r_p}\right)^{-(7-p)} + \left(\frac{d\rho}{dr}\right)^2\right)}{\rho}. \label{eq:rho-Dp}
\end{equation}
 This can for example be used to derive the asymptotic form of solution in the limit $r\to\infty$ and the corresponding UV divergences in the entanglement entropy. For our purpose, we will use it in the perturbative calculations in section~\ref{app:perturb}.
 
For the case $0<r_b< R$, the area functional for the surface segment inside the shell is~\eqref{eq:yyyyx}. The equation of motion takes the form
\begin{equation}
\frac{d^2 r}{d\rho^2} = \left(\frac{r_p}{R}\right)^{7-p}\,\frac{\left((8-p)\,\left(\frac{R}{r_p}\right)^{7-p}\,\rho - (p-1)\,r\,\frac{dr}{d\rho}\right)\left(\left(\frac{R}{r_p}\right)^{7-p} + \left(\frac{dr}{d\rho}\right)^{2}\right)}{\rho\, r}  \label{eq:r-flat-eom} 
\end{equation}
By the same argument as above, the requirement that the second derivative is finite at $\rho=0$ implies $d r/d\rho  \sim \hat c \rho $ and thus $r\sim r_b + \hat c \rho^2/2 $. Solving the equation of motion with this ansatz gives  
\begin{equation}
        r =r_b + \frac{(8-p) }{ 2\,  p }\left(\frac{R}{r_p}\right)^{7-p} \frac{\rho^2}{r_b}+\cdots .
\end{equation}
As above, the solution is fixed by the choice of $r_b$. 

Meanwhile, when $r_b=0$, finiteness of the second derivative requires $r \sim \tilde c \, \rho$. In this case, the two terms in the leftmost factor in the numerator in~\eqref{eq:r-flat-eom} must cancel against one another to keep the left hand side finite. Solving the equation of motion with this ansatz gives
\begin{equation}
   r= \sqrt{\frac{8-p}{p-1}} \left(\frac{R}{r_p}\right)^{\frac{7-p}{2}} \rho ,
\end{equation}
which is in fact an exact solution to~\eqref{eq:r-flat-eom}.

For the surfaces reaching the origin $r=0$ at $\rho=\rho_0>0$, the area functional is also~\eqref{eq:yyyyx}. However, we consider the equation of motion for $\rho(r)$, which takes the form
\begin{equation}
    \frac{d^2 \rho}{dr^2}= -\left(\frac{r_p}{R}\right)^{7-p}\frac{\left( (8-p)\,\left(\frac{R}{r_p}\right)^{7-p}\,\rho\,\frac{d\rho}{dr}-(p-1)\,r \right)\left(1 + \left(\frac{R}{r_p}\right)^{7-p}\,\left(\frac{d\rho}{dr}\right)^{2}\right)}{r\,\rho} \label{eq:rho-flat-eom}
\end{equation}
In this case, the finiteness of the second derivative requires $d\rho/dr\sim b\, r$, that is $\rho \sim \rho_0 + b r^2 /2$. Solving the equation of motion with this ansatz yields
\begin{equation}
    \rho = \rho_0 +\frac{p-1}{2(9-p)} \left(\frac{r_p}{R}\right)^{7-p} \frac{r^2}{\rho_0}+\cdots.
\end{equation}
The solution is fixed once $\rho_0$ is specified.
The special case $\rho_0=0$ is precisely the case $r_b=0$ considered above.

\section{Perturbative Calculation for Spherical Entangling Surface} \label{app:perturb}
Here we provide the details of the perturbative calculation of the entanglement entropy for spherical boundary regions. The idea is to express the profile of the surface $\rho(r)$ perturbatively around a constant cylindrical, \ie $\rho(r)\simeq P$, and as shown in eq.~\eqref{eq:AregApprox}, the area of the RT surfaces is given by a power series in inverse powers of $P/r_p$.
At large radius in the D$p$-throat region, we expand the profile as a power series in $1/r$, while in the flat space bubble, the profile is a power series in $r$. The power series coefficients will depend on $P$, and, as will be explained below, at high orders in $r$ or $1/r$, the coefficients are proportional to high powers of $1/P$. Therefore, sufficiently high powers in the series expansions can be neglected for our present purpose, which is to evaluate the area to the first subleading order. 

\paragraph{ $S^2$ entangling surface ($p=3$)}
Here we work in the rescaled coordinates explained in the main text, such that $r_3=R=L$. To start, we express the profile of the minimal surface in the AdS region as a power series in $L^2/r$ around $r=\infty$. As we will explain below, if we want to obtain the area up to order $P^{-2}$, it turns out we will need to account for terms up to order $r^{-10}$. Inserting a generic power series ansatz for $\rho(r)$ into eq.~\eqref{eq:rho-Dp} and solving for the first ten coefficients, we find that the expansion takes the form
\begin{equation}
\begin{split}
    \rho_{AdS}(r)=& P -\frac{L^4}{2 P\,  r^2} + \frac{C_0 \,L^8}{r^4}+\left(\frac{1}{8 P^5} +\frac{3 C_0}{2 P^2}\right)\frac{L^{12}}{r^6}\\
  & + \left(\frac{17}{128 \,P^2} +\frac{15 C_0}{16 \,P^4}+\frac{7 C_0^2}{2 \,P}\right)\frac{L^  {16}}{r^8}\\
  & + \left(\frac{73}{1280 \,P^9}-\frac{101\,C_0}{160 \, P^6}-\frac{201 \, C_0^2}{20\, P^3}+\frac{16\,C_0^3}{5} \right)\frac{L^  {20}}{r^{10}}+O\left(r^{-12}\right)
  \label{eq:AdSPS}
\end{split}
\end{equation}
where the first two terms are responsible for the usual universal divergence in the area. The integration constant $C_0$ parameterizes a one-parameter family of surfaces with the given boundary radius $P$. Note that $C_0$ has dimensions $\textit{length}^{-3}$, and this dictates what combinations of powers of $C_0$ and $P$ can occur in the power series coefficients. (Note that powers of $L$ only appear in the combination $L^2/r$. This is due to the fact that, in the usual Poincare coordinate $z=L^2/r$, there are no factors of $L$ in the equation of motion.) In pure AdS, $C_0$ is fixed by the requirement that the solution closes off smoothly in the bulk, in which case $C_0=-1/(2 P)^3$, giving the usual hemisphere surface~\eqref{eq:prof5}. Below, $C_0$ will instead be fixed by the matching condition at the brane shell.

To get the full profile of the surface, we also need to perform the perturbative expansion in the flat region as well. The is most conveniently done by expanding in power series around $r=0$ using eq.~\eqref{eq:rho-flat-eom}. Here, we have to keep track of terms up to $r^{4}$. Solving the equations of motion order by order gives an expansion of the form
\begin{equation}
    \rho_{flat}(r)= \rho_0 + \frac{r^2}{6 \rho_0}- \frac{r^4}{108 \rho_0^3} +O \left(r^{6}\right),
    \label{eq:flatPS}
\end{equation}
where $\rho_0$ is the radius of the surface at $r=0$, see figure \ref{fig:spherediagram}. We have thus obtained the two expansions,~\eqref{eq:AdSPS} and~\eqref{eq:flatPS}, which contain three paramaters $P$, $\rho_0$ and $C_0$. We fix the latter two in terms of the boundary radius $P$ by imposing the matching condition at the brane shell, namely that profile $\rho(r)$ and its first derivative are continuous,  
\begin{equation}
        \rho_{AdS}(L)=\rho_{flat}(L),\qquad
        \dot\rho_{AdS}(L)=\dot\rho_{flat}(L)\,.
    \label{eq:matching}
\end{equation}
In order to solve for $C_0$ and $\rho_0$, we first note that the cylindrical surface $\rho(r)\simeq P$ solves the equations of motion~\eqref{eq:rho-Dp} and~\eqref{eq:rho-flat-eom} when $P\to \infty$. In this limit, we therefore expect $\rho_0 \simeq P$ and $C_0\simeq 0$. This  motivates the ansatz 
\begin{equation}
    C_0 = \sum_{n=1}^\infty c_n P^{-n},\qquad \rho_0 = P+ \sum_{n=0}^\infty d_n P^{-n}.
    \label{eq:paramser0}
\end{equation}
Substituting this into eq.~\eqref{eq:matching} and solving for the coefficients order by order gives 
\begin{equation}
    \begin{split}
        &C_0=\frac{1}{6L^2\, P} - \frac{1}{4 P^3} + O\left(P^{-5}\right)\,,\\
        & \rho_0 =  P  - \frac{L^2}{2 P} - \frac{169 L^4}{1080 P^3} + O\left(P^{-5}\right)\,.
    \end{split}
    \label{eq:paramser}
\end{equation}
It was necessary to include terms up to $r^{-10}$ and $r^{4}$ in eqs.~\eqref{eq:AdSPS} and~\eqref{eq:flatPS}, respectively, in order to fix these parameters to order $P^{-3}$. The easiest way to see this is to use dimensional analysis to determine what combinations of $C_0$ and $P$ can appear at a given order of $r$ in~\eqref{eq:AdSPS}. Given that the leading term in $C_0$ goes like $(L^2P)^{-1}$, the leading order of $P$ appearing in the coefficient of $r^{- n}$ comes from the term with the highest power of $C_0$ (this term uses as many factors of $(L^2P)^{-1}$ as possible to achieve the right dimensionality, avoiding the alternative factors of $P^{-3}$) . Thus, the leading term is $C_0^{\lfloor \frac{n-1}{3} \rfloor} P^{1-n + 3\lfloor \frac{n-1}{3} \rfloor}\sim P^{1-n+2\lfloor \frac{n-1}{3} \rfloor}$. Finally, it is straightforward to confirm that we need $n> 10$ for the exponent to satisfy ${1-n+2\lfloor \frac{n-1}{3} \rfloor} < -3$. For eq.~\eqref{eq:flatPS}, it is clear that terms of higher order than $r^4$ contain $\rho_0^{-k}$, $k>3$, which, given the form~\eqref{eq:paramser}, only contribute terms of orders below $P^{-3}$. In other words, including terms up to $r^{-10}$ and $r^{4}$ in eqs.~\eqref{eq:AdSPS} and~\eqref{eq:flatPS}, respectively, is required for the matching conditions~\eqref{eq:matching} to fix the coefficients in these series up to order $P^{-3}$.

Why was it important to determine $C_0$ and $\rho_0$ to order $P^{-3}$ in order to obtain the result in eq.~\eqref{eq:AregApprox}? Too see this, recall that we want to use the expansions~\eqref{eq:AdSPS} and~\eqref{eq:flatPS} to evaluate the area functional to up to order $P^{-2}$ and note that the area elements in the AdS region~\eqref{eq:zzzzx} and the flat region~\eqref{eq:yyyyx} both take the following form for the appropriate choice of $a$ and $b$,
\begin{equation}
    r^a \rho(r)^2 \sqrt{1+\left(\frac{r}{L}\right)^{b}\dot\rho(r)^2}\simeq r^a \rho(r)^2 \left(1+\frac{1}{2}\left(\frac{r}{L}\right)^b\dot\rho(r)^2-\frac{1}{8}\left(\frac{r}{L}\right)^{2b}\dot\rho(r)^4\right)+\dots,
    \label{eq:Aelemser}
\end{equation}
where we have used the fact that $\dot\rho$ goes to zero when $P$ is large. The values of $a$ and $b$ are not important here. Keeping terms up to $P^{-3}$ in the expansion of $\rho(r)$ then accomplishes the following: First, since $\rho(r)$ contains terms of order $P^{-m}$, where $m\geq -1 $, it ensures that we are keeping track of all terms up to order $P^{-2}$ coming from the $\rho(r)^2$ factor. Second, since $\dot\rho(r)$ contains terms of order $P^{-l}$, where $l\geq 1 $, it ensures that we are correctly accounting for terms up to order $P^{-4}$ inside the parentheses on the right hand side, coming from powers of $\dot\rho(r)$. It is then straightforward to confirm that this correctly accounts for all terms up to order $P^{-2}$ when multiplying out the right hand side of eq.~\eqref{eq:Aelemser}.

Having carefully kept track of all terms contributing up to order $P^{-2}$ in the area, we can finally evaluate the area functional. Expanding the area functionals~\eqref{eq:zzzzx} and~\eqref{eq:yyyyx} to order $P^{-2}$ using the above equations and integrating gives the final result 
\begin{equation}
\begin{split}
     A\simeq&\, 4\pi^2 L^{8} \int_{L}^{\ruv} dr \,\frac{P^2 r}{L^4}-\left(\frac{L}{2 r}+\frac{L^2}{3 \,r ^3}   - \frac{2 \,L^4}{9\,r^5   } \right)+ \frac{L^2}{P^2}\left(\frac{L^{2}}{8\, r^{3}}
- \frac{2 L^{4}}{3\, r^{5}}
+ \frac{5 L^{6}}{6\, r^{7}}
- \frac{56 L^{8}}{135\, r^{9}}
+ \frac{2 L^{10}}{27\, r^{11}}\right) \\
&+ 4\pi^2 L^{8} \int_{0}^L dr \frac{P^2 r^5}{L^8}-\frac{r^{5}\bigl(18L^{2}-7r^{2}\bigr)}{18\,L^{8}}
\;+\;
\frac{r^{5}\bigl(-68L^{4}+15r^{4}\bigr)}
{1080\,L^{8}\,P^{2}}+\mathcal{O}(P^{-4})\\
=& \,A_{\text{UV}} + 4\pi^4L^8 \left(-\frac{P^2}{3L^2}\, +\frac{1}{2} \log{\frac{P}{L}}- \frac{11}{48}-\frac{61}{3240} \frac{L^2}{P^2}\right)+\mathcal{O}(P^{-4})\,,
\end{split}
\end{equation}
where $A_{\text{UV}}$ is the universal divergence defined in eq.~\eqref{eq:zz22}. Subtracting off this universal divergent piece yields the final result~\eqref{eq:AregApprox}.

\paragraph{$S^1$ entangling surface ($p=2$)} The calculation follows along the same lines as for $p=3$, and is therefore discussed in somewhat less detail. The expansion of the D$2$-brane throat eqation of motion~\eqref{eq:rho-Dp} around $r=\infty$ is 
\begin{equation}
    \rho_{D2}(r)=P - \frac{r_2 ^5}{6\,P \, r^3} + G_0\frac{r_2 ^{10}}{r^5} +\mathcal{O}(r^{-6}),
\end{equation}
where it turns out that expanding to order $r^{-5}$ is sufficient to obtain the first subleading term in the area. The parameter $G_0$ is a free variable, not fixed by the perturbative expansion. This parameter is to be fixed by the matching condition at the shell, where the solution is glued to the perturbative expansion around $r=0$.

Expanding the flat region equation of motion~\eqref{eq:rho-flat-eom} around $r=0$ gives
\begin{equation}
    \rho_{flat}(r)=\rho_0 + \frac{r^2}{14 \rho_0} \left(\frac{r_2}{R}\right)^5+\mathcal{O}(r^4),
\end{equation}
where expanding to second order is sufficient to compute the first subleading term in the area. The parameter $\rho_0$ is to fixed by the gluing condition at the shell. 

The two free parameters are then assumed to take the power series form in eq.~\eqref{eq:paramser0} and solved order by order using the matching condition~\eqref{eq:matching}. The result is
\begin{equation}
    \begin{split}
        &G_0=\frac{R^2 r_2^3}{14\, P}  + O\left(P^{-3}\right)\,,\\
        & \rho_0 =  P  - \frac{r_2^5}{6 R^3 P}  + O\left(P^{-3}\right)\,.
    \end{split}
    \label{eq:paramser}
\end{equation}
Subsequently, substituting the above series expansions into the area functionals in eq.~\eqref{eq:zzzzx} and \eqref{eq:yyyyx} yields
\begin{equation}
\begin{split}
        A \simeq& \frac{32 \pi ^4}{15} r_2^8 \int_R^{\ruv}dr\left[ \frac{P r}{r_2^3}+\frac{r_2^2}{P}\left(-\frac{1}{24 \,r^2}-\frac{3 R^2 }{28\,r^4}+\frac{25 R^4  }{392 \, r^6}\right)\right] \\
        &+\frac{32 \pi ^4}{15} r_2^8 \int_0^{R}dr\,r^6 \left[\frac{P}{R^5 r_2^3 }+\frac{r_2^2}{P} \left(-\frac{1}{6 R^8} +\frac{4 r^2}{49 R^{10}}\right)\right]+\mathcal{O}(P^{-3})\\
         &= A_{\mt{UV}} - \frac{32 \pi ^4}{15} r_2^8\left( \frac{5 R^2 P}{14 r_2 ^3}+\frac{5 r_2^2}{63 R P }\right)+\mathcal{O}(P^{-3}),
\end{split}
\end{equation}
where $A_{\mt{UV}}$ is the divergent contribution defined in eq.~\eqref{CircDiv}.
Finally, subtracting off the UV divergent piece yields the result eq.~\eqref{eq:p2-A-pert}.

\section{Monotonicity of Internal RT Area}
\label{sec:mon}

In this appendix, we comment on the behavior of the area of surfaces with $r_c \geq R$, \ie surfaces that close in the brane region (which we called the AdS region for $p=3$). As already mentioned, these were considered in \cite{Das:2022njy}, where they concluded that  the area is monotonically decreasing in $r_c$ and thus surfaces with large $r_c$ are always favored when there are multiple candidate surfaces for a given $\theta(\ruv)=\thbdy$. We add some more details to this discussion, showing that the area appears to have several saddlepoints with respect to $r_c$, which was not previously noted. That is, the area appears to monotonic, but not strictly monotonic, at least to the accuracy probed by our numerical calculations.  The overall conclusion that surfaces with large $r_c$ are favored still holds.  To explore this behavior, we consider the regime $\ruv \gg r_c$. Expanding~\eqref{eq:DpbraneArea} and making use of~\eqref{eq:Dptheta0}, the area takes the form
\begin{equation}
\begin{split}
        A_\mt{brane}\simeq   &V_p\,\Omega_{7-p} r_p^{\frac{7-p}{2}}\Bigg[\frac{2 \, \ruv ^{\frac{9-p}{2 }}}{9-p}-C_3^p \,r_c^{\frac{9-p}{2}}\\
        &+\frac{(9-p)r_c^{\frac{9-p}{2}} }{16}\bigg[\left((C_1^p)^2 +2 \tfrac{\sqrt{31+2p-p^2}}{(9-p) } C_1^p C_2^p -  (C_2^p)^2 \right)\cos\left(\tfrac{\sqrt{31+2p-p^2}}{2} \log \tfrac{\ruv}{r_c}\right) \\
        & -\left( \tfrac{\sqrt{31+2p-p^2}}{(9-p) }(C_1^p)^2 - 2 C_1^p C_2^p - \tfrac{\sqrt{31+2p-p^2}}{(9-p) } (C_2^p)^2  \right)\sin \left(\tfrac{\sqrt{31+2p-p^2}}{2} \log \tfrac{\ruv}{r_c}\right)\bigg]\Bigg]
        \label{eq:DpA}
\end{split}
\end{equation} 
where the first term is the area of the $\theta=0$ surface in the pure brane geometry. Note that at the first subleading order there is one term that depends on $\ruv$ in an oscillatory manner and one that is proportional to a constant $C_3^p$, which is independent of $\ruv$ and $r_c$. We note that the magnitude of the oscillating term is $\sqrt{(7-p)}((C_1^p)^2+C^2_2)/4$. We then numerically evaluate $C_3^p$ and compare with the amplitude of the oscillating term, see table \ref{tab:Cs}. Since $C_3^p$ is positive and greater in magnitude than the amplitude of the oscillating term, we see that the subleading term is strictly negative overall and thus the area is strictly less than that of the $\theta=0$ surface in this regime. We then differentiate~\eqref{eq:DpA} with respect to $r_c$, yielding
\begin{equation}
\begin{split}
       \frac{d A_\mt{brane}}{dr_c}\simeq &V_p\,\Omega_{7-p} r_p^{\frac{7-p}{2}}\Bigg[\frac{(9-p)C_3^p}{2} r_c^{\frac{7-p}{2}}\\
        & \frac{(5-p)^2 r_c^{\frac{7-p}{2}}}{16}\bigg[-\left((C_1^p)^2 +\tfrac{2(9-p) {\sqrt{31+2p-p^2}}}{(5-p)^2} C_1^p C_2^p -   (C_2^p)^2 \right)\cos\left(\tfrac{\sqrt{31+2p-p^2}}{2} \log \tfrac{\ruv}{r_c}\right) \\
        & + \left(\tfrac{(9-p) \sqrt{31+2p-p^2} }{(5-p)^2} (C_1^p)^2 -2 C_1^p C_2^p - \tfrac{(9-p) \sqrt{31+2p-p^2}}{(5-p)^2} (C_2^p)^2 \right)\sin\left(\tfrac{\sqrt{31+2p-p^2}}{2} \log \tfrac{\ruv}{r_c}\right)\bigg]\Bigg]
        \label{eq:DpdA}
\end{split}
\end{equation}
The amplitude of the oscillatory term takes the form $(7-p)((C_1^p)^2 +(C_2^p)^2)/2$. Comparing with the magnitude of the non-oscillating term, see the 2nd and 4th columns of table \ref{tab:Cs}, we see that they match at least to the 6th decimal place. Therefore, to a high degree of accuracy, there seem to be a large number of approximate local saddlepoints in \(r_c\) -- in the pure brane geometry it would be an infinite number. Note that this seeming equality, if exact, is quite surprising, as $C_3^p$ should naively depend on the full profile of $\theta_0$, not just the behavior near $\theta=0$ captured by~\eqref{eq:Dptheta0}. It would be interesting to develop an analytic argument clarifying why this equality holds or, alternatively, why it does not.

\begin{table}[]
    \centering
    \begin{tabular}{c|c|c|c}
        $p$& $C_3^p $ & $\frac{\sqrt{7-p}}{4}((C_1^p)^2+(C_2^p)^2)$ &$ \frac{7-p}{9-p}((C_1^p)^2+(C_2^p)^2)$\\
        \hline
        $0$& $0.982457$ &$ 0.835501$ &$0.982457$ \\
        $1$&$0.667303$ & $ 0.544851$ & $0.667303$\\
        $2$& $ 0.489632$ &$ 0.383197$& $0.489632$\\
        $3$&$0.385984 $&$ 0.289488$ &$0.385984$ \\
        $4$&$ 0.326864 $& $0.235894$&$0.326864$ \\
    \end{tabular}
    \caption{The constant $C_3^p$ and the amplitude of the oscillating term in~\eqref{eq:DpA} and in~\eqref{eq:DpdA} rounded to the 6th decimal place. Note the match between the second and fourth column.
    }
    \label{tab:Cs}
\end{table}

\section{Internal RT surface as a classical particle}\label{sec:particle}
In this appendix we briefly give a physical intuition for why the surfaces discussed in section \ref{sec:target} converge to the equator as they approach the asymptotic boundary. To see this, we can change to the coordinate $u\equiv  \log r/r_p$, for which the area functional takes the form 
\begin{equation}
    A_{\text{brane}} = V_p\, \Omega_{7-p} \, r_p^{8-p} \int \, e^{\frac{9-p}{2}u} \cos^{7-p}\theta \,\sqrt{d u^2+ d \theta ^2}\,. 
\end{equation}
We can parametrize $u$ and $\theta$ in terms of a fictitious time coordinate $\sigma$. Introducing the einbein $\varepsilon$, the area functional can be put in the form
\begin{equation}
    A_{\text{brane}} = V_p\, \Omega_{7-p} \, r_p^{8-p} \int d\sigma \, \frac{1}{2\varepsilon}\left(\dot u^2+ \dot \theta ^2\right) + \frac{\varepsilon}{2 }  \left(e^{\frac{9-p}{2}u} \cos^{7-p}\theta\right)^2\, .
    \label{actact}
\end{equation}
The action is diffeomorphism invariant\footnote{That is, the action~\eqref{actact} is left unchanged under a transformation of the ``time" coordinate $\tilde{\sigma}=f(\sigma)$ if the einbein transforms $\tilde{\varepsilon} = \tfrac{d\tilde{\sigma}}{d\sigma}\,\varepsilon$.} and so we fix a specific gauge by choosing $\varepsilon$ in a particular way. 
We then pick the convenient gauge with $\varepsilon=1$, for which the area functional~\eqref{actact} takes the form of a particle Lagrangian with the effective potential
\begin{equation}
    U=-\frac{e^{(9-p) u}\cos^{2(7-p)}\theta}{2}
    \label{Ueff}
\end{equation}
and the equations of motion reduce to 
$\ddot u = - \partial_u U$ and $\ddot \theta =- \partial_\theta U$.
Further, the variation of eq.~\eqref{actact} with respect to the einbein yields
 the following constraint 
\begin{equation}
  \dot u^2+\dot \theta^2-  e^{(9-p)u} \cos^{2(7-p)}\theta=0\,,
  \label{eq:zeroE}
\end{equation}
which corresponds to zero total energy for the particle trajectory.

Given this construction, we are examining the motion of the particle in a valley-shaped potential~\eqref{Ueff} that deepens as $u$ increases, \ie as the surface approaches the boundary. Each zero-energy trajectory corresponds to an extremal surface (but we must also impose that the trajectories do not self intersect). 

To understand why all surfaces converge to $\theta=0$ as they approach the boundary, consider a particle rolling down the potential valley towards $u=\infty$. First, note that the particle never climbs up to $\theta=\pm\pi/2$ as it moves outwards, since this would require $\dot u =0$ by~\eqref{eq:zeroE} while $\dot u>0$ and $\ddot u>0$ for an outward falling particle. The question is then whether the particle can oscillate in the $\theta$-direction indefinitely, or if it settles down to $\theta=0$. Physically, there are two competing effects determining if the former option is possible: the particle gains kinetic energy as it falls, but the potential well deepens, making it ever more more difficult to climb up towards large $\theta$ values. To see which effect dominates, consider an oscillating solution which has $\dot \theta=0$ at some set of turning points $\theta_i$. The condition~\eqref{eq:zeroE} gives
\begin{equation}
    \cos^{2(7-p)}\theta_i = \frac{\dot u ^2}{e^{(9-p) u}}.
\end{equation}
Differentiating the right hand side and using the equations of motion and the gauge constraint~\eqref{eq:zeroE} gives the inequality
\begin{equation}
    \frac{d}{d\sigma}\left(\frac{\dot u ^2}{e^{(9-p) u}}\right)=\frac{\dot u }{e^{(9-p) u}}\left(2 \ddot u - (9-p) \dot u ^2\right)= \frac{ (9-p)\dot u \dot \theta^2}{e^{(9-p) u}}\geq0.
\end{equation}
In other words, $\cos^{2(7-p)}\theta_i$ grows between each oscillation, meaning that $\theta_i$ is approaching the equator with each oscillation. In other words, oscillating solutions must settle down toward the equator. 

Note that this argument did not require us to linearize around $\theta=0$. However, knowing that solution approaches $\theta=0$, one can then solve the linearized equations of motion around this point to show the general behavior of the surface as it approaches the boundary. In the pure brane geometry, this behavior is captured by eq.~\eqref{eq:Dptheta0}, which shows that these surfaces oscillate with damped amplitude as they approaches the boundary.

We can briefly compare this to the analogous result in flat space. In this case the equations of motion and the constraint take essentially the same form as above, but with the new potential
\begin{equation}
    U_{flat}=-\frac{e^{2(8-p) u}\cos^{2(7-p)}\theta}{2}\,.
\end{equation}
This potential has a qualitatively similar shape to that of the brane geometry, but the coefficient for $u$ in the exponential is larger. Following the analysis above, one again finds that the particle trajectory converges towards $\theta=0$ as $u$ becomes large. Solving the linearized equations of motion is then straightforward and shows that the solutions approach $\theta=0$ as a power law with real exponent, \ie without oscillations. This is, of course, expected by considering the simple examples in section \ref{sec:start} which correspond to planes at constant $z$.

Finally, we can describe two simple types of surfaces in the language of the free classical particle. One is a particle that starts off sitting at rest at $\theta=\pm \pi/2$ and subsequently rolls down into the potential valley. This corresponds to a surface that closes off in the bulk. In the pure brane geometry these are the surfaces described by $\theta_0$ in sections \ref{sec:target}. In flat space, they correspond to constant-$z$ planes. Another simple example is solutions where the particle comes in from the boundary, scatters off the potential and falls back out towards the boundary. These surfaces are anchored on two lines of latitude at the UV-cutoff. In the pure throat geometry, these surfaces will generally have self-intersections due to the oscillatory behaviour at large $u$ and so are ruled out. However, this can be avoided if the radius where the surface turns around is chosen sufficiently close to the cutoff $\ruv$. In flat space, these are (higher dimensional) catenoids \cite{flatstuff}.

\bibliographystyle{JHEP}
\bibliography{references}

\end{document}